\documentclass[letterpaper,english,american,prr,twocolumn,amsmath,amssymb,superscriptaddress,nofootinbib]{revtex4-2}
\usepackage[T1]{fontenc}
\usepackage[active]{srcltx}
\usepackage{babel}
\usepackage{prettyref}
\usepackage{units}
\usepackage{mathtools}
\usepackage{amsmath}
\usepackage{amssymb}
\usepackage{graphicx}
\usepackage[unicode=true,
 bookmarks=true,bookmarksnumbered=false,bookmarksopen=false,
 breaklinks=false,pdfborder={0 0 1},backref=false,colorlinks=false]
 {hyperref}
\hypersetup{pdftitle={Decomposing neural networks as mappings of correlation functions},
 pdfauthor={Kirsten Fischer, Alexandre Rene, Christian Keup, Moritz Layer, David Dahmen, Moritz Helias},
 colorlinks=true,linkcolor=black,citecolor=black,urlcolor=black,filecolor=black}

\makeatletter

\newcommand{\noun}[1]{\textsc{#1}}

\usepackage[latin1]{inputenc}

\newrefformat{eq}{\hyperref[#1]{Eq.~(\ref{#1})}}
\newrefformat{cap}{\hyperref[#1]{Fig.~\ref{#1}}}
\newrefformat{fig}{\hyperref[#1]{Fig.~\ref{#1}}}
\newrefformat{tab}{\hyperref[#1]{Table ~\ref{#1}}}
\newrefformat{sec}{\hyperref[#1]{Section~\ref{#1}}}
\newrefformat{subsec}{\hyperref[#1]{Section~\ref{#1}}}
\newrefformat{cha}{\hyperref[#1]{Chapter~\ref{#1}}}
\newrefformat{app}{\hyperref[#1]{Appendix~\ref{#1}}}

\usepackage{MnSymbol}
\DeclareMathAlphabet\mathbb{U}{msb}{m}{n}

\usepackage{booktabs}
\usepackage[title]{appendix}
\usepackage{aligned-overset}

\usepackage[force]{feynmp-auto}

\makeatother

\begin{document}
 \renewcommand{\figurename}{FIG.} \renewcommand{\tablename}{TABLE} \renewcommand{\appendixname}{APPENDIX}
\global\long\def\R{\mathbb{R}}%
\global\long\def\llangle{\langle\!\langle}%
\global\long\def\rrangle{\rangle\!\rangle}%
\global\long\def\T{\mathrm{\mathrm{T}}}%
\global\long\def\tran{\mathrm{^{\mathrm{T}}}}%
\global\long\def\lowerindex{\vphantom{X^{N}}}%
\global\long\def\extralowerindex{\vphantom{X^{N^{N}}}}%
\foreignlanguage{english}{}
\global\long\def\order{\mathcal{O}}%
\foreignlanguage{english}{}
\global\long\def\N{\mathcal{N}}%
\foreignlanguage{english}{}
\global\long\def\tr{\mathrm{tr}}%

\title{Decomposing neural networks as mappings of correlation functions}
\author{Kirsten Fischer}
\email{ki.fischer@fz-juelich.de}

\affiliation{Institute of Neuroscience and Medicine (INM-6) and Institute for Advanced
Simulation (IAS-6) and JARA-Institute Brain Structure-Function Relationships
(INM-10), Jülich Research Centre, Jülich, Germany}
\affiliation{RWTH Aachen University, Aachen, Germany}
\author{Alexandre Ren\'e}
\affiliation{Institute of Neuroscience and Medicine (INM-6) and Institute for Advanced
Simulation (IAS-6) and JARA-Institute Brain Structure-Function Relationships
(INM-10), Jülich Research Centre, Jülich, Germany}
\affiliation{Department of Physics, University of Ottawa, Ottawa, Canada}
\affiliation{Department of Physics, Faculty 1, RWTH Aachen University, Aachen,
Germany}
\author{Christian Keup}
\affiliation{Institute of Neuroscience and Medicine (INM-6) and Institute for Advanced
Simulation (IAS-6) and JARA-Institute Brain Structure-Function Relationships
(INM-10), Jülich Research Centre, Jülich, Germany}
\affiliation{RWTH Aachen University, Aachen, Germany}
\author{Moritz Layer}
\affiliation{Institute of Neuroscience and Medicine (INM-6) and Institute for Advanced
Simulation (IAS-6) and JARA-Institute Brain Structure-Function Relationships
(INM-10), Jülich Research Centre, Jülich, Germany}
\affiliation{RWTH Aachen University, Aachen, Germany}
\author{David Dahmen}
\affiliation{Institute of Neuroscience and Medicine (INM-6) and Institute for Advanced
Simulation (IAS-6) and JARA-Institute Brain Structure-Function Relationships
(INM-10), Jülich Research Centre, Jülich, Germany}
\author{Moritz Helias}
\affiliation{Institute of Neuroscience and Medicine (INM-6) and Institute for Advanced
Simulation (IAS-6) and JARA-Institute Brain Structure-Function Relationships
(INM-10), Jülich Research Centre, Jülich, Germany}
\affiliation{Department of Physics, Faculty 1, RWTH Aachen University, Aachen,
Germany}
\date{\today}
\begin{abstract}

Understanding the functional principles of information processing
in deep neural networks continues to be a challenge, in particular
for networks with trained and thus non-random weights. To address
this issue, we study the mapping between probability distributions
implemented by a deep feed-forward network. We characterize this
mapping as an iterated transformation of distributions, where the
non-linearity in each layer transfers information between different
orders of correlation functions. This allows us to identify essential
statistics in the data, as well as different information representations
that can be used by neural networks. Applied to an XOR task and to
MNIST, we show that correlations up to second order predominantly
capture the information processing in the internal layers, while the
input layer also extracts higher-order correlations from the data.
This analysis provides a quantitative and explainable perspective
on classification.
\end{abstract}
\maketitle

\section{Introduction\label{sec:introduction}}

Recent years have shown a great success of deep neural networks in
solving a wide range of tasks, from image recognition \citep{Krizhevsky12_1097}
to playing Go \citep{Silver16_484}. One major branch is supervised
learning, where input-output mappings are learned from examples. In
many common problems the target output values are given by a finite
set, defining a classification task \citep{Bishop06}. The objective
then is to minimize an error measure between the correct class label
and the prediction made by the neural network with respect to the
joint probability distribution of data samples and class labels \citep{Bahri20_501}.
Thus, training dynamics, and consequently the solution strategy implemented
by the network, depend on this probability distribution and the information
it encodes. In this view, a network implements a transformation of
the input distribution with the objective to concentrate the output
distribution around the assigned target values for each class. How
such a transformation is achieved and how the network training depends
on the statistics of the presented data is, however, still mostly
unknown.

To render the decision-making process of neural networks transparent,
a profound understanding regarding their functional principles and
extraction of meaningful features from given data is required. Over
the past years the discrepancy between success in applications and
limited understanding has lead to an increased interest also in the
theoretical community \citep{Lin17_1223,Shwartz-Ziv17_arxiv,Jacot18_8580,Saxe19_11537,Bahri20_501,Cohen21_023034}.
An important line of theoretical research investigates ensembles of
neural networks in the limit of infinite width for which the central
limit theorem implies an exact equivalence to Gaussian processes (GPs)
\citep{Neal96,Williams98_1203,Lee18,Garriga19}. While this approach
is informative with respect to how relations between data samples
are transformed by the network, it does not reveal how the internal
structure of data samples is processed. As an example, for image
classification the Gaussian process view takes into account the relation
between all corresponding pixels $x_{i}^{\alpha},x_{i}^{\beta}$ of
any pair of images $\alpha,\beta$ in the form of a scalar product
$\sum_{i}x_{i}^{\alpha}x_{i}^{\beta}$. Even though the data statistics
shape the eigenfunctions of the GP's covariance matrix \citep[Sec. 4.3]{WilliamsRasmussen06},
it is not obvious which role is played by the structure within individual
images determined by e.g. correlations between pixel values $x_{i}^{\alpha}$
and $x_{j}^{\alpha}$. In particular, due to the rotational symmetry
of the scalar product, the GP view gives identical results when image
pixels are shuffled consistently across all images. However, clearly
the internal structure of data samples also contains important information
that may be employed to solve a given task. The focus of the present
study is to investigate how this information is extracted from the
data and utilized by the network to perform classification.

Other approaches \citep{Poole16_3360,Raghu17_2847,Schoenholz17},
similarly as GPs, focus on ensembles of networks with randomly drawn
weights. In contrast, here we study how particular realizations of
trained and untrained networks process different statistical features
of the data. Thus, we shift the perspective from distributions over
network parameters to distributions over the data. In particular,
we describe the input-output mapping implemented by deep neural networks
in terms of correlation functions. To trace the transformation of
correlation functions across layers of neural networks, we make use
of methods from statistical field theory \citep{Kleinert89,ZinnJustin96,Hertz16_033001,Helias20_970}
and obtain recursive relations in a perturbative manner by means of
Feynman diagrams. Our results yield a characterization of the network
as a non-linear mapping of correlation functions, where each layer
exchanges information between different statistical orders. Re-expressing
the loss function in terms of data correlations allows us to study
their role in the training process, and to link the transformation
of data correlations to the solution strategies found by the network.
For the particular example of the mean squared error loss function,
we show that network training relies exclusively on the first two
cumulants of the output (mean and covariance), while these, in turn,
are predominantly determined by means and covariances of network activations
in previous layers. Furthermore, we show that corrections from higher-order
correlations to mean and covariance, which are readily computable
with the proposed generic field-theoretical framework, are of greatest
importance in the first layer, where these corrections effectuate
the information flow from higher-order correlations to mean and covariance.

The structure of this study is as follows: \prettyref{sec:Theoretical-Background}
provides theoretical background on the definition and architecture
of deep neural networks (\prettyref{subsec:notation_network_arch}),
on empirical risk minimization in the context of classification (\prettyref{subsec:learning_theory}),
and on field-theoretical descriptions of probability distributions
in terms of cumulants and their generating function (\prettyref{subsec:Parameterizations-of-probability}).
In \prettyref{sec:theory_decomp_network} we decompose the network
mapping into correlation functions, tracing their transformations
backwards through the network. We start by relating the loss to the
first- and second-order correlations of the network outputs (\prettyref{subsec:loss_reformulation}),
then discuss the mapping of correlations by individual hidden layers
(\prettyref{subsec:trafo_data_statistics}), and end with the extraction
of data correlations by the input layer (\prettyref{subsec:training_data_statistics}).
\prettyref{sec:exp_results} applies these theoretical tools to several
example data sets. We start with an adaptation of the XOR problem,
where the input statistics are fully known and selectively presented
to the network to study different encoding and processing schemes
of class identities (\prettyref{subsec:Multiple-encodings-of}). We
proceed with an application to the MNIST data set \citep{lecun2010mnist},
where we show that classification performance is largely based on
the transformation of means and covariances across layers (\prettyref{subsec:mnist_correlations}).
Finally, we showcase the importance of higher-order correlations and
their extraction in the input layer by constructing a data set where
information on class identity is only encoded in correlations of third
and higher order (\prettyref{subsec:problem_alternating_hills}).
In \prettyref{sec:conclusion} we discuss our results and provide
an outlook.

\section{Theoretical Background\label{sec:Theoretical-Background}}

\subsection{Feed-forward network architecture\label{subsec:notation_network_arch}}

\begin{figure}
\centering{}\includegraphics{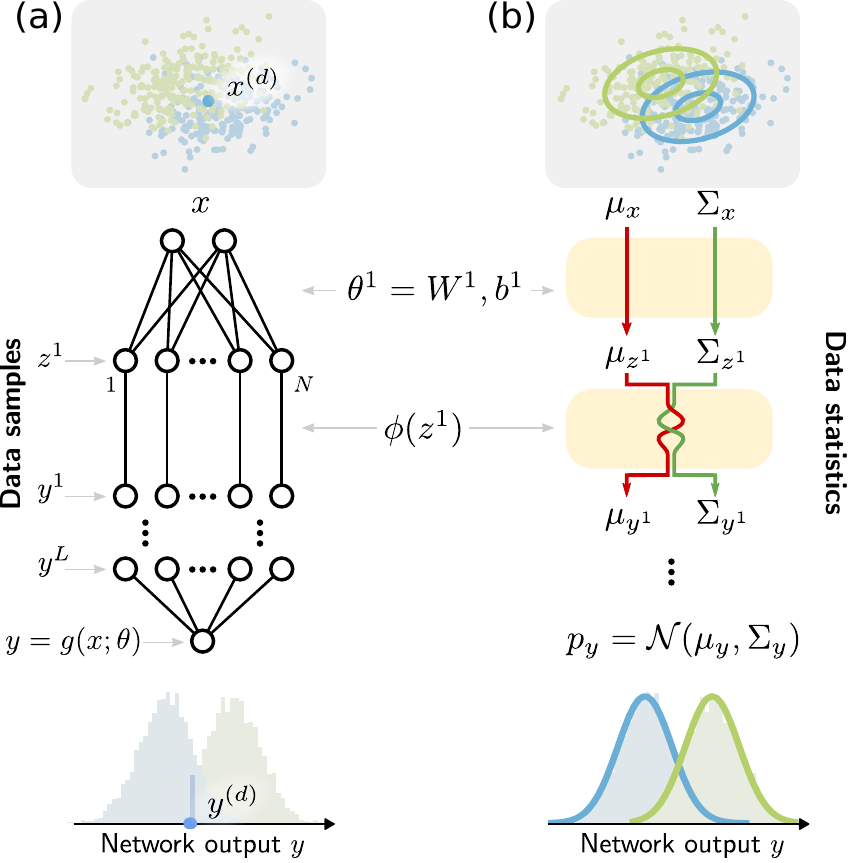}\caption{\textbf{(a)} \textbf{Network analysis based on data samples }considers
each data sample $x^{(d)}$ separately as it passes through the network,
producing a single corresponding output $y^{(d)}$. Each layer consists
of an affine transformation ($W^{l},b^{l})$ followed by a non-linearity
$\phi$ applied componentwise. \textbf{(b) Network analysis based
on data statistics }considers how the entire data distribution $p(x)$
is transformed by the network. At each step, the intermediate distribution
is parameterized by its cumulants, the most important of which are
the mean $\mu$ and the covariance $\Sigma$. The affine step transforms
$\mu$ and $\Sigma$ independently, while the non-linearity $\phi$
causes a non-trivial interaction of the two.\label{fig:Schematic-representation}}
\end{figure}

We consider fully-connected neural networks with $L$ layers of $N_{l}$
neurons each, and one additional linear readout layer, as shown in
\prettyref{fig:Schematic-representation}\textbf{a}.
 Each layer $l=1,\ldots,L$ consists of an affine transformation 
\begin{equation}
z_{i}^{l}=\sum_{j=1}^{N_{l-1}}\,W_{ij}^{l}\,y_{j}^{l-1}+b_{i}^{l}\label{eq:pre_activations}
\end{equation}
parameterized by a weight matrix $W^{l}\in\R^{N_{l}\times N_{l-1}}$
and bias vector $b^{l}\in\R^{N_{l}}$. This step is followed by the
pointwise application of a non-linear activation function $\phi$,
yielding
\begin{equation}
y_{i}^{l}=\phi\big(z_{i}^{l}\big)=\phi\left(\sum_{j=1}^{N_{l-1}}\,W_{ij}^{l}\,y_{j}^{l-1}+b_{i}^{l}\right).\label{eq:post_activations}
\end{equation}
Here $y^{0}=x\in\R^{N_{0}}$ denotes the input data of dimension
$N_{0}$. The readout layer produces the network output $y\in\R^{d_{\text{out}}}$,
specifically $y_{i}=z_{i}^{L+1}$. The network mapping $y=g(x;\theta)$
is given by iterating over network layers and characterized by parameters
$\theta:=\{W^{l},b^{l}\}_{l=1,\dots,L+1}$. 

We initialize all network parameters randomly from i.i.d. centered
Gaussians $\mbox{\ensuremath{W_{ij}^{l}\overset{\mathrm{i.i.d.}}{\sim}}\ensuremath{\mathcal{N}\left(0,\nicefrac{\sigma_{w}^{2}}{N_{l-1}}\right)}}$
and $b_{i}^{l}\overset{\mathrm{i.i.d.}}{\sim}\mathcal{N}\left(0,\sigma_{b}^{2}\right)$
before training. The scaling of the variance $\nicefrac{\sigma_{w}^{2}}{N_{l-1}}$
is chosen such that the covariance of $z^{l}$ (see \prettyref{eq:pre_activations})
is independent of the layer width.

\subsection{Learning theory: empirical risk minimization\label{subsec:learning_theory}}

The fundamental assumption underlying classification\footnote{Since this work focuses on classification tasks, we tailor the presentation
of empirical risk minimization to that context.} is the existence of a joint distribution $p(x,t)$ of data samples
$x$ and class labels $t$ that is the same for training and evaluation
\citep{Bishop06}. By Bayes' theorem, the distribution of the input
data can be treated as a mixture model $p(x)=\sum_{t}p(t)\,p(x\vert t)\,.$
The network's task is then to implement a mapping $g:x\mapsto y$
that minimizes the expectation of a loss $\ell(y,t)$ between the
network outputs $y=g(x;\theta)$ and the labels $t$.

This mapping, in turn, induces a mapping of the probability distributions
\begin{equation}
p(x|t)\mapsto p(y|t;\theta)=\int\delta(y-g(x;\theta))\,p(x\vert t)\,dx\label{eq:p_in-p_out}
\end{equation}
for each label $t$, where $\delta(.)$ refers to the Dirac delta
distribution. The unconditioned output distribution is then the weighted
sum $p(y)=\sum_{t}p(t)\,p(y\vert t;\theta).$ Ideally, the network
output $y$ matches the true label $t$ so that the target distribution
is given by $p(y\vert t)=\delta(y-t)$.

Training algorithms seek to minimize the expected loss or risk functional
\citep{Vapnik92_831}
\begin{align}
R(\theta) & =\langle\ell(y,t)\rangle_{y|\theta}=\sum_{t}p(t)\,\langle\ell(y,t)\rangle_{y\vert t;\theta}\,,\label{eq:loss_generic_class_expectations}
\end{align}
where the expectation value $\langle\cdot\rangle_{y\vert t;\theta}$
is taken with regard to the class-conditional output distributions
$p(y|t;\theta)$. In general, neither the mixture components of the
input distribution $p(x\vert t)$ nor the induced class-conditional
output distributions $p(y|t;\theta)$ are known. Instead, the expected
loss is replaced by the empirical loss or risk
\begin{equation}
R_{\text{emp}}(\theta)=\frac{1}{D}\sum_{d=1}^{D}\,\ell(g(x^{(d)};\theta),t^{(d)})\,,\label{eq:loss_emp}
\end{equation}
evaluated for a training set $\{(x^{(d)},t^{(d)})\}_{d}$, with $D$
being its size and $d$ the respective sample index. The \emph{empirical
risk minimization principle} then assumes the following: the mapping
$g(\,\cdot\,;\theta^{\ast})$ that minimizes the empirical risk $\theta^{\ast}=\text{argmin}_{\theta}\,R_{\text{emp}}(\theta)$
yields an expected risk $R(\theta^{\ast})$ that is close to its minimum
$\min_{\theta}R(\theta)$ \citep{Vapnik92_831}.

\subsection{Parameterization of probability distributions in terms of cumulants\label{subsec:Parameterizations-of-probability}}

This section contains the framework to track data correlations (cumulants)
of arbitrary order through the network. Large parts of the main text
deal with the first two orders, the Gaussian approximation. Readers
who want to obtain an overview of the main results may skip the remainder
of this section at first read. This part, however, becomes essential
when including non-Gaussian corrections and as a means to obtain an
intuitive picture of how non-linear transformations couple the different
statistical orders.

Neural networks can be regarded as complex systems that generate many
interactions between data components. A common approach to investigate
such systems is by studying generating functions of moments or cumulants
rather than the probability distributions themselves. Cumulants often
provide a more convenient parameterization of probability distributions
as they are additive with respect to addition of independent variables,
leading to simpler expressions for the transformation of statistics
across layers. Such approaches are common in statistical physics and
in mathematical statistics.

The network mapping $g:x\mapsto y$ relates the cumulant generating
function of network outputs $y$ to the statistics of the input $x$:
\begin{align}
\mathcal{W}_{y|t;\theta}(j) & =\ln\;\left\langle \exp\big(j^{\mathsf{T}}y\big)\right\rangle _{y|t;\theta}\\
 & =\ln\;\left\langle \exp\Big(j^{\mathsf{T}}g(x;\theta)\Big)\right\rangle _{x|t}.\label{eq:cum_gen_func}
\end{align}
The cumulant generating function is considered per class $t$ as the
data statistics are expected to differ between classes. The class-conditional
output cumulant of order $n$ denoted by $G_{y\vert t;\theta}^{(n)}$
is then defined as
\[
G_{y\vert t;\theta}^{(n)}=\left.\frac{d^{n}\mathcal{W}_{y|t;\theta}(j)}{dj^{n}}\right|_{j=0}\,.
\]

Evaluating \prettyref{eq:cum_gen_func} would in principle allow
one to relate $G_{y\vert t;\theta}^{(n)}$ to the input cumulants
$G_{x\vert t;\theta}^{(n^{\prime})}$. However, one intricacy is that
the network mapping $g(x;\theta)$ is given via the iterations in
\prettyref{eq:post_activations}. Their iterative nonlinear nature
makes deep neural networks powerful as universal function approximators,
but complicates their analysis in terms of data processing. Yet, we
can study the transformation of cumulants from input to output by
considering layers individually.

Since pre-activations $z^{l}$ are determined by affine linear transformations,
the cumulant generating function of pre-activations $z^{l}$ in layer
$l$ is trivially related to the cumulant generating function of post-activations
$y^{l-1}$ of layer $l-1$ as 
\begin{align}
\mathcal{W}_{z^{l}}(j) & =\ln\;\left\langle \exp\big(j^{\mathsf{T}}z^{l}\big)\right\rangle _{z^{l}}\nonumber \\
 & =\ln\;\left\langle \exp\big(j^{\mathsf{T}}W^{l}\,y^{l-1}+j^{\mathsf{T}}b^{l}\big)\right\rangle _{y^{l-1}}\nonumber \\
 & =\mathcal{W}_{y^{l-1}}\big(\big(W^{l}\big)^{\mathsf{T}}j\big)+j^{\mathsf{T}}b^{l}\,,\label{eq:cum_affine}
\end{align}
yielding for the first order cumulant ($n=1$)
\begin{equation}
G_{z^{l}}^{(1)}=W^{l}\,G_{y^{l-1}}^{(1)}+b^{l}\,,
\end{equation}
and for second- and higher-order cumulants ($n\ge2$)
\begin{equation}
G_{z^{l},\,\left(r_{1},\dots,r_{n}\right)}^{(n)}=\sum\limits _{s_{1},\dots,s_{n}}W_{r_{1}\,s_{1}}^{l}\dots W_{r_{n}\,s_{n}}^{l}\,G_{y^{l-1},\,\left(s_{1},\dots,s_{n}\right)}^{(n)}\,.\label{eq:cum_tensor_trafo}
\end{equation}
Each index $s_{i}$ is hence contracted with one factor $W_{r_{k}s_{i}}^{l}$
to produce the index $r_{k}$ of the resulting cumulant. Consequently,
cumulants of pre-activations $z^{l}$ are linear tensor transformations
of cumulants of post-activations $y^{l-1}$ of the same order.
\begin{table}
\centering{}\begin{tabular}{l@{\hskip 10pt}l@{\hskip 10pt}l}
\midrule\midrule
Meaning & \begin{tabular}{@{}l} Algebraic \\ term \end{tabular} & \begin{tabular}{@{}l} Graphical \\ representation \end{tabular}\\
\midrule
External line & $j_r \, \delta_{rs}$ &
\raisebox{-.5\height}{
\begin{fmffile}{ext_line}
\fmfset{decor_size}{4mm}
\begin{fmfgraph*}(20, 20)
\fmfstraight
\fmftopn{l}{2}
\fmffreeze
\fmfshift{(0,-.6h)}{l1,l2}
\fmfshift{(-.2w, 0.)}{l1}
\fmfshift{(-.1w, 0.)}{l2}
\fmf{plain}{l1,l2}
\fmfv{label=$j_r$, l.a=90, l.d=.1w}{l1}
\fmfv{label=$z^l_s$, l.a=90, l.d=.1w}{l2}
\end{fmfgraph*}
\end{fmffile}}
\\[10pt]

\begin{tabular}{@{}l} Cumulant vertex \\ with $n$ internal lines \end{tabular} & $G_{z^l,\,(r_{1},\ldots,\,r_{n})}^{(n)}$ & 
\raisebox{-.2\height}{
\begin{fmffile}{gen_cum}
\fmfset{decor_size}{4mm}
\parbox[c][60pt][c]{50pt}{
\begin{fmfgraph*}(20, 20)
\fmfstraight
\fmftopn{l}{1}
\fmfrightn{r}{2}
\fmfbottom{b1}
\fmfleft{t1}
\fmffreeze
\fmfshift{(-.5w, .2h)}{l1,r1,r2,b1,t1}
\fmfshift{(-.5w,-.5h)}{l1}
\fmfshift{(-.5w,-.4h)}{r1}
\fmfshift{(-.2w,-.2h)}{r2}
\fmfshift{(.6w,-.6h)}{t1}
\fmfshift{(-1.w,-0.2h)}{b1}
\fmf{plain}{l1,b1}
\fmf{plain}{l1,r2}
\fmf{plain}{l1,t1}
\fmfv{d.s=circle, d.filled=empty}{l1}
\fmfv{label=$\ldots$, l.a=60, l.d=0.05}{r1}
\fmfv{label=$z^l_{r_1}$, l.a=60, l.d=0.05}{r2}
\fmfv{label=$z^l_{r_n}$, l.a=-30, l.d=0.1w}{b1}
\end{fmfgraph*}}
\end{fmffile}}
\\[10pt]

\begin{tabular}{@{}l} $\phi$-vertex with \\ $m$ internal lines and\\ one external line \end{tabular} & \begin{tabular}{@{}l} $j_r \,\frac{1}{m!} \,\phi^{(m)}\big|_{z^{l}=0}$ \\ $\quad \times \, \delta_{ri_1}\, \dots \, \delta_{ri_m}$ \end{tabular} & 
\raisebox{-0.1\height}{
\begin{fmffile}{phi_vertex}
\fmfset{decor_size}{4mm}
\parbox[c][60pt][c]{50pt}{
\begin{fmfgraph*}(20, 20)
\fmfstraight
\fmftopn{l}{2}
\fmfrightn{r}{2}
\fmfleftn{t}{1}
\fmffreeze
\fmfshift{(-.5w,.2h)}{l1,l2,r1,r2,t1}
\fmfshift{(0.,.5h)}{l1,l2,r1,r2}
\fmfshift{(-.1w,0)}{l1}
\fmfshift{(0.,-1.h)}{l1,l2}
\fmfshift{(-.75w,0.)}{l2}
\fmfshift{(-.5w,0)}{l1}
\fmfshift{(0,-0.5h)}{r1}
\fmfshift{(0,-.5h)}{r2}
\fmfshift{(1.4w,-0.1h)}{t1}
\fmf{plain}{l1,l2}
\fmf{plain}{l2,r1}
\fmf{plain}{l2,r2}
\fmfv{d.s=circle, d.filled=shaded}{l2}
\fmfv{label=$j_r$, l.a=90, l.d=.1w}{l1}
\fmfv{label=$z^l_{i_m}$, l.a=-10, l.d=.05w}{r1}
\fmfv{label=$z^l_{i_1}$, l.a=60, l.d=.05w}{r2}
\fmfv{label=$\dots$, l.a=90, l.d=.1w}{t1}
\end{fmfgraph*}}
\end{fmffile}}\\[-10pt]
\midrule\midrule
\end{tabular}\caption{Diagrammatic elements for the perturbative expansion of $\mathcal{W}_{y^{l}}\left(j\right)$
for $y^{l}=\phi(z^{l})$.\label{tab:feyn_rules}}
\end{table}

The non-linear activation function $\phi$ in each layer $l$ then
relates the pre-activations $z^{l}$ to the corresponding post-activations
$y^{l}$:
\begin{align}
\mathcal{W}_{y^{l}}(j) & =\ln\;\left\langle \exp\big(j^{\mathsf{T}}y^{l}\big)\right\rangle _{y^{l}}\nonumber \\
 & =\ln\;\left\langle \exp\big(j^{\mathsf{T}}\phi(z^{l})\big)\right\rangle _{z^{l}}\,.\label{eq:W_y^l}
\end{align}

This cumulant generating function of the post-activations $y^{l}$
cannot, in general, be computed exactly. One common approximation
technique is a perturbative expansion \citep{Helias20_970}, which
we here recast in the following way: by replacing $\phi(z^{l})$ with
its Taylor expansion $\sum_{m}\frac{\phi^{(m)}\vert_{z^{l}=0}}{m!}(z^{l})^{m}$
in \prettyref{eq:W_y^l} and treating nonlinear terms ($m>1$) as
perturbations, we can construct cumulants $G_{y^{l}}^{(n)}$ as series
of Feynman diagrams composed of the graphical elements shown in \prettyref{tab:feyn_rules}.
For example, for the mean of the first layer we get the following
diagrams:\foreignlanguage{english}{\begin{fmffile}{ex_diag_rules}
\fmfset{thin}{0.75pt}
\fmfset{decor_size}{4mm}
\begin{equation}
	\label{eq:examples_diagrams}
	\begin{tabular}{ccccccc}
		$G^{(1)}_{y^1,\,i}$ 
					&$=$ &\parbox{20pt}{
							\begin{fmfgraph*}(35, 20)
							\fmfstraight
							\fmftopn{l}{3}
							\fmffreeze
							\fmfshift{(-0.35w,-.5h)}{l1,l2,l3}
							\fmfshift{(-0.05w,0)}{l2}
							\fmf{plain}{l1,l2}
							\fmf{plain}{l2,l3}
							\fmfv{decor.shape=circle, decor.filled=shaded}{l2}
							\fmfv{decor.shape=circle, decor.filled=empty}{l3}
							\end{fmfgraph*}
						} &$+$
						&\parbox{20pt}{
							\begin{fmfgraph*}(20, 20)
							\fmfstraight
							\fmftopn{l}{3}
							\fmffreeze
							\fmfshift{(-.5w.,-.5h)}{l1,l2,l3}
							\fmfshift{(-.3w,0)}{l1}
							\fmfshift{(.5w,0)}{l3}
							\fmf{plain}{l1,l2}
							\fmf{plain,tension=0.8,right=1.}{l2,l3}
							\fmf{plain,tension=0.8,right=1.}{l3,l2}
							\fmfv{d.s=circle, d.filled=shaded}{l2}
							\fmfv{d.s=circle, d.filled=empty}{l3}
							\end{fmfgraph*}
						} & $+$ & $\dots$\\
					\phantom{$G^{(1)}_{z,\,i}$} & &\phantom{$G^{(1)}_{z,\,i} \qquad$}
					& &\phantom{$\quad\frac{\phi^{(2)}}{2!}\,G^{(2)}_{z,\,ii}$}
					& &\\[-5pt]
			&$=$ &$\frac{\phi^{(1)}\big|_{x=0}}{1!} \, G^{(1)}_{x,\,i}$ &$+$ &$\frac{\phi^{(2)}\big|_{x=0}}{2!}\,G^{(2)}_{x,\,ii}$ & $+$ & $\dots$
	\end{tabular}
\end{equation}
\end{fmffile}
}We find that in general these expressions involve two types of factors,
which we represent with two types of vertices: empty circles with
internal lines, representing cumulants $G_{z^{l}}^{(n)}$ of pre-activations
$z^{l}$, and hatched circles with one external line $j$ that stem
from Taylor coefficients $\frac{\phi^{(m)}\big|_{z^{l}=0}}{m!}$ of
the non-linearity.

For constructing a cumulant $G_{y^{l}}^{(n)}$ of the post-activations
$y^{l}$ of order $n$, we need to determine all diagrams with $n$
external lines. External lines occur on cumulant vertices as well
as on hatched vertices. Furthermore, they always need to be connected
to a cumulant vertex, but cannot be connected to one another. Finally,
due to the \emph{linked cluster theorem}, only connected diagrams
need to be considered, since others do not contribute to cumulants.
When evaluating the generated diagrams, all permutations of indices
$(r_{1},\dots,r_{n})$ for both internal and external lines need to
be taken into account. Symmetries within diagrams result in their
repeated occurrence, which is reflected in combinatorial pre-factors
(for more details, see \citep{Helias20_970}).

Using this perturbative approach for determining the cumulants $G_{y^{l}}^{(n)}$
of the post-activations $y^{l}$ has two main advantages: First, it
provides a principled way to go beyond Gaussian statistics and include
higher-order cumulants. Second, the availability of a diagrammatic
language allows us to graphically represent the information transfer
from cumulants $G_{z^{l}}^{(n)}$ of the pre-activations $z^{l}$
to cumulants $G_{y^{l}}^{(m)}$ of the post-activations $y^{l}$.

The diagrammatic representation introduced above assumes that the
activation function $\phi$ can be expanded as a Taylor series. For
non-differentiable functions such as $\text{ReLU}$, this approach
can be adapted by using a Gram-Charlier expansion of the probability
distribution $p(z^{l})$. The expectation value in \prettyref{eq:W_y^l}
then becomes a sum of Gaussian integrals, which can be calculated
either analytically (see \prettyref{app:stats_trafos} for $\text{ReLU}$
as an example) or numerically.

\section{Decomposing deep neural networks into correlation functions\label{sec:theory_decomp_network}}

Analyzing how deep networks process data is difficult due to their
iterative, parameter-dependent definition. Statistical learning theory
studies the expected error \citep{Vapnik98_learningtheory}, thus
shifting from the transformation of data samples to that of data distributions.
We follow this idea here by studying how data correlations of the
input are iteratively transformed by deep networks, as illustrated
in \prettyref{fig:Schematic-representation}, and how they shape the
expected loss.

\subsection{Data correlations drive network training\label{subsec:loss_reformulation}}

We here discuss the dependence of the expected loss in \prettyref{eq:loss_generic_class_expectations}
on the data correlations. In general, the expected risk $R(\theta)$
is a function of the class labels $t$ and the class-conditional cumulants
$G_{y\vert t;\theta}^{(n)}$ of arbitrary orders $n$:
\begin{align*}
R(\theta) & =\sum_{t}\int\mathrm{d}y\,\ell(y,t)\,p(y\vert t;\theta)\\
 & \eqqcolon\sum_{t}\sigma_{t}(\{G_{y\vert t;\theta}^{(n)}\}_{n})\\
 & \eqqcolon\sigma(\{\{G_{y\vert t;\theta}^{(n)}\}_{n};t\}_{t})\,.
\end{align*}
However, for the often employed mean squared error $\ell_{\text{MSE}}(y,t)=\Vert y-t\Vert^{2}$,
$R(\theta)$ only depends on the mean $\mu_{y}^{t}$ and variance
$\Sigma_{y}^{t}$ of outputs of each class $t$ as
\begin{equation}
R_{\text{MSE}}(\{\mu_{y}^{t},\Sigma_{y}^{t};t\}_{t})=\sum_{t}p(t)\,\big(\mathrm{tr}\,\Sigma_{y}^{t}+\Vert\mu_{y}^{t}-t\Vert^{2}\big).\label{eq:loss_stat_mse}
\end{equation}
Training therefore aims to match class means and labels, while minimizing
the variance of each class's output.\footnote{\prettyref{eq:loss_stat_mse} should not be confused with the bias-variance
decomposition \citep{Kohavi96_275}, where the expectation over finite
datasets of fixed size is taken instead of the expectation over the
input distribution $p(x)$ itself.}

In this case, the first- and second-order cumulants (mean and covariance)
of the last layer alone drive network training, thus singling these
out as the relevant statistics. This result has two implications:
1.) In deep feed-forward networks, only non-Gaussian statistics that
appear in network layers before the final layer can contribute to
the learned information processing by influencing the first two cumulants
in the final layer. 2.) If networks produce non-Gaussian statistics
in the final layer, these do not serve a functional role per se; rather
they may arise as a by-product of earlier layers operating on higher-order
statistics.

Thus, understanding the network mapping reduces to understanding how
the Gaussian statistics $(\mu_{y}^{t},\Sigma_{y}^{t})$ of the output
arise from the presented data distribution across multiple network
layers. Network training and the resulting information processing
within the network is therefore directly linked to how data correlations
are transformed by the network.

\subsection{Propagation of data correlations within the network\label{subsec:trafo_data_statistics}}

To understand how the extraction of information from the input and
its internal processing shape the first- and second-order cumulants
of the output, we follow these two quantities backwards through the
network. According to \prettyref{eq:cum_affine}-\prettyref{eq:cum_tensor_trafo},
the affine transformation in each layer implies for the pre-activations
$z^{l}$:
\begin{align}
\mu_{z^{l}} & =W^{l}\,\mu_{y^{l-1}}+b^{l}\,,\quad\Sigma_{z^{l}}=W^{l}\,\Sigma_{y^{l-1}}(W^{l})^{\mathsf{T}},\label{eq:stat_pre_act}
\end{align}
showing that the two quantities are transformed independently of each
other in this step (\prettyref{fig:Schematic-representation}).

In general, the non-linear activation function $\phi:z^{l}\mapsto y^{l}$
makes the statistics of the post-activations $y^{l}$ dependent on
cumulants of arbitrary orders in the pre-activations $z^{l}$ through
(cf. \prettyref{subsec:Parameterizations-of-probability})
\begin{align}
\mu_{y^{l}} & =\left.\frac{d\mathcal{W}_{y^{l}}(j)}{dj}\right|_{j=0}=\langle\phi(z^{l})\rangle_{z^{l}}\,,\label{eq:mu_y}\\
\Sigma_{y^{l}} & =\left.\frac{d^{2}\mathcal{W}_{y^{l}}(j)}{dj\,dj^{\mathsf{T}}}\right|_{j=0}=\langle\phi(z^{l})\,\phi(z^{l})^{\mathsf{T}}\rangle_{z^{l}}-\mu_{y^{l}}\mu_{y^{l}}^{\mathsf{T}}.\label{eq:sigma_y}
\end{align}
However, due to the central limit theorem, initializing the weights
independently causes the affine transformation $y^{l-1}\mapsto z^{l}$
to mainly pass on the Gaussian part of the statistics, since higher-order
cumulants $G_{z^{l},\,\left(i_{1},\dots,i_{n}\right)}^{(n)}=\llangle z_{i_{1}}^{l}z_{i_{2}}^{l}\ldots z_{i_{n}}^{l}\rrangle\sim\mathcal{O}((N_{l-1})^{-\frac{n}{2}+1})$
are suppressed by the layer width $N_{l-1}$ for $n>2$. In \prettyref{app:Weakly-correlated}
we derive sufficient conditions under which the Gaussian approximation
remains valid also for wide trained networks. In brief, we find that
it suffices to have a natural scaling of weights $w\sim\order(N^{-\frac{1}{2}})$
as well as an approximate orthogonal decomposition of the sending
layer's covariance matrix by the row vectors of the connectivity to
the next layer, \prettyref{eq:weak_corr_pre-act}. These conditions
are in particular different from those of the lazy (kernel or neural
tangent kernel) regimes, where weights only change marginally. Under
these conditions, in the limit of infinitely wide networks, expectations
over pre-activations $\langle\cdot\rangle_{z^{l}}$ can be taken with
respect to Gaussian distributions $z^{l}\sim\mathcal{N}(\mu_{z^{l}},\Sigma_{z^{l}})$,
and we obtain that the mean and covariance of post-activations are
non-linear functions of only mean and covariance of pre-activations
\begin{alignat}{1}
\mu_{y^{l}} & =f_{\mu}(\mu_{z^{l}},\Sigma_{z^{l}})\,,\quad\Sigma_{y^{l}}=f_{\Sigma}(\mu_{z^{l}},\Sigma_{z^{l}})\,.\label{eq:stat_post_act}
\end{alignat}
These functions mediate interactions between first- and second-order
cumulants.

Applying this argument iteratively to the network layers $l=L+1,\,L,\dots,\,2$,
it follows that the information processing in the \emph{internal}
network layers is largely determined by an iterated, non-linear mapping
of mean and covariance. The interaction functions $f_{\mu}$ and
$f_{\Sigma}$ can be calculated numerically for arbitrary activation
functions. In particular, $\phi$ need not be differentiable. Analytic
expressions can be obtained for various activation functions $\phi$;
we provide expressions for $\phi=\text{ReLU}$ and $\text{\ensuremath{\phi(z)=z+\alpha\,z^{2}}}$
in \prettyref{app:stats_trafos} \prettyref{tab:interaction_functions}.
The latter, minimally nonlinear activation function yields especially
interpretable interaction functions that are constructed from the
following diagrams:\begin{fmffile}{quad_act_diagr}
\fmfset{thin}{0.75pt}
\fmfset{decor_size}{4mm}
\begin{subequations}
\begin{alignat}{2}
	\label{eq:diagrams_mean_quad_act}
		\mu_{y^l,\,i}
		&=\quad \parbox{20pt}{
							\begin{fmfgraph*}(20, 20)
							\fmfstraight
							\fmftopn{l}{2}
							\fmffreeze
							\fmfshift{(0,-.5h)}{l1,l2}
							\fmf{plain}{l1,l2}
							\fmfv{decor.shape=circle, decor.filled=empty}{l2}
							\end{fmfgraph*}
						}
						&&+\; \parbox{20pt}{
							\begin{fmfgraph*}(20, 20)
							\fmfstraight
							\fmftopn{t}{1}
							\fmfbottomn{b}{3}
							\fmffreeze
							\fmfshift{(0,-.25h)}{t1}
							\fmfshift{(-.5w,0)}{b1}
							\fmfshift{(.5w,0)}{b3}
							\fmfshift{(.8w,.5h)}{t1,b1,b2,b3}
							\fmf{plain}{t1,b2}
							\fmf{plain}{b1,b2}
							\fmf{plain}{b2,b3}
							\fmfv{d.s=circle, d.filled=shaded}{b2}
							\fmfv{d.s=circle, d.filled=empty}{b1,b3}
							\end{fmfgraph*}
						} \notag \\[5pt]
		&               &&\;+\; \parbox{20pt}{
							\begin{fmfgraph*}(20, 20)
							\fmfstraight
							\fmftopn{l}{3}
							\fmffreeze
							\fmfshift{(.3w.,-.5h)}{l1,l2,l3}
							\fmfshift{(-.3w,0)}{l1}
							\fmfshift{(.5w,0)}{l3}
							\fmf{plain}{l1,l2}
							\fmf{plain,tension=0.8,right=1.}{l2,l3}
							\fmf{plain,tension=0.8,right=1.}{l3,l2}
							\fmfv{d.s=circle, d.filled=shaded}{l2}
							\fmfv{d.s=circle, d.filled=empty}{l3}
							\end{fmfgraph*}
						} \notag \\[5pt]
		&=\quad\, \mu_{z^l,\,i}\quad &&\;+  \alpha\,(\mu_{z^l,\,i})^{2} \notag \\
		&               &&\;+  \alpha\,\Sigma_{z^l,\,ii}\, , \\[10pt]
		\Sigma_{y^l,\,ij} 
		&=\quad \parbox{20pt}{
							\begin{fmfgraph*}(20, 20)
							\fmfstraight
							\fmftopn{l}{3}
							\fmffreeze
							\fmfshift{(0,-.5h)}{l1,l2,l3}
							\fmfshift{(-.3w,0)}{l1}
							\fmfshift{(.3w,0)}{l3}
							\fmf{plain}{l1,l2}
							\fmf{plain}{l2,l3}
							\fmfv{decor.s=circle, decor.filled=empty}{l2}
							\end{fmfgraph*}
						}
						&&\;+\; \parbox{20pt}{
							\begin{fmfgraph*}(20, 20)
							\fmfstraight
							\fmftop{l1,l2}
							\fmfbottomn{b}{5}
							\fmffreeze
							\fmfshift{(1.8w,.7h)}{b1,b2,b3,b4,b5,l1,l2}
							\fmfshift{(0,-.2h)}{b1,b2,b3,b4,b5}
							\fmfshift{(0,-.45h)}{l1,l2}
							\fmfshift{(-.5w,0)}{l1}
							\fmfshift{(.5w,0)}{l2}
							\fmfshift{(-1.5w,0)}{b1}
							\fmfshift{(-.75w,0)}{b2}
							\fmfshift{(.75w,0)}{b4}
							\fmfshift{(1.5w,0)}{b5}
							\fmf{plain}{b1,b2}
							\fmf{plain}{b2,b3}
							\fmf{plain}{b3,b4}
							\fmf{plain}{b4,b5}
							\fmf{plain}{b2,l1}
							\fmf{plain}{b4,l2}
							\fmfv{d.s=circle, d.filled=empty}{b1,b3,b5}
							\fmfv{d.s=circle, d.filled=shaded}{b2,b4}
							\end{fmfgraph*}
						} \notag \\[5pt]
		&\;\; +\; \parbox{20pt}{
							\begin{fmfgraph*}(20, 20)
							\fmfstraight
							\fmftop{l1,l2}
							\fmfbottomn{b}{4}
							\fmffreeze
							\fmfshift{(1.w,.75h)}{b1,b2,b3,b4,l1,l2}
							\fmfshift{(0,-.25h)}{b1,b2,b3,b4}
							\fmfshift{(0,-.75h)}{l1,l2}
							\fmfshift{(-.3w,-1.h)}{l2}
							\fmfshift{(.4w,0)}{l1}
							\fmfshift{(-.3w,0)}{l2}
							\fmfshift{(-1.w,0)}{b1}
							\fmfshift{(-.5w,0)}{b2}
							\fmfshift{(.3w,0)}{b3}
							\fmfshift{(.8w,0)}{b4}
							\fmf{plain}{b1,b2}
							\fmf{plain}{b3,b4}
							\fmf{plain,tension=0.8,left=.5}{b2,l1}
							\fmf{plain,tension=0.8,left=.5}{l1,b3}
							\fmf{plain,tension=0.8,right=.5}{b2,l2}
							\fmf{plain,tension=0.8,right=.5}{l2,b3}
							\fmfv{d.s=circle, d.filled=shaded}{b2,b3}
							\fmfv{d.s=circle, d.filled=empty}{l1,l2}
							\end{fmfgraph*}
						}
						&&\;+\; \parbox{20pt}{
							\begin{fmfgraph*}(20, 20)
							\fmfstraight
							\fmftop{l1}
							\fmfbottomn{b}{5}
							\fmffreeze
							\fmfshift{(1.3w,.7h)}{b1,b2,b3,b4,b5,l1}
							\fmfshift{(0,-.2h)}{b1,b2,b3,b4,b5}
							\fmfshift{(0,-.45h)}{l1}
							\fmfshift{(-1.3w,0)}{b1}
							\fmfshift{(-.75w,0)}{b2}
							\fmfshift{(.75w,0)}{b4}
							\fmfshift{(1.3w,0)}{b5}
							\fmf{plain}{b1,b2}
							\fmf{plain}{b2,b3}
							\fmf{plain}{b3,b4}
							\fmf{plain}{b3,l1}
							\fmfv{d.s=circle, d.filled=shaded}{b3}
							\fmfv{d.s=circle, d.filled=empty}{b2,b4}
							\end{fmfgraph*}} \notag \\[5pt]
		&=\;\;\; \Sigma_{z^l,\,ij} &&\;+ 4\,\alpha^{2}\,\mu_{z^l,\,i}\,\Sigma_{z^l,\,ij}\,\mu_{z^l,\,j} \notag \\
		&\;\; + 2\,\alpha^{2}\,(\Sigma_{z^l,\,ij})^{2} &&\;+ 2\,\alpha\,\Sigma_{z^l,\,ij}\,\left(\mu_{z^l,\,i}+\mu_{z^l,\,j}\right).
\end{alignat}
\end{subequations}
\end{fmffile}\vspace{2pt}\foreignlanguage{english}{The last diagram contributing to $\Sigma_{y^{l},\,ij}$
corresponds to an expression containing two terms. These terms result
from the permutation of the indices $(i,j)$ (see \prettyref{subsec:Parameterizations-of-probability}).}

Training introduces correlations between weights, thus violating the
independence assumption of the central limit theorem. Also the sufficient
conditions for the Gaussian approximation to be consistent (\prettyref{app:Weakly-correlated})
are not necessary conditions; for example pairs of neurons may be
perfectly correlated without violating a Gaussian description. We
will therefore show in the following that empirically the first- and
second-order cumulants provide a useful approximation for the information
propagation \emph{within} the network.

\subsection{Information extraction in the input layer\label{subsec:training_data_statistics}}

So far we have studied the internal network layers. Here, we discuss
the role of the input layer in extracting information from higher-order
correlations of the input data. Since the pre-activations of this
layer $z_{i}^{1}=\sum_{j=1}^{N_{0}}\,W_{ij}^{1}\,x_{j}+b_{i}^{1}$
involve a sum over the input dimension $N_{0}$ instead of the network
width $N$, higher-order cumulants $G_{z^{1}}^{(n>2)}$ scale with
$N_{0}^{1-\frac{n}{2}}$ and need to be taken into account for smaller
input dimension $N_{0}$. In consequence, cumulants of multiple orders
$n$ contribute to the mean and covariance of the post-activations
$y^{1}$: 
\begin{equation}
\mu_{y^{1}}=h_{\mu}(\{G_{z^{1}}^{(n)}\}_{n}),\quad\Sigma_{y^{1}}=h_{\Sigma}(\{G_{z^{1}}^{(n)}\}_{n})\,.\label{eq:cum_postact_input}
\end{equation}
These mean and covariance are then passed on through the entire network.

The interaction functions $h_{\mu}$ and $h_{\Sigma}$ can be systematically
approximated for any activation function, either by the diagrammatic
techniques discussed in \prettyref{subsec:Parameterizations-of-probability}
in the case of differentiable functions or alternatively by a Gram-Charlier
expansion for non-differentiable functions (see \prettyref{app:stats_trafos}
for $\text{ReLU}$ as an example). Analytically simple and exact
expressions can be computed for a quadratic non-linearity (see Eq. \eqref{eq:diagrams_mean_quad_act});
in this case, the expression for the mean does not get any contribution
from $G_{z^{1}}^{(n>2)}$, while the covariance gets additional contributions
from third- and fourth-order input correlations:\foreignlanguage{english}{\begin{fmffile}{cov_quad_act_corr}
\fmfset{thin}{0.75pt}
\fmfset{decor_size}{4mm}
\begin{equation*}
	\label{eq:diagrams_cov_quad_act_corr}
	\begin{tabular}{llll}
		$\left. \Sigma_{y^l,\,ij}\right.\vert_{\text{add.}}$ 
							&$=$ &\parbox{20pt}{
							\begin{fmfgraph*}(20, 20)
							\fmfstraight
							\fmftopn{t}{4}
							\fmffreeze
							\fmfshift{(1.55w,-.5h)}{t1,t2,t3,t4}
							\fmfshift{(-1.w,0)}{t1}
							\fmfshift{(-.5w,0)}{t2}
							\fmfshift{(.5w,0)}{t4}
							\fmf{plain}{t3,t4}
							\fmf{plain}{t1,t2}
							\fmf{plain,tension=0.8,right=1.}{t2,t3}
							\fmf{plain,tension=0.8,right=1.}{t3,t2}
							\fmfv{d.s=circle, d.filled=empty}{t3}
							\fmfv{d.s=circle, d.filled=shaded}{t2}
							\end{fmfgraph*}
						}\\[10pt]
					& &$+$ &\parbox{20pt}{
							\begin{fmfgraph*}(20, 20)
							\fmfstraight
							\fmftopn{t}{5}
							\fmfbottom{b1}
							\fmffreeze
							\fmfshift{(1.1w,0)}{t1,t2,t3,t4,t5,b1}
							\fmfshift{(0,-.5h)}{t1,t2,t3,t4,t5}
							\fmfshift{(-1.75w,0)}{t1}
							\fmfshift{(-1.15w,0)}{t2}
							\fmfshift{(-.55w,0)}{t3}
							\fmfshift{(.25w,0)}{t4}
							\fmfshift{(1.w,0)}{t5}
							\fmfshift{(.5w,1.25h)}{b1}
							\fmf{plain}{t4,t5}
							\fmf{plain}{t1,t2}
							\fmf{plain}{t4,t3}
							\fmf{plain}{t4,b1}
							\fmf{plain,tension=0.8,right=1.}{t3,t2}
							\fmf{plain,tension=0.8,right=1.}{t2,t3}
							\fmfv{d.s=circle, d.filled=empty}{t3,t5}
							\fmfv{d.s=circle, d.filled=shaded}{t2,t4}
							\end{fmfgraph*}
						}\\[10pt]
					& &$+$ &\parbox{20pt}{
							\begin{fmfgraph*}(20, 20)
							\fmfstraight
							\fmftopn{t}{5}
							\fmffreeze
							\fmfshift{(1.1w,-.5h)}{t1,t2,t3,t4,t5}
							\fmfshift{(-1.75w,0)}{t1}
							\fmfshift{(-.6w,0)}{t3}
							\fmfshift{(-1.2w,0)}{t2}
							\fmfshift{(.55w,0)}{t5}
							\fmf{plain}{t1,t2}
							\fmf{plain}{t4,t5}
							\fmf{plain,tension=0.8,right=1.}{t2,t3}
							\fmf{plain,tension=0.8,right=1.}{t3,t2}
							\fmf{plain,tension=0.8,right=1.}{t3,t4}
							\fmf{plain,tension=0.8,right=1.}{t4,t3}
							\fmfv{d.s=circle, d.filled=shaded}{t2,t4}
							\fmfv{d.s=circle, d.filled=empty}{t3}
							\end{fmfgraph*}
						}\\[15pt]
				&$=$ & &$\mkern-30mu \alpha \, \Big(G_{z^l, \,(i,\,j,\,j)}^{(3)}+G_{z^l, \, (j,\,i,\,i)}^{(3)}\Big)$\\[10pt]
		& &$+$ &$\mkern-30mu 2\alpha^{2} \, \Big(G_{z^l,\,(i)}^{(1)} \, G_{z^l, \, (i,\,j,\,j)}^{(3)} +G_{z^l,\,(j)}^{(1)} \, G_{z^l,\,(j,\,i,\,i)}^{(3)}\Big)$\\[10pt]
		& &$+$ &$\mkern-30mu \alpha^{2}\,G_{z^l,\,(i,\,i,\,j,\,j)}^{(4)}$\\
	\end{tabular}
\end{equation*}
\end{fmffile}As in the previous section, there are two diagrams that each correspond
to an expression containing multiple terms. These terms result from
the permutation of the indices $(i,j)$ (see \prettyref{subsec:Parameterizations-of-probability}).}

The cumulants of the pre-activations $G_{z^{1}}^{(n)}$ are linked
to the cumulants of the input data $G_{x}^{(n)}$ by a mapping between
corresponding orders $n$ as $G_{x}^{(n)}\rightarrow G_{z^{1}}^{(n)}$
(see \prettyref{eq:cum_tensor_trafo}), yielding
\begin{align*}
\mu_{y^{1}} & =\tilde{h}_{\mu}(\{G_{x}^{(n)}\}_{n};\{W^{1},b^{1}\})\,,\\
\Sigma_{y^{1}} & =\tilde{h}_{\Sigma}(\{G_{x}^{(n)}\}_{n};\{W^{1},b^{1}\})\,.
\end{align*}
Thus, the input layer effectively extracts information from higher-order
correlations of the input data $G_{x}^{(n>2)}$.

\subsection{Statistical model of a feed-forward network\label{subsec:stat_model}}

Putting together all previous sections, we introduce the \emph{statistical
model} corresponding to a given network model. This model represents
the information processing performed by the network in terms of the
data correlations $\{G_{x}^{(n)}\}_{n}$.

By iterating \prettyref{eq:stat_pre_act}, \prettyref{eq:cum_postact_input},
and \prettyref{eq:stat_post_act}, respectively, across layers, one
obtains the mean and covariance of the network output $y=g(x;\theta)$
as functions of the statistics of $x$:
\begin{align}
\mu_{y} & =W^{L+1}(f_{\mu}(\dots\{\tilde{h}_{\nu}(\{G_{x}^{(n)}\}_{n})\}_{\nu=\mu,\Sigma}\dots))+b^{L+1}\nonumber \\
 & \eqqcolon g_{\mu}(\{G_{x}^{(n)}\}_{n};\theta,\phi)\,,\label{eq:stat_model_mean}\\
\Sigma_{y} & =W^{L+1}(f_{\Sigma}(\dots\{\tilde{h}_{\nu}(\{G_{x}^{(n)}\}_{n})\}_{\nu=\mu,\Sigma}\dots))(W^{L+1})^{\T}\nonumber \\
 & \eqqcolon g_{\Sigma}(\{G_{x}^{(n)}\}_{n};\theta,\phi)\,.\label{eq:stat_model_cov}
\end{align}
Since the network decomposes into a mapping for each class label $t$,
one obtains the distribution of the network output as a Gaussian mixture
$p(y)=\sum_{t}p(t)\,\mathcal{N}(\mu_{y}^{t},\Sigma_{y}^{t})(y)$.
The parameters $(\mu_{y}^{t},\Sigma_{y}^{t})$ are determined by the
propagation of data correlations $\{G_{x}^{(n),\,t}\}_{\lowerindex n,\,t}$
through the network \prettyref{eq:stat_model_mean}-\eqref{eq:stat_model_cov}.
Note that these are generally not exact due to the Gaussian approximation
of pre-activations $z^{l}$ at each intermediate layer. In the following,
we call the mapping 
\begin{equation}
g_{\mathrm{stat}}:(\{G_{x}^{(n),\,t}\}_{\lowerindex n,\,t},\theta,\phi)\mapsto p(y)\label{eq:statistical_model}
\end{equation}
the \emph{statistical model} of the network. One important feature
is that the statistical model shares the parameter structure $\theta=\{W^{l},b^{l}\}_{l=1,\dots,L+1}$
with the corresponding network model. In consequence, there is a one-to-one
correspondence between the statistical model \prettyref{eq:statistical_model}
and the network model $g\colon(x;\theta)\mapsto y$ given a fixed
set of parameters $\theta$.

Beyond empirically comparing these two models, the statistical model
can be used to assess the relevance of data correlations $\{G_{x}^{(n),\,t}\}_{\lowerindex n,\,t}$
for solving a particular task. We have shown in \prettyref{subsec:loss_reformulation}
that the expected mean squared error loss $R_{\text{MSE}}(\{\mu_{y}^{t},\Sigma_{y}^{t}\}_{t})$
is given by \prettyref{eq:loss_stat_mse}, so that it depends solely
on mean and covariance of the output. By the statistical model \prettyref{eq:statistical_model},
the mean squared error $R_{\text{MSE}}$ thus can be approximated
as a function of the data correlations $\{G_{x}^{(n),\,t}\}_{\lowerindex n,\,t}$
and the network parameters $\theta$:
\begin{align}
 & R_{\text{MSE}}\bigl(\{\mu_{y}^{t},\Sigma_{y}^{t}\}_{t}\bigr)\\
 & \approx R_{\text{MSE}}\bigl(\{G_{x}^{(n),\,t}\}_{\lowerindex n,t};\theta\bigr)\,.
\end{align}
Minimizing this loss then yields optimal parameters $\theta^{*}$
for the statistical model. The corresponding network model $g(x;\theta^{*})$
is then dependent on the given set of data correlations $\{G_{x}^{(n),\,t}\}_{\lowerindex n,\,t}$,
allowing the investigation of their relevance in solving a particular
network task.

\section{Experimental results\label{sec:exp_results}}

We now apply the developed methods to the XOR problem and the MNIST
dataset. We use the network architecture defined in \prettyref{subsec:notation_network_arch}
with fixed network width $N_{l}=N$ for $l\ge1$ and either the $\mathrm{ReLU}$
activation function or a minimal non-linearity, namely the quadratic
activation function $\text{\ensuremath{\phi(z)=z+\alpha\,z^{2}}}$
with $\alpha=0.5$.

\subsection{Training details}

For initialization of network parameters $\theta$, we use $\sigma_{w}^{2}=\sigma_{b}^{2}=0.75$.
Following the standard procedure, networks are trained by optimizing
the empirical risk per data batch $\{(x^{(b)},\,t^{(b)})\}_{b}$ of
the expected MSE loss:
\begin{equation}
R_{\text{emp,\,MSE}}(\theta)=\frac{1}{B}\sum_{b=1}^{B}\,\ell_{\text{MSE}}(g(x^{(b)};\theta),t^{(b)})\,.\label{eq:loss_data}
\end{equation}
The batch size $B$ is set to $10$ on XOR and $100$ on MNIST. For
optimization, we use \noun{Adam} \citep{Kingma15_iclr,Loshchilov19}
with learning rate $10^{-3}$, momenta $\beta_{1}=0.9$ and $\beta_{2}=0.999$,
$\epsilon=10^{-8}$, and $\lambda=0$. The choice of the optimizer
does not affect the above presented derivations. Network implementations
were done in \textsc{\noun{PyTorch}} \citep{Paszke19_8024}.

\subsection{Multiple information encodings of the XOR problem\label{subsec:Multiple-encodings-of}}

We first study an adaptation of the XOR problem as a non-linearly
separable Gaussian mixture distribution. We make use of two conceptual
advantages of this XOR task: First, knowing the exact input distribution
allows us to focus on the internal information processing within the
network. Second, the fact that each class is itself a mixture distribution
allows us to trace the class-conditional correlations in two alternative
forms, corresponding to two different statistical representations
of class membership, which isolate different statistics of the input
-- respectively the mean and the covariance. We find that while the
task can be solved for both representations, they correspond to different
local minima of the empirical loss landscape.

\subsubsection{Problem setup as a Gaussian mixture\label{subsec:XOR-problem-setup}}

\begin{figure*}
\centering{}\includegraphics{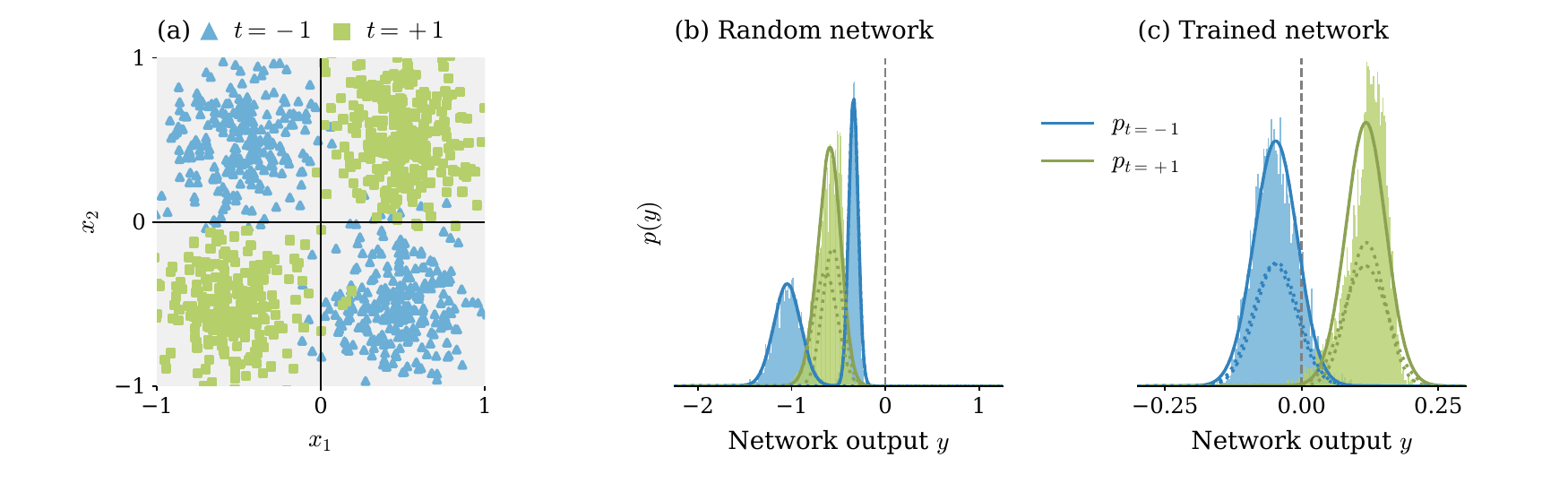}\caption{\textbf{Information propagation in $\mathrm{ReLU}$ networks for the
XOR problem. (a) }The distribution of input data is modeled as a Gaussian
mixture. Data samples $x^{(d)}$ (blue and green dots) are assigned
to class labels $t=\pm1$ based on the respective mixture component.
\textbf{(b},\textbf{c)} Distribution of the network output for random
(b) and trained (c) parameters. Class-conditional distributions (solid
curves) are determined as a superposition of the propagated mixture
components (dashed curves) as in \prettyref{eq:output_dist_theory}
and empirical estimates (blue and green histograms) are obtained from
the test data. Since networks are trained on class labels $t=\pm1$,
the classification threshold is set to $y=0$ (gray lines). Other
parameters: $\phi=\text{ReLU}$, depth $L=1$, width $N=10$; trained
network in (c) achieves $P=93.82\%$ performance compared to $P_{\text{opt}}=97.5\%$.
\label{fig:xor_illustration_prob_dist_mapping}}
\end{figure*}

Our adaptation of the XOR problem uses real-valued instead of binary
inputs and describes the input distribution as a Gaussian mixture
of four components, illustrated in \prettyref{fig:xor_illustration_prob_dist_mapping}\textbf{a}.
For the class label $t=\text{\ensuremath{+1}}$, we choose the mean
values of its two components $\pm$ as $\mu_{x}^{t=+1,\,\pm}=\pm(0.5,0.5)^{\intercal}$;
for $t=-1$, we use $\mu_{x}^{t=-1,\,\pm}=\pm(-0.5,0.5)^{\intercal}.$
Covariances are isotropic throughout with $\Sigma_{x}^{t,\,\pm}=0.05\,\mathbb{I}$
and the input distribution 
\[
p(x,t)=p(t)\,\sum_{\pm}\,p_{\pm}\,\mathcal{N}(\mu_{x}^{t,\,\pm},\Sigma_{x}^{t,\,\pm})(x)
\]
weighs all components equally $p(t)=p_{\pm}=\frac{1}{2}$. A data
sample $x^{(d)}$ is assigned a target label $t^{(d)}\in\{\pm1\}$
based on the mixture component it is drawn from. From the geometry
of the problem follows that the optimal decision boundaries coincide
with the axes in data space ( \prettyref{fig:xor_illustration_prob_dist_mapping}\textbf{a}),
allowing us to calculate the optimal performance $P_{\text{opt}}=97.5\%$.
We use training and test data sets of sizes $n_{\text{train}}=10^{5}$
and $n_{\text{test}}=10^{4}$, respectively.

\subsubsection{Accuracy of internal information processing in terms of correlation
functions}

Given the exact input distribution for this problem, we trace the
transformation of mean and covariance predicted by \prettyref{eq:stat_model_mean}
and \prettyref{eq:stat_model_cov} for each mixture component $(t,\,\pm)$
separately, obtaining 
\begin{equation}
p_{\text{theo.}}(y)=\sum_{t}p(t)\,\sum_{\pm}\,p_{\pm}\,\mathcal{N}(\mu_{y}^{t,\,\pm},\Sigma_{y}^{t,\,\pm})(y)\,,\label{eq:output_dist_theory}
\end{equation}
where $\mu_{y}^{t,\,\pm}=g_{\mu}(\mu_{x}^{t,\,\pm},\Sigma_{x}^{t,\,\pm};\theta,\phi)$
and $\Sigma_{y}^{t,\,\pm}=g_{\Sigma}(\mu_{x}^{t,\,\pm},\Sigma_{x}^{t,\,\pm};\theta,\phi)$
are functions of the input statistics, the network parameters, and
depend on the choice of activation function. In \prettyref{fig:xor_illustration_prob_dist_mapping}
we compare this theoretical result to an empirical estimate of the
output distribution $p_{\text{emp.}}(y)$, given as a histogram obtained
from the test data. We test the validity of the statistical model
for both, an untrained network with random weight initialization (\prettyref{fig:xor_illustration_prob_dist_mapping}\textbf{b})
and a trained network (\prettyref{fig:xor_illustration_prob_dist_mapping}\textbf{c}).
\begin{figure}
\centering{}\includegraphics{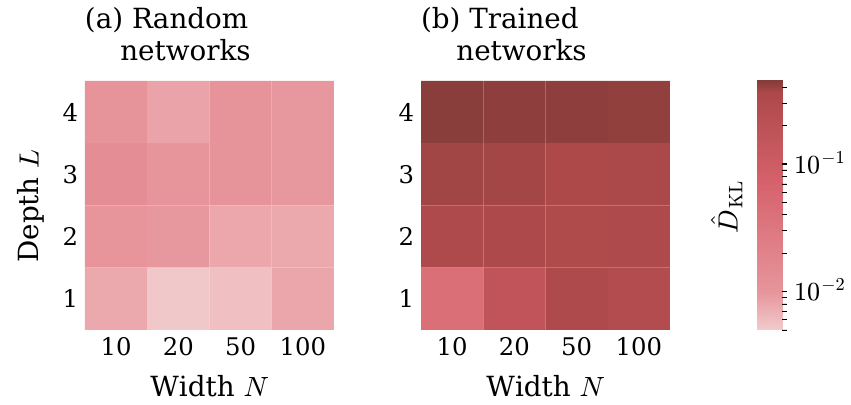}\caption{\textbf{Deviation between theoretical and empirical output distribution
for (a) random and (b) trained networks}, measured across $50$ different
network realizations using the normalized Kullback-Leibler divergence
$\hat{D}_{\mathrm{KL}}(p_{\text{emp.}}\Vert p_{\text{theo.}})$. On
average, the trained networks achieve performance values of $P=97.00\%\pm0.05\%$
compared to $P_{\text{opt}}=97.5\%$.\label{fig:xor_kl_div} Networks
were trained to perform the XOR task described in \prettyref{subsec:XOR-problem-setup}.
Other parameters: $\phi=\text{ReLU}$.}
\end{figure}

The untrained network produces an output distribution of complex shape
composed of superimposed close-to Gaussian distributions, each corresponding
to one component, as shown in \prettyref{fig:xor_illustration_prob_dist_mapping}\textbf{b}.
Training the network reshapes the output distribution such that the
class-conditional distributions $p(y\vert t)$ become well separated
by the threshold at $y=0$, as shown in \prettyref{fig:xor_illustration_prob_dist_mapping}\textbf{c}.
The overlap between these two distributions around the threshold corresponds
to the classification error. Qualitatively, theory and simulation
agree well for both random and trained networks. These results apply
for different activation functions $\phi$ (see \prettyref{fig:xor_illustration_prob_dist_mapping_quad_app}
in \prettyref{app:xor_illustration_quad} for $\ensuremath{\phi(z)=z+\alpha\,z^{2}}$).

To quantify the alignment of theory and simulation, we compute the
Kullback-Leibler divergence $D_{\mathrm{KL}}$ between the empirical
estimate $p_{\text{emp.}}(y)$ and the theoretical result $p_{\text{theo.}}(y)$,
considering the empirical distribution $p_{\text{emp.}}(y)$ as the
reference. To account for the variability of output distributions
across different network realizations, this quantity is normalized
by the entropy $H$ of the empirical distribution $p_{\text{emp.}}(y)$,
yielding $\hat{D}_{\mathrm{KL}}(p_{\text{emp.}}\Vert p_{\text{theo.}})=\nicefrac{D_{\mathrm{KL}}(p_{\text{emp.}}\Vert p_{\text{theo.}})}{H(p_{\text{emp.}})}.$

We average $\hat{D}_{\mathrm{KL}}(p_{\text{emp.}}\Vert p_{\text{theo.}})$
across $50$ different network realizations, for random (\prettyref{fig:xor_kl_div}\textbf{a})
and trained (\prettyref{fig:xor_kl_div}\textbf{b}) networks. In
both cases, the deviation between theory and simulation is generally
small, but increases mildly with the network depth $L$ as approximation
errors accumulate across network layers. For random networks, the
deviations are generally small with a slight decrease of the deviation
for wider networks, in agreement with the central limit theorem as
discussed in \prettyref{subsec:trafo_data_statistics}. For trained
networks with thus correlated parameters, there is an overall increase
of deviations between theory and simulation. Nonetheless, this increase
remains modest, showing that the theory continues to be applicable
for networks with trained, and thus non-random, parameters. Again
these results apply for different activation functions $\phi$ (see
\prettyref{fig:xor_kl_div_quad} in \prettyref{app:xor_illustration_quad}
for $\ensuremath{\phi(z)=z+\alpha\,z^{2}}$). When evaluating the
expressions for $\mathrm{ReLU}$, one needs to be careful with the
numerics due to the appearing error functions.
\begin{figure*}[!t]
\centering{}\includegraphics{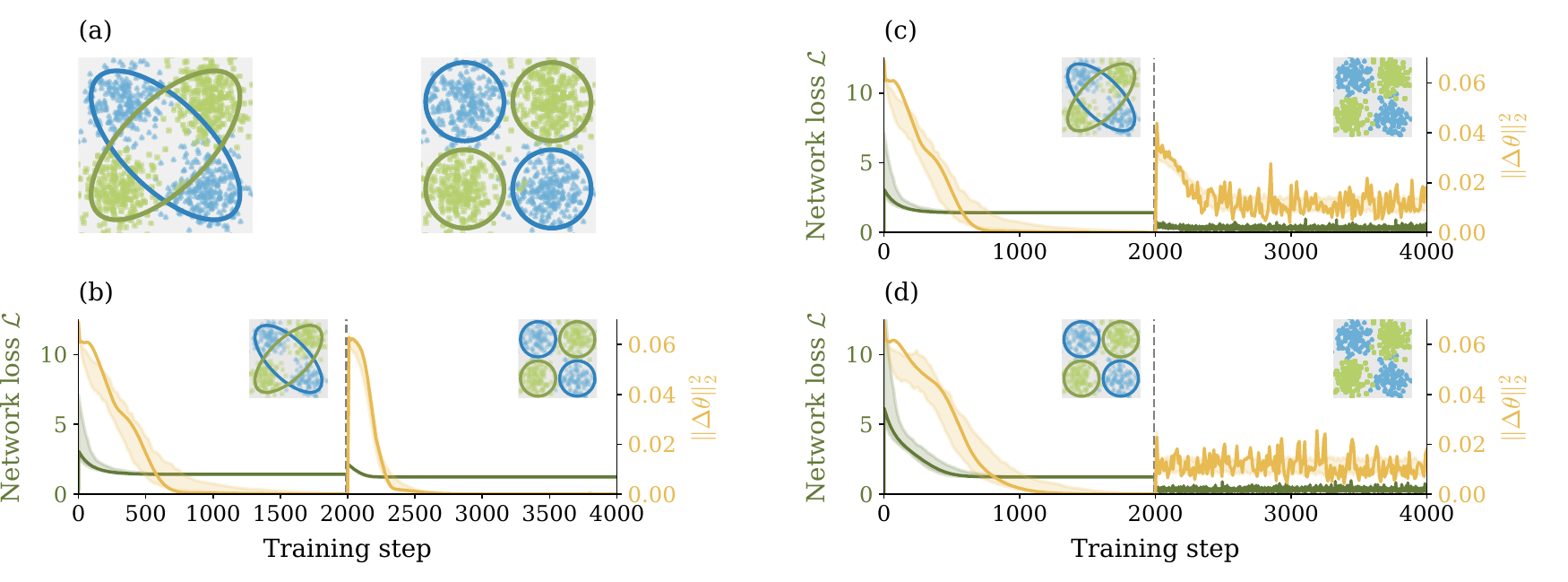}\caption{\textbf{Mean and covariance coding.} \textbf{(a)} Data distributions
of XOR task. Left: \textbf{Covariance coding}; class membership (blue
and green ellipses) is encoded in the covariance alone. Right: \textbf{Mean
coding}; the two individual Gaussian components of each class (both
blue circles / both green circles) differ in their means, while having
the same covariance. \textbf{(b-d)} Evolution of network loss $R$
($R_{\text{emp,\,MSE}}$ \prettyref{eq:loss_data} for training the
network model, $R_{\text{MSE}}$ \prettyref{eq:loss_stat_mse} for
the statistical model) and change of network parameters $||\Delta\theta||_{2}^{2}$
in each training step: The first $T_{1}=2000$ training steps train
one model, starting from random parameters $\theta$. The model representation
is changed at $T_{1}$, starting from parameters $\theta(T_{1})$
obtained in the preceding period. The change of network parameters
is evaluated every $10$ training steps $\Vert\Delta\theta(T)\Vert_{2}^{2}=\Vert\theta(T)-\theta(T-10)\Vert_{2}^{2}$.
Shaded areas show the typical range, based on lower and upper quartiles
across $10^{2}$ network realizations. Solid curves show the behavior
of a single network realization. \textbf{(b)} First period: statistical
model with covariance coding; second period: statistical model with
mean coding.\textbf{ (c)} First period: statistical model with covariance
coding; second period: network model. \textbf{(d)} First period: statistical
model with mean coding; second period: network model. Training parameters:
$n_{\text{train}}=10^{4},$ $2$ epochs. Other parameters: $\ensuremath{\phi(z)=z+\alpha\,z^{2}}$,
depth $L=1$, width $N=10$.\label{fig:xor_comp_coding_paradigms}}
\end{figure*}

\subsubsection{Different information coding paradigms and their relations}

In the previous section, we have shown that the mapping implemented
by the network can be described as a mapping of correlation functions
(see \prettyref{eq:statistical_model}). On the level of data correlations,
it directly follows that the network's expressivity with respect to
a given task depends on two properties: (1) the ability of the network
architecture to implement a desired mapping of data correlations from
its input to its output; (2) the way in which information about class
membership is represented by data correlations in the input.

A complete study of the first property would be provided by fully
describing the space of possible mappings, which is challenging in
general. However, the forward mapping of cumulants we have obtained
in \prettyref{subsec:trafo_data_statistics} allows us to probe this
space experimentally, and it provides a path to more systematic studies
of network expressivity -- see our remarks on statistical receptive
fields in \prettyref{sec:conclusion}.

In this section, we study the second property by investigating two
different information representations: (A) the class membership is
represented by different means between classes, while the covariances
and all higher-order cumulants are identical; (B) the class membership
is represented by different covariances between classes, while the
means and all higher-order cumulants are identical. Accordingly, these
two representations are called \emph{mean coding} (A) and \emph{covariance
coding} (B) in the following. While each of these two settings confines
the class membership to one particular cumulant order, the more general
case is that class membership is represented by various orders of
statistical moments. In that case, the network may make use of this
duplicate information to maximize performance.

To be able to compare these settings, in either case we train models
on a single task defined via a single data distribution, but present
different statistical representations of the data. We use the statistical
model corresponding to the network described in \prettyref{subsec:stat_model},
limiting input correlations to mean and covariance by setting higher-order
cumulants to zero. We take the binary XOR problem (see \prettyref{subsec:XOR-problem-setup})
which can be cast into either information representation in a natural
way: For mean coding (A), we provide to the network both the class
labels $t$ and the specific mixture component $\pm$ from which a
sample was drawn, yielding four sets of statistics $\{\mu_{x}^{m},\,\Sigma_{x}^{m}\}_{m=(t,\pm)}$
with different means but identical covariances. For covariance coding
(B), only the class label $t$ is provided to the network, yielding
two sets of statistics $\{\mu_{x}^{m},\,\Sigma_{x}^{m}\}_{m=(t)}$,
for which the covariances $\Sigma_{x}^{t=\pm 1}=\big(\begin{smallmatrix}0.3&\pm0.25\\\pm0.25&0.3\end{smallmatrix}\big)$
differ between the two classes, while their means are the same (see
\prettyref{fig:xor_comp_coding_paradigms}\textbf{a}). In both cases,
all higher-order cumulants of the component distributions $m=(t,\pm)$
and $m=(t)$, respectively, are set to zero. Note that for mean coding
the class distributions $p(x,t=\pm1)=\sum_{\pm}\,p_{\pm}\,\mathcal{N}(\mu_{x}^{t,\,\pm},\Sigma_{x}^{t,\,\pm})(x)$
indeed include higher-order cumulants. The different sets of input
statistics $\{\mu_{x}^{m},\,\Sigma_{x}^{m}\}_{m=(t,\pm)}$ (A) and
$\{\mu_{x}^{m},\,\Sigma_{x}^{m}\}_{m=(t)}$ (B), respectively, define
different statistical models for mean and covariance coding.

We compare these two statistical representations A and B of the network
to the network trained directly on batches of samples; the latter
we refer to as \emph{sample coding} in the following. Sample coding
can be considered as the case where potentially all statistical moments
of the data are accessible to the network. Our goal is to address
the following questions: First, which statistical representation most
closely matches the information representation used by a network trained
on data samples? Second, is there a difference in performance between
information representations; in particular, can the network equivalently
use the information provided by either mean or covariance coding?
Finally, does the network make use of duplicate information in different
cumulant orders to improve performance in the case of sample coding?

To answer these questions, we optimize models until convergence using
either representation. We then switch to a different representation,
continuing optimization for the same number of steps, and observe
the stability of the previously found solution. Each experimental
setup is repeated with $10^{2}$ different weight initializations.
Results are shown in \prettyref{fig:xor_comp_coding_paradigms} for
three different coding combinations, where we plot both the loss and
the magnitude of change $||\Delta\theta||_{2}^{2}$ of the model parameters.

We find that after initial optimization all three models correspond
to networks with at least $P=91\%$ performance, so training converges
in all cases and the networks implement viable solutions before the
switch. Thus, the behavior after the switch indicates how the found
solution is affected by changing the statistical representation. Furthermore,
we observe that immediately after the switch from covariance to mean
coding, $||\Delta\theta||_{2}^{2}$ jumps to values similar to the
initial training steps (\prettyref{fig:xor_comp_coding_paradigms}\textbf{b}).
This indicates a near complete change of the model, which suggests
that mean and covariance coding induce fundamentally different solutions.
In contrast, the jump is modest when switching from covariance to
sample coding (\prettyref{fig:xor_comp_coding_paradigms}\textbf{c}),
and non-existent when switching from mean to sample coding (\prettyref{fig:xor_comp_coding_paradigms}\textbf{d})
-- suggesting that those different solutions coexist in the true
loss landscape of the network model. Thus, we find that the network
utilizes the presented information in different ways for the two representations,
as expected based on the information flow in these networks, derived
in \prettyref{subsec:trafo_data_statistics}.

In particular, the case of covariance coding highlights the importance
of a non-linear activation function when the discriminating information
is not contained in the class means. Since classification is based
on different mean values in the network output, the difference in
covariance for each class needs to be transferred to the mean. This
information transfer is mediated by the non-linearity $\phi$; for
the case $\ensuremath{\phi(z)=z+\alpha\,z^{2}}$ used in \prettyref{fig:xor_comp_coding_paradigms},
we have the particularly simple transfer function 
\begin{equation}
\mu_{y^{l},\,i}=\mu_{z^{l},\,i}+\alpha\,(\mu_{z^{l},\,i})^{2}+\alpha\,\Sigma_{z^{l},\,ii}
\end{equation}
from covariances to means. Here, only diagonal entries of the covariance
enter, while the input covariances $\Sigma_{x}^{t=\pm 1}=\big(\begin{smallmatrix}0.3&\pm0.25\\\pm0.25&0.3\end{smallmatrix}\big)$
differ in their off-diagonal entries. The information transfer from
off-diagonal to diagonal entries is mediated by the affine transformation
(see \prettyref{eq:stat_pre_act}) prior to the activation function.
In this way, we can track how information flows into the mean as it
is transformed by successive network layers.

In summary, we find that for this task the network can effectively
utilize the information presented by either mean or covariance coding,
both representations leading to different solutions with comparable
performance. Sample coding tends to yield similar solutions as mean
coding, implying that the network makes use of duplicate information
present in higher-order moments of the data samples from each class.

\subsection{Essential data correlations of the MNIST data set\label{subsec:mnist_correlations}}

\begin{figure}
\centering{}\includegraphics{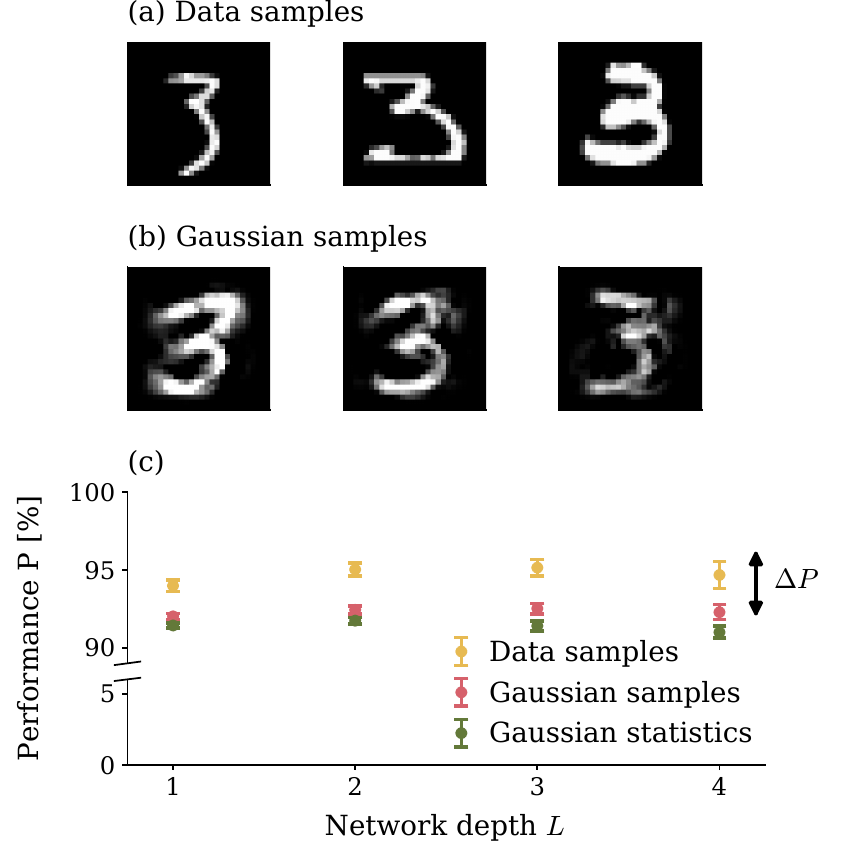}\caption{\textbf{First- and second-order correlations of MNIST. (a)} Three
example data samples showing the digit $3$ from the MNIST training
data set. \textbf{(b)} Data samples showing the digit $3$, drawn
from the Gaussian approximation of the input distribution. \textbf{(c)}
Classification performance on the MNIST test data set for different
input encodings. Network training consistently achieves performance
values of $P\approx94\%$ or more (yellow), while performance of the
corresponding statistical model is $3.3\%\pm0.6\%$ lower (green).
Training networks on Gaussian input samples yields a comparable performance
difference (red). In all cases, performance is evaluated on the MNIST
test data set. Error bars show the standard deviation across $10^{2}$
different network realizations. Other parameters: $\phi(z)=z+\alpha\,z^{2}$,
depth $L\in[1,2,3,4]$, width $N=100$.\label{fig:mnist_input_performance}}
\end{figure}
We consider in this section the MNIST data set \citep{lecun2010mnist},
consisting of $10$ classes of $28\times28$ images. This data set
is highly structured: if one approximates each class by a multivariate
Gaussian, the resulting samples are already visually recognizable
(\prettyref{fig:mnist_input_performance}\textbf{a,b}; see \prettyref{app:mnist_sample_generation}
for further details). Our goal is to use the theory developed in previous
sections to quantify this observation, in a matter which can be generalized
to different data sets and different sets of input cumulants. We also
argue that truncation of cumulants in the input layer has the largest
impact, and in the process validate our theory on a non-trivial task.

Concretely, we proceed as follows: when optimizing the parameters
$\theta^{*}$ of the statistical model, we restrict the data statistics
to a particular set of cumulants $\{G_{x}^{(n)}\}_{n=1,\dots,\hat{n}}$
and compare the achieved performance to that of the network trained
on samples $y=g(x;\theta^{*})$. The difference in performance is
then indicative of the importance of the cumulants we kept. We employ
one-hot encoding, making the network output $d_{\text{out}}=10$ dimensional.

As a baseline, we first train network models on both the MNIST data
set (\prettyref{fig:mnist_input_performance}\textbf{a}) and the corresponding
Gaussian samples (\prettyref{fig:mnist_input_performance}\textbf{b}).
The latter case limits the information that can be extracted by the
input layer to the class-conditional means $\mu_{x}^{t}$ and covariances
$\Sigma_{x}^{t}$. In both cases, networks are trained with the standard
empirical loss (\prettyref{eq:loss_data}); in particular, this allows
inner network layers to make use of cumulants of any order. With respect
to classification performance, we find that training on Gaussian samples
yields a performance that is lower by $\Delta P\simeq2.4\%\pm0.7\%$
(\prettyref{fig:mnist_input_performance}\textbf{c}): a difference
we can ascribe to the removal of higher-order cumulants in the data
distribution. Based on the modest magnitude of this difference, we
conclude that data mean and covariance are already highly informative
for these data and account for about $P\approx91\%$.

We next train the corresponding statistical model (\prettyref{eq:statistical_model})
on the Gaussian approximation of MNIST (\prettyref{fig:mnist_input_performance}\textbf{b}).
Compared to the network model trained on the Gaussian samples corresponding
to the same data distribution, we find only slightly lower performance
-- by about $0.9\pm0.4\%$ (\prettyref{fig:mnist_input_performance}\textbf{c})
-- suggesting that the statistical model given by \prettyref{eq:statistical_model}
is a good representation for the information processing in internal
network layers. The fact that most of the performance drop with respect
to standard training on MNIST is due to the Gaussian approximation
of the input data indicates the importance of processing higher-order
cumulants by the input layer. In the next section, we show with an
illustrative example how these can be included into the theory.

\subsection{Including higher-order correlation functions in the input layer\label{subsec:problem_alternating_hills}}

\begin{figure}
\centering{}\includegraphics{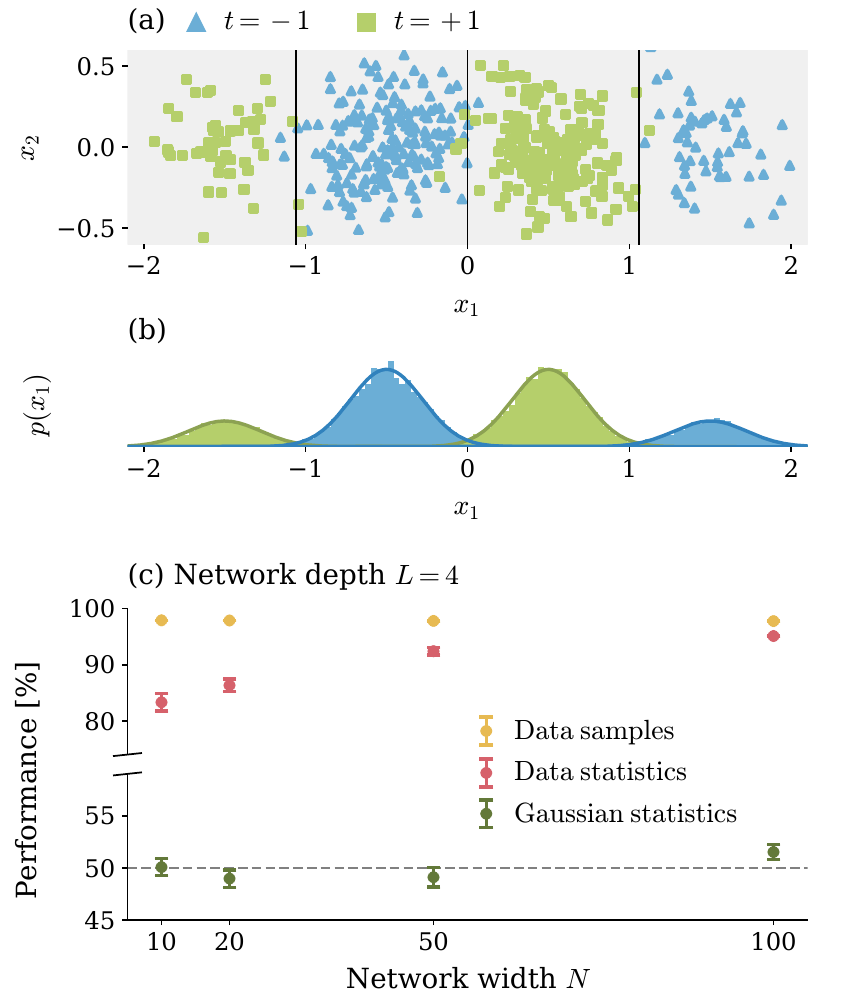}\caption{\textbf{Information extracted from higher-order correlations.} \textbf{(a)
}The distribution of input data is modeled as a Gaussian mixture.
Data samples $x^{(d)}$ (blue and green dots) are assigned to class
labels $t=\pm1$ based on their respective mixture component. The
two classes have zero mean and the same covariance. \textbf{(b)} Projection
of data samples to the $x_{1}$-axis (histograms), which corresponds
to the marginalization of the input distribution with respect to $x_{2}$
(solid lines), illustrating the different weighing of the mixture
components. \textbf{(c)} Classification performance for different
model choices. Network training consistently achieves performance
values of $P\approx96\%$ or more (yellow). Optimizing the statistical
model $g_{\mathrm{stat}}(\{G_{x}^{(n)}\}_{n=1,2},\theta)$ that considers
only the first- and second-order correlations (green) results in performance
values corresponding to chance level (dotted line). However, including
the third-order correlations into the statistical model $\tilde{g}_{\mathrm{stat}}(\{G_{x}^{(n)}\}_{n=1,2,3,4},\theta)$
nearly bridges this gap (red). In all cases, performance is evaluated
on a test data set. Error bars show the error of the mean of performance
across $10^{2}$ different network realizations. Other parameters:
$\phi(z)=z+\alpha\,z^{2}$, depth $L=4$, width $N=[10,\,20,\,50,\,100]$.\label{fig:problem_alternating_hills}}
\end{figure}

So far we have studied class-conditional means $\mu_{x}^{t}$ and
covariances $\Sigma_{x}^{t}$ of the input data; however, these two
statistics may not always be informative. It is in fact easy to construct
a low-dimensional task with two classes $t=\pm1$, where both class-conditional
means and covariances of the data are identical -- $\mu_{x}^{t=-1}=\mu_{x}^{t=+1}$,
$\Sigma_{x}^{t=-1}=\Sigma_{x}^{t=+1}$ -- thereby conveying no information
regarding the class membership (\prettyref{fig:problem_alternating_hills}\textbf{a},\textbf{b}).
Classification in such cases must therefore rely on higher-order statistics.
For the example in \prettyref{fig:problem_alternating_hills}, since
third-order cumulants differ between classes ($G_{x}^{(3),\,t=-1}=-G_{x}^{(3),\,t=+1}$),
we expect their inclusion into the statistical model to be sufficient
for solving the task. We here demonstrate that such higher-order cumulants
can indeed be treated by our approach -- in particular, we validate
the statement made in \prettyref{subsec:training_data_statistics}
that it suffices to consider higher-order cumulants in only the first
layer.

The input distribution for this task is defined as a Gaussian mixture
of four components, illustrated in \prettyref{fig:problem_alternating_hills}\textbf{a},\textbf{b}
(details in \prettyref{app:supplement_problem_alternating_hills}).
As expected, training the network model yields near-optimal performance
values, while a statistical model $g_{\mathrm{stat}}(\{G_{x}^{(n)}\}_{n=1,2},\theta)$
that considers only the class-conditional means $\mu_{x}^{t}$ and
covariances $\Sigma_{x}^{t}$ fails to solve the task, yielding chance-level
performance (\prettyref{fig:problem_alternating_hills}\textbf{c}).
This performance gap is nearly bridged when we include the third-order
input cumulants $G_{x}^{(3)}$ (via \prettyref{eq:contrib_higher_orders_quad_act})
in the first layer of the statistical model $\tilde{g}_{\mathrm{stat}}(\{G_{x}^{(n)}\}_{n=1,2,3,4},\theta)$.
The activation function $\phi$ allows information in $G_{x}^{(3)}$
to be transferred to lower-order cumulants, which are then processed
by subsequent layers in the manner described in previous sections
--  facilitating different means $G_{y}^{(1)}$ in the output of
the statistical model.

\subsection{High dimensionality of input data justifies Gaussian description
of fully-connected deep networks}

We study the CIFAR-10 data set \citep{Krizhevsky09_thesis}, consisting
of $10$ classes of $32\times32$ images with $3$ color channels.
Compared to MNIST, we expect two antagonistic effects. On the one
hand, since images within one class of CIFAR-10 are significantly
more heterogeneous, we expect the class-conditional distributions
to be more complex, and consequently to require higher-order cumulants
to accurately represent its statistical structure. On the other hand,
due to the larger input dimensionality $N_{0}=3072$ compared to $N_{0}=784$
for MNIST, higher-order cumulants are more strongly suppressed in
the input layer (see \prettyref{subsec:trafo_data_statistics}). To
check how these two effects interplay in feed-forward networks, we
employ the methods presented in previous sections to restrict training
to certain cumulants, similar as in \prettyref{subsec:mnist_correlations}.

\begin{figure}
\centering{}\includegraphics{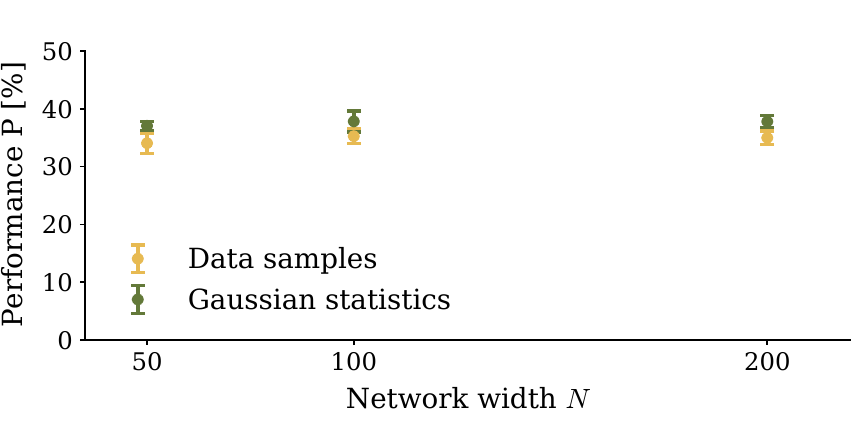}\caption{\textbf{Classification performance on CIFAR-10.} Training the statistical
model on Gaussian statistics (green) consistently achieves higher
performance values than training network models on data samples (yellow).
In both cases, performance is evaluated on the CIFAR-10 test data
set. Error bars show the standard deviation across $10$ different
network realizations. Other parameters: $\phi(z)=z+\alpha\,z^{2}$,
depth $L=2$, width $N\in[50,100,200]$.\label{fig:cifar10_performance}}
\end{figure}

We train network models on the CIFAR-10 data set and compare these
to the statistical model trained on the Gaussian approximation of
CIFAR-10 (\prettyref{fig:cifar10_performance}). In both cases, performance
is evaluated on the CIFAR-10 test data set. We find that network models
trained on data samples achieve performance values of $P=34.8\%\pm1.4\%$.
In contrast to MNIST, the statistical model trained on the Gaussian
statistics consistently achieves \emph{higher} performance values
of $P=37.6\%\pm1.3\%$. These results are directly linked to the two
aforementioned effects: They indicate that due to the large input
dimensionality, networks predominantly process only the Gaussian statistics
-- the statistical model therefore continues to provide a good representation
of the network. Moreover, estimates of the Gaussian statistics are
more accurate in the statistical model (averaged over the full training
set of $50,000$ images) compared to training on data samples (averaged
over mini-batches of $100$ images), possibly explaining the slightly
higher performance values. Importantly, although the achieved performance
values are far below values reported for other architectures such
as convolutional ResNets \citep{Zagoruyko16_87}, they are representative
for fully-connected feed-forward networks \citep{Lee18}. The difference
between the architectures lies in the extracted statistical information.
For high-dimensional input data, the here presented theory predicts
that fully-connected feed-forward networks are limited to Gaussian
statistics, which can only partly capture the statistical structure
of more complex data sets such as CIFAR-10. Hence, the presented decomposition
of a network in terms of cumulants allows us to relate the power of
network architectures to the processing of statistical information
contained in the data.

\section{Discussion\label{sec:conclusion}}

The question of how neural networks process data is fundamentally
the question of how information is encoded in the data distribution
and subsequently transformed by the network. We here present an
analytical approach, based on methods from statistical physics, to
study the mapping of data distributions implemented by deep feed-forward
neural networks: we parameterize the data distribution in terms of
correlation functions and derive their successive transformations
across layers. We show that the initial network layer effectuates
the extraction of information from higher-order correlations in the
data; for subsequent layers, a restriction to first- and second-order
correlation functions (mean and covariance) already captures the
main properties of the network computation. This reduction of the
bulk of the network to a non-linear mapping of a few correlation functions
provides an attractive view for further analyses. It relies on the
assumption of sufficiently wide layers to apply the central limit
theorem, but, in practice, we find that the approximations are useful
even for narrow networks.

We validate these results for different data sets. We first investigate
an adaptation of the XOR problem that is purely based on first- and
second-order cumulants. Despite the non-linear transformations in
each layer giving rise to higher-order correlations, the network solutions
to this task can largely be described in terms of transformations
solely between mean and covariance of each class. We then consider
the MNIST database: we show that network solutions based on empirical
estimates for mean and covariance of each class capture a large amount
of the variability within the data set, but still exhibit a non-negligible
performance gap in comparison to solutions based on the actual data
set. We discuss how this performance difference results from the omission
of higher-order correlations. We then introduce an example task where
higher-order correlations exclusively encode class membership, which
allows us to explore their role in isolation. Finally, we show that
for high-dimensional input data such as CIFAR-10, the first layer
of fully-connected networks predominantly extracts the Gaussian statistics.
As a consequence, the information processing in these networks is
well described by the Gaussian theory.

\paragraph*{Limitations}

 The dimensionality $N_{0}$ of the data may limit the applicability
of the presented approach to low orders $n$, since cumulants of
order $n$ are tensors with $N_{0}^{n}$ entries. We note, however,
that there exist methods to ease the computational cost of higher-order
cumulants in large dimensions: for example, one can make use of the
inherent symmetries in these tensors, as well as in the theory itself.
The application of such methods to our framework remains a point for
future work. A parameterization of a probability distribution in terms
of cumulants, moreover, needs to be chosen such that it maintains
positivity of the probability density function. Conserving this property
implies constraints for truncating cumulant orders, which require
further investigations.

The presented framework and its perturbative methods naturally apply
to polynomial approximations of activation functions. Although networks
with polynomial non-linearity are, in principle, not capable of universal
function approximation \citep{Cybenko1989,Leshno93_861,Pinkus99_143},
this is not an issue for the classification tasks we consider. To
obtain illustrative analytical expressions for the mixing of correlation
functions, we chose to demonstrate the approach with a quadratic activation
function. Non-polynomial and even non-differentiable activation functions
can, however, also be dealt with in our framework using Gram-Charlier
expansions that are detailed for the example of the ReLU activation
in the \prettyref{app:stats_trafos}. While we here mostly focus on
the mean and covariance, we also show how to generalize the results
to higher-order cumulants.

\paragraph*{Relation to kernel limit of deep networks}

In this paper we study \textit{individual} networks with specific
parameters $\theta$. There is a complementary approach that studies
\textit{ensembles} of (infinitely) wide networks with random parameters:
\citet{Poole16_3360} expose a relation between the Lyapunov exponents
and the depth to which information propagates in randomly initialized
deep networks. They find the regime close to chaos beneficial for
information propagation. We similarly find that the depth scale of
information propagation controls the propagation of the Gaussian statistics
across data samples studied in the current work, if network parameters
are drawn randomly (see \prettyref{app:Depth-scales}, i.p. \prettyref{fig:Depth-scale}).
Furthermore, random network parameters are central to studying training
as Bayesian inference \citep{Mackay2003}: independent Gaussian priors
on the network parameters render Bayesian inference exact on the resulting
Gaussian process \citep{Williams98,Williams06,Lee18,Jacot18_8580}.
The works \citep{Dyer20_ICLR,Naveh21_064301,Yaida20,Cohen21_023034}
use methods similar to ours to compute finite-width corrections and
corrections arising from training with stochastic gradient descent.
These approaches consider distributions over network parameters $\theta$.
The statistics of the data in this view enters in the form of the
pairwise overlaps $\sum_{i}x_{i}\,x_{i}^{\prime}$ between pairs of
patterns $x$ and $x^{\prime}$. In the large data limit, the data
statistics can moreover be described by a density $p(x)$, whose properties
shape the eigenfunctions $\phi_{i}$ of the kernel $k$ in the form
$\int\,k(x,x^{\prime})\,\phi_{i}(x^{\prime})\,p(x^{\prime})\,dx^{\prime}=\lambda_{i}\,\phi_{i}(x)$
\citep[Sec. 4.3]{WilliamsRasmussen06}. In contrast, in the present
work we study the transformation of an input distribution $p(x)$
by a network with fixed parameters $\theta$. The focus on individual
networks rather than ensembles allows us to directly take into account
the internal statistical structure of data samples, for example in
the form of the mean $\mu_{x,i}$ and covariances $\Sigma_{x,ij}$
for individual pixels $i$ and $j$ in images.

\paragraph*{Related works}

Describing data and network activity in terms of correlations was
initially explored by \citet{Deco94} on the particular architecture
of volume-preserving networks. They derived expressions of the output
in terms of its correlations as well as training rules that aim to
decorrelate given input data. The work we present here differs in
that our goal is not to impose a specific statistical structure on
the network output, but to relate the correlations of the input and
output distributions and thereby obtain a description of the information
processing within the network.

While we do show that these distributions are not exactly Gaussian,
that the networks can utilize higher-order correlations in the hidden
layers, and how these contributions could in principle be computed,
we focus mostly on self-consistently tracking the distributions in
Gaussian approximation. This is because, as we show, this approximation
is tractable while staying accurate also for trained networks and
capturing the majority of the test accuracy in our examples. That
a Gaussian approximation is surprisingly effective has also been argued
in a recent line of works using teacher-student models with realistic
data structure \citep{Goldt20_prx,Goldt22_426,Loureiro22_114001}.
We also derive conditions under which a Gaussian approximation of
the activity in the inner layers of a deep network is consistent in
the limit of wide layers: Scaling of weight amplitudes $w_{ij}\propto N^{-\frac{1}{2}}$,
weak pairwise correlations $c_{ij}=\order(N^{-1})$ as well as an
approximate pairwise orthogonal decomposition of the previous layer's
covariance matrix by the row vectors of the following layer. Under
these conditions we show that cumulants of order higher than two are
at most $\order(N^{-\frac{1}{2}})$. Our approach is inspired by and
analogous to the Gaussian equivalence property proposed by \citep{Goldt20_prx};
in particular, we also use as the central argument an expansion of
higher order cumulants caused by weak pairwise correlations. Our result
differs, though, by us treating layered networks instead of random
feature maps embedding a low dimensional manifold in \citep{Goldt20_prx}.
Other works which are based on a Gaussian approximation of the representation
in each layer are: \citep{Yang21_11727} using general deep networks,
\citep{Fang21_1887} focusing on Res-nets, and \citep{Seddik20_8573}
considering the case of GANs. Going beyond random weights, other works
study dimensionality reduction and decorrelation in both random deep
networks and trained deep belief networks \citep{Huang18_062313},
explicitly analyzing the effects of weak correlations among weights
\citep{Zhou21_012315}. Finally, a pedagogical text focusing on field-theory
for deep neural networks has recently been published \citep{Roberts22}.

\paragraph*{Outlook}

Tracing transformations of data correlations through layers of a neural
network allows the investigation of mechanisms for both information
encoding and processing; in this manner, it presents a handle towards
interpretability of deep networks. The availability of tractable expressions
describing the transformations of data correlations within neural
networks is therefore an interesting prospect for future work seeking
to dissect how networks learn and perform tasks. In this context,
the theory we propose assumes data statistics of the input distribution
$p(x)$ to be known and exposes how statistical features of the data
are transformed to generate the output, with the goal of shedding
light onto the networks' functioning principles.

Another natural application of the proposed framework is the identification
of essential correlations in the data. In that scenario, we do not
need the exact distribution $p(x)$, but only sufficiently accurate
estimates of some statistics of $x$ that can be obtained from the
training data. By manipulating the information available to the model
during training, we expose different information encodings the network
can employ to solve the same task. We believe this approach could
be used to identify data statistics required to solve a given task.

More complex data sets, such as CIFAR-10, require richer network architectures
than fully-connected feed-forward networks to achieve high performance.
For example, applying the presented approach to ResNet-50 \citep{He16_CVPR}
would require the extension to convolutional network layers and skip
connections. However, since these are equivalent to linear layers
with weight matrices of a particular shape \citep{Garriga19}, they
can straightforwardly be included in the framework.

Another future direction targets expressivity of deep networks: by
reversely tracing the data correlations through the network, from
target to data, one may ask which input distributions are mapped to
a given output distribution -- in effect constructing layer-resolved,
statistical receptive fields for each target. Expressing these receptive
fields in terms of data correlations may also be useful for studying
how the complexity of data distributions is reduced by deep neural
networks.
\begin{acknowledgments}
We are grateful to Claudia Merger and Anno Kurth for helpful discussions.
We thank Peter Bouss for feedback on an earlier version of the manuscript.
This work was partly supported by the German Federal Ministry for
Education and Research (01IS19077A and 01IS19077B), the Excellence
Initiative of the German federal and state governments (ERS PF-JARA-SDS005),
and the Helmholtz Association Initiative and Networking Fund under
project number SO-092 (Advanced Computing Architectures, ACA).
\end{acknowledgments}

\begin{appendices}

\section{Higher-order cumulants of post-activations caused by weakly correlated
pre-activations\label{app:Higher-order-correlations-caused-by-weak-Gaussian}}

\setcounter{equation}{0} \renewcommand\theequation{A\arabic{equation}}We
study how weak correlations between pre-activations affect higher-order
cumulants of the post-activations. Assume pre-activations $x,y$ are
zero-mean Gaussian distributed and weakly correlated. Let $\phi$
be a piece-wise differentiable activation function. The covariance
matrix of $x$ and $y$ be $C=\big(\begin{smallmatrix}a&c\\c&a\end{smallmatrix}\big)$.
For simplicity, we denote $\langle\circ\rangle\coloneqq\langle\circ\rangle_{(x,y)\sim\mathcal{N}(0,C)}$.
Then by Price's theorem \citep[Appendix A]{Price58_69,PapoulisProb4th,Schuecker16b_arxiv}
\begin{align*}
\frac{\partial}{\partial c}\langle f(x)g(y)\rangle & =\langle f^{\prime}(x)g^{\prime}(y)\rangle.
\end{align*}
This can be used to expand $\langle f(x)g(y)\rangle$ for small $c$
as
\begin{align}
\langle f(x)g(y)\rangle & =\langle f(x)g(y)\rangle_{c=0}+\langle f^{\prime}(x)g^{\prime}(y)\rangle_{c=0}\,c+\order(c^{2})\nonumber \\
 & =\langle f(x)\rangle\langle g(y)\rangle+\langle f^{\prime}(x)\rangle\langle g^{\prime}(y)\rangle\,c+\order(c^{2}).\label{eq:expand_low_corr}
\end{align}
This expression corresponds to Eq. (A4) in \citep{Goldt20_prx}, but
Goldt et al. use a different approach than Price's theorem. In the
expression in \citep{Goldt20_prx} one needs to replace $\langle uf(u)\rangle=\langle f^{\prime}(u)\rangle$),
which holds since they assume $\langle u^{2}\rangle=1$.

Next, we consider the centered variables
\begin{align*}
\tilde{f}(x) & :=f(x)-\langle f(x)\rangle
\end{align*}
and correspondingly for $g$, one gets
\begin{align*}
\langle\tilde{f}(x)\tilde{g}(y)\rangle & =\langle f^{\prime}(x)\rangle\langle g^{\prime}(y)\rangle\,c+\order(c^{2}).
\end{align*}
We may generalize this property to expectation values of more than
two functions $f$, $g$
\begin{align}
F_{n}(x) & :=\Big\langle\prod_{k=1}^{n}\tilde{f}_{k}(x_{k})\Big\rangle.\label{eq:exp_higher_order}
\end{align}
By the marginalization property of Gaussian distributions, the joint
distribution of any subset of $x_{i}$ is Gaussian distributed, too,
where the covariance matrix is the corresponding sector of the matrix
$C_{ij}=\llangle x_{i}x_{j}\rrangle$. Therefore for any $i,j$ we
define the function 
\begin{align*}
F_{n}(x\backslash\{x_{i},x_{j}\}) & =\Big\langle\prod_{k=1}^{n}\tilde{f}_{k}(x_{k})\Big\rangle_{(x_{i},x_{j})}\\
 & =\langle\tilde{f}_{i}(x_{i})\tilde{f}_{j}(x_{j})\rangle_{(x_{i},x_{j})}\;\prod_{k\backslash\{i,j\}}\tilde{f}_{k}(x_{k}).
\end{align*}
Applying \eqref{eq:expand_low_corr} to the first term yields
\begin{equation}
\begin{aligned}F_{n}(x\backslash\{x_{i},x_{j}\}) & =\big[c_{ij}\,\langle f_{i}^{\prime}(x_{i})f_{j}^{\prime}(x_{j})\rangle_{(x_{i},x_{j}),c_{ij}=0}+\order(c_{ij}^{2})\big]\\
 & \quad\times\prod_{k\backslash\{i,j\}}\tilde{f}_{k}(x_{k}).
\end{aligned}
\label{eq:expand_one_pair}
\end{equation}
Now take the expectation also across the remaining variables $x\backslash\{x_{i},x_{j}\}$
with probability $p(x\backslash\{x_{i},x_{j}\})$. We may consider
$p(x_{1},\ldots,x_{N})=p(x_{i},x_{j}|x\backslash\{x_{i},x_{j}\})\,p(x\backslash\{x_{i},x_{j}\})$
and use \eqref{eq:expand_one_pair} for the conditional expectation
value over $x_{i},x_{j}$ with regard to $p(x_{i},x_{j}|x\backslash\{x_{i},x_{j}\})$,
so that it follows
\begin{flalign*}
 & \big\langle F_{n}(x\backslash\{x_{i},x_{j}\})\big\rangle_{x\backslash\{x_{i},x_{j}\}}\\
 & =\Big\langle\big[c_{ij}\,f_{i}^{\prime}(x_{i})f_{j}^{\prime}(x_{j})+\order(c_{ij}^{2})\big]\;\prod_{k\backslash\{i,j\}}\tilde{f}_{k}(x_{k})\Big\rangle_{x,c_{ij}=0}
\end{flalign*}
The pair $(i,j)$ has been chosen arbitrary. The remaining factors
$\prod_{k\backslash\{i,j\}}\tilde{f}_{k}(x_{k})$ can now be expanded
in a similar manner, where all remaining $k\backslash\{i,j\}$ need
to be paired. Any such pairing yields non-zero contributions. Together
one therefore has
\begin{align}
\big\langle F_{n}(x)\big\rangle_{x} & =\sum_{\sigma\in\Pi}\,c_{\sigma(1)\sigma(2)}\,\langle f_{\sigma(1)}^{\prime}(x_{\sigma(1)})\rangle\langle f_{\sigma(2)}^{\prime}(x_{\sigma(2)})\rangle\nonumber \\
 & \phantom{=\sum_{\sigma\in\Pi}\,}\cdots c_{\sigma(n-1)\sigma(n)}\,\langle f_{\sigma(n-1)}^{\prime}(x_{\sigma(n-1)})\rangle\langle f_{\sigma(n)}^{\prime}(x_{\sigma(n)})\rangle\nonumber \\
 & \phantom{=}+\order(c_{\circ\circ}^{\frac{n}{2}+1}),\label{eq:expansion_weak_corr}
\end{align}
where $\sum_{\sigma\in\Pi}$ sums over all disjoint pairings of indices
(This expression corresponds to A16 in \citep{Goldt20_prx}, apart
from minor typos; the factors $b_{i}$ seem to be missing, $p$ should
be $m$, and we interpret the upper case of their A16 to be meant
as $b_{1}\cdots b_{m}\,\sum_{\sigma\pi\Pi}\,m_{\sigma_{1}\sigma_{2}}\cdots m_{\sigma_{m-1}\sigma_{m}}$).
This expression is also consistent with Wick's theorem, to which it
needs to reduce in the case of an identity mapping $f(x)=x$.

The expansion \eqref{eq:expansion_weak_corr} holds for arbitrary
$n$. For any $n$, the result is correct up to terms of order $\order(c^{\frac{n}{2}})$.
All cumulants of order $n\ge3$ thus vanish at the given order $\order(c^{\frac{n}{2}})$.
This can be exemplified on the fourth order (dropping the arguments
$x$ for brevity)
\begin{align}
\llangle\tilde{f}_{1}\,\tilde{f}_{2}\,\tilde{f}_{3}\,\tilde{f}_{4}\rrangle & =\langle\tilde{f}_{1}\,\tilde{f}_{2}\,\tilde{f}_{3}\,\tilde{f}_{4}\rangle\nonumber \\
 & -\langle\tilde{f}_{1}\,\tilde{f}_{2}\rangle\,\langle\tilde{f}_{3}\,\tilde{f}_{4}\rangle\nonumber \\
 & -\langle\tilde{f}_{1}\,\tilde{f}_{3}\rangle\,\langle\tilde{f}_{2}\,\tilde{f}_{4}\rangle\nonumber \\
 & -\langle\tilde{f}_{1}\,\tilde{f}_{4}\rangle\,\langle\tilde{f}_{2}\,\tilde{f}_{3}\rangle.\label{eq:cum_mom}
\end{align}
The first line on the right hand side, according to \eqref{eq:expansion_weak_corr}
is 
\begin{align*}
\langle\tilde{f}_{1}\,\tilde{f}_{2}\,\tilde{f}_{3}\,\tilde{f}_{4}\rangle & =c_{12}\,\langle f_{1}^{\prime}\rangle\,\langle f_{2}^{\prime}\rangle\,c_{34}\,\langle f_{3}^{\prime}\rangle\,\langle f_{4}^{\prime}\rangle\\
 & +c_{13}\,\langle f_{1}^{\prime}\rangle\,\langle f_{3}^{\prime}\rangle\,c_{24}\,\langle f_{2}^{\prime}\rangle\,\langle f_{4}^{\prime}\rangle\\
 & +c_{14}\,\langle f_{1}^{\prime}\rangle\,\langle f_{4}^{\prime}\rangle\,c_{23}\,\langle f_{2}^{\prime}\rangle\,\langle f_{3}^{\prime}\rangle+\order(c^{3}).
\end{align*}
Expanding each of the three negative terms on the right hand side
of \eqref{eq:cum_mom} with help of \eqref{eq:expansion_weak_corr}
yields, for example for the first of them
\begin{align*}
-\langle\tilde{f}_{1}\,\tilde{f}_{2}\rangle\,\langle\tilde{f}_{3}\,\tilde{f}_{4}\rangle= & -c_{12}\,\langle f_{1}^{\prime}\rangle\,\langle f_{2}^{\prime}\rangle\,c_{34}\,\langle f_{3}^{\prime}\rangle\,\langle f_{4}^{\prime}\rangle+\order(c^{3}),
\end{align*}
which precisely cancels the corresponding term in \eqref{eq:cum_mom}
at the given accuracy $\order(c^{3})$. Analogous results hold at
any even order (odd orders vanish for the centered variables), so
that we find
\begin{align}
\Big\langle\negthinspace\Big\langle\prod_{k=1}^{n}\tilde{f}_{k}(x_{k})\Big\rangle\negthinspace\Big\rangle & =\order\big(c^{\frac{n}{2}+1}\big).\label{eq:higher_cumulants_scaling}
\end{align}

We also need to consider the case that indices in \eqref{eq:cum_mom}
repeat, for example $\llangle\tilde{f}_{1}\,\tilde{f}_{1}\,\tilde{f}_{2}\,\tilde{f}_{2}\rrangle$.
In general, assume we have $r$ different indices $j_{1},\ldots,j_{r}$
among the $n$ indices and want to compute the $n$-th cumulant for
$n>r$. Within a set of repeated variables correlations are of order
$\order(1)$ instead of $\order(c)$. In the expansion \eqref{eq:expansion_weak_corr}
variables with repeated indices must be treated as a single variable.
For the given example, define $g_{i}:=\tilde{f}_{i}^{2},\;i=1,2$
and centered variables $\tilde{g}_{i}:=\tilde{f}_{i}^{2}-\langle\tilde{f}_{i}^{2}\rangle$.
One then has with \eqref{eq:expansion_weak_corr}

\begin{align}
\langle\tilde{g}_{1}\,\tilde{g}_{2}\rangle & =c_{12}\,\langle g_{1}^{\prime}\rangle\,\langle g_{2}^{\prime}\rangle+\order(c^{2}),\label{eq:forth_repeated}
\end{align}
which is also the second cumulant, because the $\tilde{g}$ are centered.
We then expand the forth cumulant with repeated indices analogous
to \eqref{eq:cum_mom} as
\begin{align}
\llangle\tilde{f}_{1}\,\tilde{f}_{1}\,\tilde{f}_{2}\,\tilde{f}_{2}\rrangle & =\langle\tilde{f}_{1}\,\tilde{f}_{1}\,\tilde{f}_{2}\,\tilde{f}_{2}\rangle\nonumber \\
 & -\langle\tilde{f}_{1}\,\tilde{f}_{1}\rangle\,\langle\tilde{f}_{2}\,\tilde{f}_{2}\rangle\nonumber \\
 & -\langle\tilde{f}_{1}\,\tilde{f}_{2}\rangle\,\langle\tilde{f}_{1}\,\tilde{f}_{2}\rangle\nonumber \\
 & -\langle\tilde{f}_{1}\,\tilde{f}_{2}\rangle\,\langle\tilde{f}_{1}\,\tilde{f}_{2}\rangle.\label{eq:forth_cum_repeated}
\end{align}
The fourth moment in the first line, using the definitions of $g$
and $\tilde{g}$ above as well as \eqref{eq:forth_repeated}, is
\begin{align*}
\langle\tilde{f}_{1}\,\tilde{f}_{1}\,\tilde{f}_{2}\,\tilde{f}_{2}\rangle & \stackrel{\phantom{(\ref{eq:forth_repeated})}}{=}\langle g_{1}\,g_{2}\rangle\\
 & \stackrel{\phantom{(\ref{eq:forth_repeated})}}{=}\langle\tilde{g}_{1}\,\tilde{g}_{2}\rangle+\langle g_{1}\rangle\langle g_{2}\rangle\\
 & \stackrel{(\ref{eq:forth_repeated})}{=}c_{12}\,\langle g_{1}^{\prime}\rangle\,\langle g_{2}^{\prime}\rangle+\langle g_{1}\rangle\langle g_{2}\rangle+\order(c^{2}).
\end{align*}
Combined with \eqref{eq:forth_cum_repeated} one has
\begin{align*}
\llangle\tilde{f}_{1}\,\tilde{f}_{1}\,\tilde{f}_{2}\,\tilde{f}_{2}\rrangle & =c_{12}\,\langle g_{1}^{\prime}\rangle\,\langle g_{2}^{\prime}\rangle+\order(c^{2}),
\end{align*}
where we dropped all terms of order $\order(c^{2})$, such as $\langle\tilde{f}_{1}\,\tilde{f}_{2}\rangle\,\langle\tilde{f}_{1}\,\tilde{f}_{2}\rangle=\order(c_{12}^{2})$
and used that $\langle\tilde{f}_{1}\,\tilde{f}_{1}\rangle\,\langle\tilde{f}_{2}\,\tilde{f}_{2}\rangle=\langle g_{1}\rangle\langle g_{2}\rangle$.

Now consider that $r$ is odd, such as in 
\begin{align}
\llangle\tilde{f}_{1}\,\tilde{f}_{2}\,\tilde{f}_{2}\,\tilde{f}_{3}\rrangle & =\langle\tilde{f}_{1}\,\tilde{f}_{2}\,\tilde{f}_{2}\,\tilde{f}_{3}\rangle\nonumber \\
 & -\langle\tilde{f}_{1}\,\tilde{f}_{2}\rangle\,\langle\tilde{f}_{2}\,\tilde{f}_{3}\rangle\nonumber \\
 & -\langle\tilde{f}_{1}\,\tilde{f}_{2}\rangle\,\langle\tilde{f}_{2}\,\tilde{f}_{3}\rangle\nonumber \\
 & -\langle\tilde{f}_{1}\,\tilde{f}_{3}\rangle\,\langle\tilde{f}_{2}\,\tilde{f}_{2}\rangle.\label{eq:forth_3_diff}
\end{align}
The fourth moment then is

\begin{align*}
\langle\tilde{f}_{1}\,\tilde{f}_{2}\,\tilde{f}_{2}\,\tilde{f}_{3}\rangle & \stackrel{\phantom{(\ref{eq:forth_repeated})}}{=}\langle\tilde{f}_{1}\,g_{2}\,\tilde{f}_{3}\rangle\\
 & \stackrel{\phantom{(\ref{eq:forth_repeated})}}{=}\underbrace{\langle\tilde{f}_{1}\,\tilde{g}_{2}\,\tilde{f}_{3}\rangle}_{\order(c^{2})}+\langle g_{2}\rangle\,\langle\tilde{f}_{1}\tilde{f}_{3}\rangle\\
 & \stackrel{(\ref{eq:forth_repeated})}{=}\langle g_{2}\rangle\,\langle f_{1}^{\prime}\rangle\langle f_{3}^{\prime}\rangle\,c_{13}+\order(c^{2}).
\end{align*}
Applied to \eqref{eq:forth_3_diff}, we have
\begin{align*}
\llangle\tilde{f}_{1}\,\tilde{f}_{2}\,\tilde{f}_{2}\,\tilde{f}_{3}\rrangle & =\langle g_{2}\rangle\,\langle f_{1}^{\prime}\rangle\langle f_{3}^{\prime}\rangle\,c_{13}\\
 & \quad-\langle\tilde{f}_{1}\,\tilde{f}_{3}\rangle\,\langle\tilde{f}_{2}\,\tilde{f}_{2}\rangle+\order(c^{2})\\
 & =\order(c^{2}),
\end{align*}
where the terms $\propto c$ cancel exactly. These two examples show
the structure of the expansion: If we have $r$ different indices
$j_{1},\ldots,j_{r}$, the $n$-th cumulant for $n>r$ of these variables
will be of the order in $c$ that equals the number of pairs to join
all different indices. So together ($r$ even or odd) we get
\begin{align}
\Big\langle\negthinspace\Big\langle\prod_{k=1}^{n}\tilde{f}_{j_{k}}(x_{j_{k}})\Big\rangle\negthinspace\Big\rangle & =\order\big(c^{\lceil\frac{r}{2}\rceil}\big).\label{eq:high_cum_repeated}
\end{align}
This expression describes the scaling of the higher-order cumulants
of post-activations with weak correlations of the pre-activations.

\section{Weakly-correlated Gaussian network mapping\label{app:Weakly-correlated}}

\setcounter{equation}{0} \renewcommand\theequation{B\arabic{equation}}The
application of a non-linear activation function $\phi$ in each network
layer generates higher-order cumulants from Gaussian distributed pre-activations,
as discussed in \prettyref{subsec:trafo_data_statistics}. We here
derive conditions under which higher-order cumulants $G_{z^{l}}^{(n)}$
of the pre-activations $z^{l}$ beyond mean and covariance on expectation
scale down with the layer width $N$. Consequently, these become negligible
for wide networks where $N\gg1$.

We apply the considerations in the previous \prettyref{app:Higher-order-correlations-caused-by-weak-Gaussian}
of weakly correlated Gaussian variables to the network mapping. Pre-activations
in layer $l$ are given by

\begin{align}
z_{i}^{l} & =\sum_{a=1}^{N}\,W_{ia}^{l}\,y_{a}^{l-1}+b_{i}^{l},\label{eq:affine_lin}
\end{align}
which then produce post-activations
\begin{align}
y_{i}^{l} & =\phi\big(z_{i}^{l}\big).\label{eq:activation}
\end{align}
Assume the pre-activations $z_{i}^{l}$ are Gaussian and weakly correlated
to order $\epsilon$
\begin{align}
\llangle z_{i}z_{j}\rrangle & \stackrel{i\neq j}{=}\order(\epsilon).\label{eq:weak_pairwise_pre}
\end{align}
We want to derive conditions under which it then follows that also
pre-activations $z_{i}^{l+1}$ in the next layer have this property.
By induction through the layer index, one then has established a condition
under which the neglect of non-Gaussian cumulants in the inner layers
of the network is justified. To this end, we define centered variables
\begin{align*}
\tilde{z} & :=z-\langle z\rangle
\end{align*}
as well as
\begin{align*}
y=f(\tilde{z}) & :=\phi\big(\langle z\rangle+\tilde{z}\big)\\
\tilde{y}=\tilde{f}(\tilde{z}) & :=f(\tilde{z})-\big\langle f(\tilde{z})\big\rangle.
\end{align*}
The variance of pre-activations should be of order unity
\begin{align*}
\langle\big(\tilde{z}_{a}^{l}\big)^{2}\rangle & =\order(1),
\end{align*}
because one aims to explore the dynamic range of the gain function,
which we assume to be of order unity (we use the dynamic range of
the gain function to define our scale). It then follows that also
the post-activations have a variance of order unity, so
\begin{align*}
\langle\big(\tilde{f}_{a}^{l}\big)\rangle & =\order(1).
\end{align*}
Such conditions are typically also enforced by batch-normalization.
If the $y_{a}^{l}$ were uncorrelated, the variance of the pre-activations
in the next layer is given by
\begin{align*}
\order(1)\stackrel{!}{=}\langle\big(\tilde{z}_{i}^{l+1}\big)^{2}\rangle & =\sum_{a}\,\big[W_{ia}^{l+1}\big]^{2}\,\langle\big(\tilde{f}_{a}^{l}\big)^{2}\rangle.
\end{align*}
For the variances on both sides to be of order unity, we need that
\begin{align}
W_{ia}^{l+1} & =\order(N^{-\frac{1}{2}}),\label{eq:scaling_weights}
\end{align}
which means that rows and columns of the matrix $W^{l+1}$ are vectors
with lengths of order unity.

Now assume the presence of correlations of order
\begin{align}
\langle\tilde{y}_{a}^{l}\tilde{y}_{b}^{l}\rangle=\langle\tilde{f}_{a}^{l}\tilde{f}_{b}^{l}\rangle=:C_{ab} & =\order(\epsilon)\label{eq:weak_pairwise}
\end{align}
between the outputs of layer $l$ across different neurons $a\neq b$.
The expression for the variance then changes to
\begin{align}
\langle\big(\tilde{z}_{i}^{l+1}\big)^{2}\rangle & =\sum_{a,b}\,W_{ia}^{l+1}W_{ib}^{l+1}\,C_{ab}.\label{eq:var_general}
\end{align}
To have low correlations in the next layer, one needs to demand that
\begin{align}
\order(\epsilon)\stackrel{!}{=}\langle\tilde{z}_{i}^{l+1}\,\tilde{z}_{j}^{l+1}\rangle & =\sum_{a,b}\,W_{ia}^{l+1}W_{jb}^{l+1}\,C_{ab}\quad\forall i\neq j.\label{eq:weak_corr_pre-act}
\end{align}
This can be interpreted as demanding that different rows $W_{i\circ}^{l+1}$
and $W_{j\circ}^{l+1}$ project out mutually nearly orthogonal sub-spaces
out of the space of principal components of $C$. This means that
different neurons $i$ and $j$ each specialize on sub-spaces that
have little mutual overlap.

Now consider higher-order correlations. It follows from \eqref{eq:higher_cumulants_scaling}
and from the condition of weak pairwise correlations \eqref{eq:weak_pairwise_pre}
that for $n\ge3$
\begin{align}
\Big\langle\negthinspace\Big\langle\prod_{i=1}^{n}y_{i}^{l}\Big\rangle\negthinspace\Big\rangle & =\Big\langle\negthinspace\Big\langle\prod_{i=1}^{n}\tilde{f}_{i}(\tilde{z}_{i}^{l})\Big\rangle\negthinspace\Big\rangle=\order\big(\epsilon^{\frac{n}{2}+1}\big).\label{eq:scaling_higher_order}
\end{align}
The cumulants of the pre-activations $z_{i}^{l+1}$ are given by those
of the post-activations as

\begin{align}
\llangle z_{i_{1}}^{l+1}\cdots z_{i_{n}}^{l+1}\rrangle & =\sum_{j_{1},\ldots,j_{n}=1}^{N}\,W_{i_{1}j_{1}}^{l+1}\cdots W_{i_{n}j_{n}}^{l+1}\,\Big\langle\negthinspace\Big\langle\prod_{k=1}^{n}\tilde{f}_{j_{k}}(\tilde{z}_{j_{k}}^{l})\Big\rangle\negthinspace\Big\rangle.\label{eq:high_order_cum}
\end{align}
We now distinguish three cases:

1.) Diagonal contributions: First consider the special case where
all indices $j_{1}=\ldots=j_{n}$ are identical. One then gets a contribution
to \eqref{eq:high_order_cum} at order $n\ge3$
\begin{align}
 & \sum_{j=1}^{N}W_{i_{1}j}^{l+1}\cdots W_{i_{n}j}^{l+1}\,\underbrace{\big\langle\negthinspace\big\langle\big(\tilde{f}_{j}\big)^{n}\big\rangle\negthinspace\big\rangle}_{\order(1)}\label{eq:correction_diagonal}\\
\overset{\phantom{n\ge3}}{=} & \order\big(N^{1-\frac{n}{2}}\big)\nonumber \\
\stackrel{n\ge3}{<} & \order\big(N^{-\frac{1}{2}}\big)\nonumber 
\end{align}
For $n\ge3$ this is suppressed by a large layer width $N$.

2.) Off-diagonal contributions with all distinct indices: Next consider
the off-diagonal terms, where all sending neurons' indices are unequal
$j_{1}\neq j_{2}\neq\ldots\neq j_{n}$, so that we can use \eqref{eq:scaling_higher_order}.
For $n$ odd, the contributions vanish, because then \eqref{eq:scaling_higher_order}
vanishes. For $n$ even we get

\begin{align}
 & \sum_{(j_{1}\neq j_{2},\ldots,\neq j_{n})=1}^{N}\,W_{i_{1}j_{1}}^{l+1}\cdots W_{i_{n}j_{n}}^{l+1}\,\Big\langle\negthinspace\Big\langle\prod_{k=1}^{n}\tilde{f}_{j_{k}}(\tilde{z}_{j_{k}}^{l})\Big\rangle\negthinspace\Big\rangle\nonumber \\
 & \stackrel{\phantom{N\gg n}}{=}\order\big(\frac{N!}{(N-n)!}\,N^{-\frac{n}{2}}\,\epsilon^{\frac{n}{2}+1}\big)\nonumber \\
 & \stackrel{N\gg n}{=}\order\big(N^{\frac{n}{2}}\,\epsilon^{\frac{n}{2}+1}\big)\label{eq:off_diag}
\end{align}
For these contributions to be suppressed for $n\ge3$ with increasing
network size, we thus need to demand that the order of pairwise correlations
$\epsilon$ is at most 
\begin{align*}
\epsilon & =\order(N^{-1}),
\end{align*}
so that the off-diagonal contribution \eqref{eq:off_diag} is
\begin{align*}
 & \order\big(N^{-1}\big),
\end{align*}
which is hence suppressed with network size also for large orders
$n$.

3.) Off-diagonal contributions with two or more equal indices: Now
consider terms for which a subset of $j_{a},j_{b},j_{c},\ldots$ assume
the same value. Let the number of disjoint indices $j_{1}\neq j_{2}\neq\ldots\neq j_{r}$
be $r<n$. Each pair of equal indices can be seen as the appearance
of one Kronecker $\delta_{j_{a}j_{b}}$, which eliminates one summation
$\sum_{j=1}^{N}$ -- hence one factor $N$ less. But at the same
time, by \eqref{eq:forth_cum_repeated}, also the moments are increased
$\langle\prod_{k=1}^{n}\tilde{f}_{j_{k}}(\tilde{z}_{j_{k}}^{l})\rangle=\order\big(\epsilon^{\lceil\frac{r}{2}\rceil}\big)$.
Together, we get a contribution
\begin{align}
 & \order\big(\frac{N!}{(N-r)!}\,N^{-\frac{n}{2}}\epsilon^{\lceil\frac{r}{2}\rceil}\big)\nonumber \\
 & \stackrel{N\gg1,\:\epsilon=\order(N^{-1})}{=}\order\big(N^{r}N^{-\frac{n}{2}}N^{-\lceil\frac{r}{2}\rceil}\big)\nonumber \\
 & =\order\big(N^{\lfloor\frac{r}{2}\rfloor-\frac{n}{2}}\big)\nonumber \\
 & <\order(N^{-\frac{1}{2}}),\label{eq:correction_partial_off}
\end{align}
where we used $r<n$ in the last step and upper bounded the expression
by the worst case, in which $n=r+1$, where $n$ is odd. So contributions
from partial diagonal terms are suppressed with network size, too.

In summary, we have shown that the Gaussian approximation with weak
pairwise correlations of order $\epsilon<\order(N^{-1})$ is consistently
maintained in the limit of wide networks $N\gg1$ if synaptic amplitudes
scale as \eqref{eq:scaling_weights} and if the rows of the connectivity
$W^{l}$ in each layer $l$ in addition obey the approximate orthonormality
condition \eqref{eq:weak_corr_pre-act}. From a functional perspective
the latter condition makes sense, because this condition assures that
the $N$ neurons in each layer are used effectively to represent the
entire variability that is present in the previous layer, avoiding
redundancy among neurons.

Finally, we note that \prettyref{eq:weak_corr_pre-act} is fulfilled
for Gaussian initialized, untrained networks. The network parameters
$\theta=\{W^{l},b^{l}\}_{l=1,\dots,L+1}$ are drawn i.i.d. from zero-mean
Gaussians $W_{rs}^{l}\overset{\mathrm{i.i.d.}}{\sim}\mathcal{N}\left(0,\nicefrac{\sigma_{w}^{2}}{N_{l-1}}\right)$
and $b_{r}^{l}\overset{\mathrm{i.i.d.}}{\sim}\mathcal{N}\left(0,\sigma_{b}^{2}\right)$.
This choice of initialization precisely preserves the magnitude of
the covariance within the network:
\begin{align*}
\langle\tilde{z}_{i}^{l+1}\,\tilde{z}_{j}^{l+1}\rangle & \overset{\phantom{N\gg1}}{=}\sum_{a,b}\,W_{ia}^{l+1}W_{jb}^{l+1}\,C_{ab}^{l}\\
 & \overset{N\gg1}{\approx}\langle\sum_{a,b}\,W_{ia}^{l+1}W_{jb}^{l+1}\,C_{ab}^{l}\rangle_{w}\\
 & \overset{\phantom{N\gg1}}{=}\delta_{ij}\,\frac{\sigma_{w}^{2}}{N}\sum_{a}C_{aa}^{l}=\mathcal{O}(\epsilon)\,.
\end{align*}
Due to the resulting covariance in the next layer being approximately
diagonal, the calculations simplify significantly in this case. The
above considerations include conditions also for trained networks
where correlations among weights cause correlations between pairs
of pre-activations $(z_{i}^{l},z_{j}^{l})$.

\section{Interaction functions for different activation functions\label{app:stats_trafos}}

\setcounter{equation}{0}\renewcommand\theequation{C\arabic{equation}}In
\prettyref{subsec:trafo_data_statistics}, we derived the interaction
functions resulting from the non-linearity $\phi$,
\begin{align*}
f_{\mu}(\mu_{z^{l}},\Sigma_{z^{l}}) & =\langle\phi(z^{l})\rangle_{\extralowerindex z^{l}}\,,\\
f_{\Sigma}(\mu_{z^{l}},\Sigma_{z^{l}}) & =\langle\phi(z^{l})\,\phi(z^{l})^{\mathsf{T}}\rangle_{\extralowerindex z^{l}}-\mu_{y^{l}}\mu_{y^{l}}^{\mathsf{T}}\,.
\end{align*}
\prettyref{tab:interaction_functions} gives these expressions for
the $\mathrm{ReLU}$ and quadratic non-linearities.
\begin{table*}[!t]
\begin{equation*}
\begin{array}{ll} \midrule\midrule
\text{Non-linearity} & \text{Interaction function}  \\
\midrule
\phi(z)=\mathrm{ReLU}(z) \quad
& 
  f_{\mu,\,i} =
  \frac{\sqrt{\Sigma_{z^{l},\,ii}}}{\sqrt{2\pi}}\,
  \exp\left(-\frac{\mu_{z^{l},\,i}^{2}}{2\Sigma_{z^{l},\,ii}}\right)
  +\frac{\mu_{z^{l},\,i}}{2}\left(1+\text{erf}\left(\frac{\mu_{z^{l},\,i}}{\sqrt{2\Sigma_{z^{l},\,ii}}}\right)\right)
\\[20pt]
& \begin{aligned}[t]
f_{\Sigma,\,ii} &= 
  \frac{\Sigma_{z^{l},\,ii}}{2}\,
  \biggl(1+\text{erf}\biggl(\frac{\mu_{z^{l},\,i}}{\sqrt{2\Sigma_{z^{l},\,ii}}}\biggr)\biggr)
  +\frac{\mu_{z^{l},\,i}^{2}}{4}-\biggl(
    \frac{\sqrt{\Sigma_{z^{l},\,ii}}}{\sqrt{2\pi}}\,
    \exp\biggl(-\frac{\mu_{z^{l},\,i}^{2}}{2\Sigma_{z^{l},\,ii}}\biggr)
    +\frac{\mu_{z^{l},\,i}}{2}\,\text{erf}\biggl(\frac{\mu_{z^{l},\,i}}{\sqrt{2\Sigma_{z^{l},\,ii}}}\biggr)
  \biggr)^{2}
\end{aligned} \\[15pt]
& \begin{aligned}
f_{\Sigma,\,ij} &=
  \frac{\sqrt{\det(\tilde{\Sigma}_{z^{l}})}}{2\pi}\,
    \exp\Bigl(-\tfrac{1}{2}\tilde{\mu}_{z^{l}}^{\mathsf{T}}\tilde{\Sigma}_{z^{l}}^{-1}\tilde{\mu}_{z^{l}}\Bigr)\\
  &\quad
    +\frac{\sqrt{\det(\tilde{\Sigma}_{z^{l}})}}{2\pi}\,
    \mu_{z^{l},\,j}
    \frac{\sqrt{\pi\,\tilde{\Sigma}_{z^{l},\,jj}^{-1}}}{\sqrt{2}}\,
    \exp\left(-\tfrac{1}{2}\tilde{\Sigma}_{z^{l},\,ii}^{-1}\,\mu_{z^{l},\,i}^{2}\right)\, \exp\Biggl(
      \frac{\bigl(\tilde{\Sigma}_{z^{l},\,ji}^{-1}\,\mu_{z^{l},\,i}\bigr)^{2}}
           {2\tilde{\Sigma}_{z^{l},\,jj}^{-1}}
    \Biggr)\,
    \left(1+\text{erf}\left(
      \frac{\Bigl(\tilde{\Sigma}_{z^{l}}^{-1}\tilde{\mu}_{z^{l}}\Bigr)_{j}}
           {\sqrt{2\tilde{\Sigma}_{z^{l},\,jj}^{-1}}}\right)\right) \\[0.5ex]
  &\quad
    +\frac{\sqrt{\det(\tilde{\Sigma}_{z^{l}})}}{2\pi}\,
    \mu_{z^{l},\,i}
    \frac{\sqrt{\pi\,\tilde{\Sigma}_{z^{l},\,ii}^{-1}}}{\sqrt{2}}\,
    \exp\left(-\tfrac{1}{2}\tilde{\Sigma}_{z^{l},\,jj}^{-1}\,\mu_{z^{l},\,j}^{2}\right)\, \exp\Biggl(
      \frac{\bigl(\tilde{\Sigma}_{z^{l},\,ij}^{-1}\,\mu_{z^{l},\,j}\bigr)^{2}}
           {2\tilde{\Sigma}_{z^{l},\,ii}^{-1}}
      \Biggr) \,
    \left(1+\text{erf}\left(
      \frac{\left(\tilde{\Sigma}_{z^{l}}^{-1}\tilde{\mu}_{z^{l}}\right)_{i}}
           {\sqrt{2\tilde{\Sigma}_{z^{l},\,ii}^{-1}}}\right)\right) \\
  & \quad
    +\biggl[\mu_{z^{l},\,i}\,\mu_{z^{l},\,j}-\tilde{\Sigma}_{z^{l},\,ij}^{-1}\,\det\left(\tilde{\Sigma}_{z^{l}}\right)\biggr]\, \left[
      \frac{1}{2}\text{erf}\left(\frac{\sqrt{2}\,\mu_{z^{l},\,i}}{\sqrt{\Sigma_{z^{l},\,ii}}}\right)
      + \frac{1}{2}\text{erf}\left(\frac{\sqrt{2}\,\mu_{z^{l},\,j}}{\sqrt{\Sigma_{z^{l},\,jj}}}\right)
      + F_{\tilde{\mu}_{z^{l}}\,,\tilde{\Sigma}_{z^{l}}}(0,0)
      \right] \\[1ex]
  & \quad
    -\left[
      \frac{\sqrt{\Sigma_{z^{l},\,ii}}}{\sqrt{2\pi}}\,
        \exp\biggl(-\frac{\mu_{z^{l},\,i}^{2}}{2\Sigma_{z^{l},\,ii}}\biggr) 
      +\frac{\mu_{z^{l},\,i}}{2}
        \biggl(1+\text{erf}\biggl(
          \frac{\mu_{z^{l},\,i}}{\sqrt{2\Sigma_{z^{l},\,ii}}}
        \biggr)\biggr)
    \right] \\
  & \quad \hphantom{-\bigl[} \times \left[
      \frac{\sqrt{\Sigma_{z^{l},\,jj}}}{\sqrt{2\pi}}\,
        \exp\biggl(-\frac{\mu_{z^{l},\,j}^{2}}{2\Sigma_{z^{l},\,jj}}\biggr)
      +\frac{\mu_{z^{l},\,j}}{2}
        \biggl(1+\text{erf}\biggl(
          \frac{\mu_{z^{l},\,j}}{\sqrt{2\Sigma_{z^{l},\,jj}}}
        \biggr)\biggr)
      \right]
\end{aligned}\\[4cm]
\phi(z)=z+\alpha\,z^{2}
& f_{\mu,\,i} = 
  \mu_{z^{l},\,i}
  +\alpha\,(\mu_{z^{l},\,i})^{2}+\alpha\,\Sigma_{z^{l},\,ii} \\[5pt]
& \begin{aligned}[t]
f_{\Sigma,\,ij} &=
  \Sigma_{z^{l},\,ij}
  +2\,\alpha\,\Sigma_{z^{l},\,ij}\,\left(\mu_{z^{l},\,i}+\mu_{z^{l},\,j}\right)
  +2\,\alpha^{2}\,(\Sigma_{z^{l},\,ij})^{2}+4\,\alpha^{2}\,\mu_{z^{l},\,i}\,\Sigma_{z^{l},\,ij}\,\mu_{z^{l},\,j}
\end{aligned} \\
\midrule\midrule
\end{array}
\end{equation*}\caption{Interaction functions for different non-linearities $\phi$. We assume
$z^{l}\sim\mathcal{N}(\mu_{z^{l}},\,\Sigma_{z^{l}})$ in both examples.
For $\mathrm{ReLU}$, $\tilde{\mu}_{z^{l}}$ and $\tilde{\Sigma}_{z^{l}}$
denote the marginalized mean and covariance with respect to $\tilde{z}^{l}=(z_{i}^{l},z_{j}^{l})^{\mathsf{T}}$,
and $F_{\tilde{\mu}_{z^{l}}\,,\tilde{\Sigma}_{z^{l}}}(x,y)$ denotes
the corresponding cumulative distribution function.\label{tab:interaction_functions}}
\end{table*}

\subsection*{Derivations for $\mathrm{ReLU}$ activations}

We here consider networks with the $\mathrm{ReLU}$ activation function
$\phi(z)=\max(0,z)$. Taking the distribution of pre-activations $z^{l}$
to be Gaussian distributed with mean $\mu_{z^{l}}$ and covariance
$\Sigma_{z^{l}}$, the mean post-activations are given by
\begin{align}
\mu_{y^{l},\,i} & =\langle\max(0,z_{i}^{l})\rangle_{\extralowerindex z^{l}\sim\mathcal{N}(\mu_{z^{l}},\,\Sigma_{z^{l}})}\\
 & =\frac{1}{\sqrt{2\pi\,\Sigma_{z^{l},\,ii}}}\int_{0}^{\infty}\,\mathrm{d}z_{i}^{l}\,z_{i}^{l}\,\exp\Biggl(-\frac{\left(z_{i}^{l}-\mu_{z^{l},\,i}\right)^{2}}{2\Sigma_{z^{l},\,ii}}\Biggr)\\
 & =-\frac{\sqrt{\,\Sigma_{z^{l},\,ii}}}{\sqrt{2\pi}}\int_{-\mu_{z^{l},\,i}}^{\infty}\,\mathrm{d}z_{i}^{l}\,\frac{-z_{i}^{l}}{\,\Sigma_{z^{l},\,ii}}\,\exp\Biggl(-\frac{(z_{i}^{l})^{2}}{2\,\Sigma_{z^{l},\,ii}}\Biggr)\nonumber \\
 & \quad+\mu_{z^{l},\,i}\,\frac{1}{\sqrt{2\pi\,\Sigma_{z^{l},\,ii}}}\int_{-\mu_{z^{l},\,i}}^{\infty}\,\mathrm{d}z_{i}^{l}\,\exp\Biggl(-\frac{(z_{i}^{l})^{2}}{2\,\Sigma_{z^{l},\,ii}}\Biggr)\\
 & =\frac{\sqrt{\,\Sigma_{z^{l},\,ii}}}{\sqrt{2\pi}}\,\exp\Biggl(-\frac{\mu_{z^{l},\,i}^{2}}{2\,\Sigma_{z^{l},\,ii}}\Biggr)\nonumber \\
 & \phantom{=}+\frac{\mu_{z^{l},\,i}}{2}\bigg(1+\text{erf}\bigg(\frac{\mu_{z^{l},\,i}}{\sqrt{2\,\Sigma_{z^{l},\,ii}}}\bigg)\bigg)\,.
\end{align}
For the covariance of post-activations, we distinguish the cases $i=j$
and $i\not=j$, starting with the former by calculating its second
moment as
\begin{align}
 & \langle\phi(z_{i}^{l})\,\phi(z_{i}^{l})\rangle_{\extralowerindex z^{l}\sim\mathcal{N}(\mu_{z^{l}},\,\Sigma_{z^{l}})}\nonumber \\
= & \frac{1}{\sqrt{2\pi\,\Sigma_{z^{l},\,ii}}}\int_{0}^{\infty}\,\mathrm{d}z_{i}^{l}\,(z_{i}^{l})^{2}\,\exp\Big(-\frac{1}{2\Sigma_{z^{l},\,ii}}(z_{i}^{l}-\mu_{z^{l},\,i})^{2}\Big)\\
= & -\frac{\sqrt{\Sigma_{z^{l},\,ii}}}{\sqrt{2\pi}}\,\mu_{z^{l},\,i}\,\exp\Big(-\frac{\mu_{z^{l},\,i}^{2}}{2\Sigma_{z^{l},\,ii}}\Big)\nonumber \\
 & +\frac{\Sigma_{z^{l},\,ii}}{2}\,\bigg(1+\text{erf}\bigg(\frac{\mu_{z^{l},\,i}}{\sqrt{2\,\Sigma_{z^{l},\,ii}}}\bigg)\bigg)\nonumber \\
 & +\sqrt{\frac{2}{\pi}}\,\mu_{z^{l},\,i}\,\sqrt{\Sigma_{z^{l},\,ii}}\,\exp\Big(-\frac{\mu_{z^{l},\,i}^{2}}{2\Sigma_{z^{l},\,ii}}\Big)\nonumber \\
 & +\frac{\mu_{z^{l},\,i}^{2}}{2}\,\bigg(1+\text{erf}\bigg(\frac{\mu_{z^{l},\,i}}{\sqrt{2\,\Sigma_{z^{l},\,ii}}}\bigg)\bigg)\\
= & \frac{\sqrt{\Sigma_{z^{l},\,ii}}\,\mu_{z^{l},\,i}}{\sqrt{2\pi}}\,\exp\Big(-\frac{\mu_{z^{l},\,i}^{2}}{2\Sigma_{z^{l},\,ii}}\Big)\nonumber \\
 & +\frac{\Sigma_{z^{l},\,ii}+\mu_{z^{l},\,i}^{2}}{2}\,\bigg(1+\text{erf}\bigg(\frac{\mu_{z^{l},\,i}}{\sqrt{2\,\Sigma_{z^{l},\,ii}}}\bigg)\bigg)\,.
\end{align}
Combining with the expression for the mean $\mu_{y^{l}}$ then yields
the diagonal terms of the covariance:
\begin{align}
\Sigma_{y^{l},\,ii}= & \langle\phi(z_{i}^{l})\,\phi(z_{i}^{l})\rangle_{\extralowerindex z^{l}\sim\mathcal{N}(\mu_{z^{l}},\,\Sigma_{z^{l}})}-\bigl(\mu_{y^{l},\,i}\bigr)^{2}\\
= & \frac{\Sigma_{z^{l},\,ii}}{2}\,\bigg(1+\text{erf}\bigg(\frac{\mu_{z^{l},\,i}}{\sqrt{2\Sigma_{z^{l},\,ii}}}\bigg)\bigg)+\frac{\mu_{z^{l},\,i}^{2}}{4}\nonumber \\
 & -\bigg(\frac{\sqrt{\Sigma_{z^{l},\,ii}}}{\sqrt{2\pi}}\,\exp\Big(-\frac{1}{2\Sigma_{z^{l},\,ii}}\mu_{z^{l},\,i}^{2}\Big)\nonumber \\
 & \qquad+\frac{\mu_{z^{l},\,i}}{2}\,\text{erf}\bigg(\frac{\mu_{z^{l},\,i}}{\sqrt{2\Sigma_{z^{l},\,ii}}}\bigg)\bigg)^{2}\,.
\end{align}
In the case $i\not=j$, we look at the joint distribution of $(z_{i},z_{j})$
and denote the marginalized mean and covariance by $\tilde{\mu}_{z}=(\mu_{z,\,i},\,\mu_{z,\,j})^{\mathsf{T}}$
and ${\scriptstyle \tilde{\Sigma}_{z}={\scriptstyle {\scriptscriptstyle \left({\scriptstyle \begin{array}{cc}
\Sigma_{z,\,ii} & \Sigma_{z,\,ij}\\
\Sigma_{z,\,ji} & \Sigma_{z,\,jj}
\end{array}}\right)}}}$. For the second moment, we obtain\allowdisplaybreaks
\begin{align}
 & \langle\phi(z_{i}^{l})\,\phi(z_{j}^{l})\rangle_{\extralowerindex z^{l}\sim\mathcal{N}(\mu_{z^{l}},\,\Sigma_{z^{l}})}\nonumber \\
= & \frac{1}{\sqrt{(2\pi)^{2}\,\det(\tilde{\Sigma}_{z})}}\,\int_{0}^{\infty}\,\mathrm{d}z_{i}^{l}\,\int_{0}^{\infty}\,\mathrm{d}z_{j}^{l}\,z_{i}^{l}\,z_{j}^{l}\nonumber \\
 & \qquad\times\exp\Big(-\frac{1}{2}(\tilde{z}^{l}-\mu_{z})^{\mathsf{T}}\tilde{\Sigma}_{z}^{-1}(\tilde{z}^{l}-\mu_{z})\Big)\\
= & \frac{\sqrt{\det(\tilde{\Sigma}_{z})}}{2\pi}\,\exp\Biggl(-\frac{\tilde{\mu}_{z}^{\T}\tilde{\Sigma}_{z}^{-1}\tilde{\mu}_{z}}{2}\Biggr)\nonumber \\
 & +\frac{\sqrt{\det(\tilde{\Sigma}_{z})}}{2\pi}\,\tilde{\Sigma}_{z^{l},\,jj}^{-1}\,\mu_{z^{l},\,j}\frac{\sqrt{\pi}}{\sqrt{2\tilde{\Sigma}_{z^{l},\,jj}^{-1}}}\nonumber \\
 & \qquad\times\exp\left(-\frac{\tilde{\Sigma}_{z^{l},\,ii}^{-1}\,\mu_{z^{l},\,i}^{2}}{2}\right)\,\exp\Biggl(\frac{\bigl(\tilde{\Sigma}_{z^{l},\,ji}^{-1}\,\mu_{z^{l},\,i}\bigr)^{2}}{2\tilde{\Sigma}_{z^{l},\,jj}^{-1}}\Biggr)\nonumber \\
 & \qquad\times\Biggl[1+\text{erf}\Biggl(\frac{\bigl(\tilde{\Sigma}_{z^{l}}^{-1}\tilde{\mu}_{z^{l}}\bigr)_{j}}{\sqrt{2\tilde{\Sigma}_{z^{l},\,jj}^{-1}}}\Biggr)\Biggr]\nonumber \\
 & +\frac{\sqrt{\det(\tilde{\Sigma}_{z^{l}})}}{2\pi}\,\tilde{\Sigma}_{z^{l},\,ii}^{-1}\,\mu_{z^{l},\,i}\frac{\sqrt{\pi}}{\sqrt{2\tilde{\Sigma}_{z^{l},\,ii}^{-1}}}\nonumber \\
 & \qquad\times\exp\left(-\frac{\tilde{\Sigma}_{z^{l},\,jj}^{-1}\,\mu_{z^{l},\,j}^{2}}{2}\right)\,\exp\Biggl(\frac{\bigl(\tilde{\Sigma}_{z^{l},\,ij}^{-1}\,\mu_{z^{l},\,j}\bigr)^{2}}{2\tilde{\Sigma}_{z^{l},\,ii}^{-1}}\Biggr)\nonumber \\
 & \qquad\times\Biggl[1+\text{erf}\Biggl(\frac{(\tilde{\Sigma}_{z^{l}}^{-1}\tilde{\mu}_{z^{l}})_{i}}{\sqrt{2\tilde{\Sigma}_{z^{l},\,ii}^{-1}}}\Biggr)\Biggr]\nonumber \\
 & +\Biggl(\mu_{z^{l},\,i}\,\mu_{z^{l},\,j}-\tilde{\Sigma}_{z^{l},\,ij}^{-1}\,\det\bigl(\tilde{\Sigma}_{z^{l}}\bigr)\Biggr)\nonumber \\
 & \qquad\times\Biggl[\frac{1}{2}\,\text{erf}\bigg(\frac{\sqrt{2}\,\mu_{z^{l},\,i}}{\sqrt{\Sigma_{z^{l},\,ii}}}\bigg)+\frac{1}{2}\,\text{erf}\bigg(\frac{\sqrt{2}\,\mu_{z^{l},\,j}}{\sqrt{\Sigma_{z^{l},\,jj}}}\bigg)+F_{\tilde{\mu}_{z^{l}}\,,\tilde{\Sigma}_{z^{l}}}(0,0)\Biggr],
\end{align}
\allowdisplaybreaks[0]where $\tilde{\mu}_{z^{l}}$ and $\tilde{\Sigma}_{z^{l}}$
denote the marginalized mean and covariance with respect to $\tilde{z}^{l}=(z_{i}^{l},z_{j}^{l})^{\mathsf{T}}$,
and $F_{\tilde{\mu}_{z^{l}},\,\tilde{\Sigma}_{z^{l}}}(x,y)$ denotes
the corresponding cumulative distribution function. $F_{\tilde{\mu}_{z^{l}},\,\tilde{\Sigma}_{z^{l}}}(0,0)$
is also known as the \emph{quadrant probability}. By subtracting $\mu_{\lowerindex y^{l}\!,\,i}\,\mu_{\lowerindex y^{l}\!,\,j}$,
we obtain the expression for the cross-covariances given in \prettyref{tab:interaction_functions}.

\subsubsection*{Contributions from higher-order correlations\label{subsec:higher_order_interact_relu}}

Using the Gram-Charlier expansion \citep{Blinnikov98_193} of the
probability density function $p_{z_{i}^{l}}(z_{i}^{l})$, we can derive
approximate expressions for the interaction of higher-order correlations
of the pre-activations $z^{l}$. As an example, we derive contributions
to the mean of the post-activations $y^{l}$ up to linear order in
$G_{z^{l},\,(i,i,i)}^{(3)}$ for $\mathrm{ReLU}$ activations. The
Gram-Charlier expansion up to third order is 
\begin{align*}
p_{z_{i}^{l}}(z_{i}^{l}) & \approx\left(1+\frac{G_{z^{l}\!,\,(i,i,i)}^{(3)}}{3!\,\sqrt{\Sigma_{z^{l}\!,\,ii}^{3}}}\Biggl[\Biggl(\frac{z_{i}^{l}-\mu_{z^{l}\!,\,i}}{\sqrt{\Sigma_{z^{l},\,ii}}}\Biggr)^{3}\!\!-3\frac{\bigl(z_{i}^{l}-\mu_{z^{l}\!,\,i}\bigr)}{\sqrt{\Sigma_{z^{l},\,ii}}}\Biggr]\right)\\
 & \quad\times\frac{1}{\sqrt{2\pi\,\Sigma_{z^{l},\,ii}}}\,\exp\bigg(-\frac{\bigl(z_{i}^{l}-\mu_{z^{l}\!,\,i}\bigr)^{2}}{2\Sigma_{z^{l},\,ii}}\bigg)\,.
\end{align*}
Inserting this into the expression for the mean $\mu_{y^{l}\!,\,i}$
of the post-activations $y^{l}$, we get\looseness=-1 \allowdisplaybreaks
\begin{align}
\mu_{y^{l}\!,\,i} & =\langle\phi(z_{i}^{l})\rangle_{\extralowerindex z^{l}\sim\mathcal{N}(\mu_{z^{l}},\,\Sigma_{z^{l}})}\\
 & =\int_{0}^{\infty}\,\mathrm{d}z_{i}^{l}\,z_{i}^{l}\,p_{z_{i}^{l}}(z_{i}^{l})\\
 & \approx\int_{0}^{\infty}\,\mathrm{d}z_{i}^{l}\,z_{i}^{l}\,\frac{1}{\sqrt{2\pi\,\Sigma_{z^{l},\,ii}}}\,\exp\bigg(-\frac{\bigl(z_{i}^{l}-\mu_{z^{l}\!,\,i}\bigr)^{2}}{2\Sigma_{z^{l},\,ii}}\bigg)\nonumber \\
 & \quad+\frac{G_{z^{l}\!,\,(i,i,i)}^{(3)}}{3!\,\sqrt{\Sigma_{z^{l}\!,\,ii}^{3}}}\,\int_{0}^{\infty}\!\!\mathrm{d}z_{i}^{l}\,z_{i}^{l}\nonumber \\
 & \quad\times\Biggl[\Biggl(\frac{z_{i}^{l}-\mu_{z^{l}\!,\,i}}{\sqrt{\Sigma_{z^{l},\,ii}}}\Biggr)^{3}\!\!-3\frac{\bigl(z_{i}^{l}-\mu_{z^{l}\!,\,i}\bigr)}{\sqrt{\Sigma_{z^{l},\,ii}}}\Biggr]\nonumber \\
 & \quad\times\frac{\exp\Bigl(-\bigl(z_{i}^{l}-\mu_{z^{l}\!,\,i}\bigr)^{2}\big/2\Sigma_{z^{l}\!,\,ii}\Bigr)}{\sqrt{2\pi\,\Sigma_{z^{l}\!,\,ii}}}\\
 & =\frac{\sqrt{\Sigma_{z^{l}\!,\,ii}}\,\mu_{z^{l}\!,\,i}}{\sqrt{2\pi}}\,\exp\Bigl(-\frac{\mu_{z^{l}\!,\,i}^{2}}{2\,\Sigma_{z^{l}\!,\,ii}}\Bigr)\nonumber \\
 & \quad+\frac{\mu_{z^{l}\!,\,i}}{2}\Biggl(1+\text{erf}\Biggl(\frac{\mu_{z^{l}\!,\,i}}{\sqrt{2\,\Sigma_{z^{l}\!,\,ii}}}\Biggr)\Biggr)\nonumber \\
 & \quad-\frac{G_{z^{l},\,(i,i,i)}^{(3)}}{2\,\Sigma_{z^{l}\!,\,ii}^{2}}\,\bigl(\Sigma_{z^{l}\!,\,ii}^{2}-1\bigr)\,\frac{1}{2}\Biggl(1+\text{erf}\Biggl(\frac{\mu_{z^{l}\!,\,i}}{\sqrt{2\,\Sigma_{z^{l}\!,\,ii}}}\Biggr)\Biggr)\nonumber \\
 & \quad+\frac{G_{z^{l}\!,\,(i,i,i)}^{(3)}}{3!\,\Sigma_{z^{l}\!,\,ii}^{3}}\,\Big(3\,\mu_{z^{l}\!,\,i}\Sigma_{z^{l}\!,\,ii}^{2}+2\,\Sigma_{z^{l}\!,\,ii}+\mu_{z^{l}\!,\,i}^{3}+\mu_{z^{l}\!,\,i}^{2}-3\Bigr)\nonumber \\
 & \quad\times\frac{\sqrt{\Sigma_{z^{l}\!,\,ii}}\,\mu_{z^{l}\!,\,i}}{\sqrt{2\pi}}\,\exp\bigg(-\frac{\mu_{z^{l}\!,\,i}^{2}}{2\Sigma_{z^{l}\!,\,ii}}\bigg)
\end{align}
\allowdisplaybreaks[0]

Alternatively, if one wishes to compute higher-order cumulants of
$y^{l}$, this can be done by first evaluating the integrals for higher-order
moments, analogously to the computations above for the first and second
moment. Cumulants can then be obtained via the relations given by
\citet{Gardiner85}.

\subsection*{Derivations for quadratic activations\label{subsec:interact_derivations_quad}}

We here consider networks with a quadratic activation function $\phi(z)=z+\alpha\,z^{2}$.
For any distribution of pre-activations $z^{l}$ with mean $\mu_{z^{l}}$
and covariance $\Sigma_{z^{l}}$, the mean post-activations are given
by
\begin{align}
\mu_{y^{l}\!,\,i} & =\langle z_{i}^{l}+\alpha\,(z_{i}^{l})^{2}\rangle_{z^{l}}\\
 & =\mu_{z^{l},\,i}+\alpha\,(\mu_{z^{l},\,i})^{2}+\alpha\,\Sigma_{z^{l},\,ii}\,.\label{eq:mean-post-act-quadratic}
\end{align}
For the covariance of post-activations, we first calculate the second
moment
\begin{align}
\langle\phi(z_{i}^{l})\,\phi(z_{j}^{l})\rangle_{z^{l}} & =\left\langle \left[z_{i}^{l}+\alpha\,(z_{i}^{l})^{2}\right]\,\left[z_{j}^{l}+\alpha\,(z_{j}^{l})^{2}\right]\right\rangle _{z^{l}}\\
 & =\Sigma_{z^{l}\!,\,ij}+\mu_{z^{l},\,i}\,\mu_{z^{l},\,j}+\alpha\,M_{z^{l}\!,\,(i,j,j)}^{(3)}\nonumber \\
 & \phantom{=}+\alpha\,M_{z^{l}\!,\,(j,i,i)}^{(3)}+\alpha^{2}\,M_{z^{l}\!,\,(i,i,j,j)}^{(4)}\,,
\end{align}
where $M_{z^{l}}^{(n)}$ denotes the $n$-th moment of pre-activations
$z^{l}$. Combining the expression \prettyref{eq:mean-post-act-quadratic}
for the mean $\mu_{y^{l}}$ then yields the covariance
\begin{align}
\Sigma_{y^{l},\,ij} & =\langle\phi(z_{i}^{l})\,\phi(z_{j}^{l})\rangle_{z^{l}}-\langle\phi(z_{i}^{l})\rangle_{z^{l}}\,\langle\phi(z_{j}^{l})\rangle_{z^{l}}\\
 & =\Sigma_{z^{l},\,ij}+2\,\alpha\,\Sigma_{z^{l},\,ij}\,\left(\mu_{z^{l},\,i}+\mu_{z^{l},\,j}\right)\nonumber \\
 & \phantom{=}+2\,\alpha^{2}\,(\Sigma_{z^{l}\!,\,ij})^{2}+4\,\alpha^{2}\,\mu_{z^{l},\,i}\,\Sigma_{z^{l},\,ij}\,\mu_{z^{l},\,j}+\Sigma_{y^{l},\,ij}\vert_{n>2}\,,
\end{align}
where $\Sigma_{y^{l},\,ij}\vert_{n>2}$ contains all terms involving
cumulants of order $n>2$. It is given by
\begin{align}
\Sigma_{y^{l},\,ij}\vert_{n>2} & =\alpha\,(1+2\alpha\,\mu_{z^{l},\,i})\,G_{z^{l},\,(i,j,j)}^{(3)}\nonumber \\
 & \phantom{=}+\alpha\,(1+2\alpha\,\mu_{z^{l},\,j})\,G_{z^{l},\,(j,i,i)}^{(3)}\nonumber \\
 & \phantom{=}+\alpha^{2}\,G_{z^{l},\,(i,i,j,j)}^{(4)}\,.\label{eq:contrib_higher_orders_quad_act}
\end{align}
In these expressions, $G_{z^{l},\,(i_{1},\,i_{2},\,\dotsc,\,i_{n})}^{(n)}$
denotes the $n$-th cumulant of pre-activations given by $\llangle z_{i_{1}}^{l}z_{i_{2}}^{l}\dotsb z_{i_{n}}^{l}\rrangle$.
If the pre-activations $z^{l}$ are Gaussian distributed $z^{l}\sim\mathcal{N}(\mu_{z^{l}},\Sigma_{z^{l}})$,
all cumulants beyond second order vanish, $G_{z^{l}}^{(n>2)}=0$,
yielding $\Sigma_{y^{l},\,ij}\vert_{n>2}=0$ and consequently the
result in \prettyref{tab:interaction_functions}.

\section{Information propagation in networks with quadratic activations\label{app:xor_illustration_quad}}

\begin{figure*}[!t]
\centering{}\includegraphics{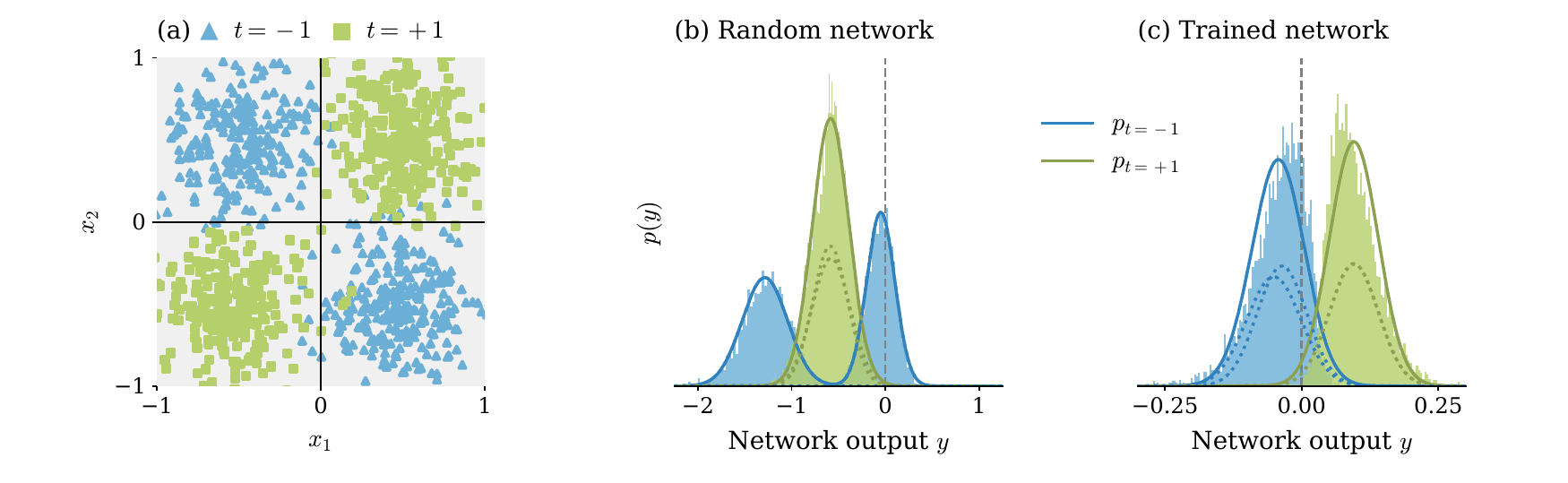}\caption{\textbf{Information propagation in networks with quadratic activation
function for the XOR problem. (a) }The distribution of input data
is modeled as a Gaussian mixture. Data samples $x^{(d)}$ (blue and
green dots) are assigned to class labels $t=\pm1$ based on the respective
mixture component. \textbf{(b,c)} Distribution of the network output
for random (b) and trained (c) parameters. Class-conditional distributions
(solid curves) are determined as a superposition of the propagated
mixture components (dashed curves) and empirical estimates (blue and
green histograms) are obtained from the test data. Since networks
are trained on class labels $t=\pm1$, the classification threshold
is set to $y=0$ (gray lines). Other parameters: $\ensuremath{\phi(z)=z+\alpha\,z^{2}}$,
network depth $L=1$, width $N=10$; trained network in (c) achieves
$P=90.46\%$ performance.\label{fig:xor_illustration_prob_dist_mapping_quad_app}}
\end{figure*}

\setcounter{equation}{0}\renewcommand\theequation{D\arabic{equation}}\prettyref{fig:xor_illustration_prob_dist_mapping}
in the main text illustrates information propagation in networks with
$\mathrm{ReLU}$ activations. For completeness, we include here as
Supplemental \prettyref{fig:xor_illustration_prob_dist_mapping_quad_app}
the analogous illustration for a network with quadratic activations.
For \prettyref{fig:xor_kl_div}, we include the results for networks
with quadratic activation function in supplemental \prettyref{fig:xor_kl_div_quad}.
\begin{figure}
\centering{}\includegraphics{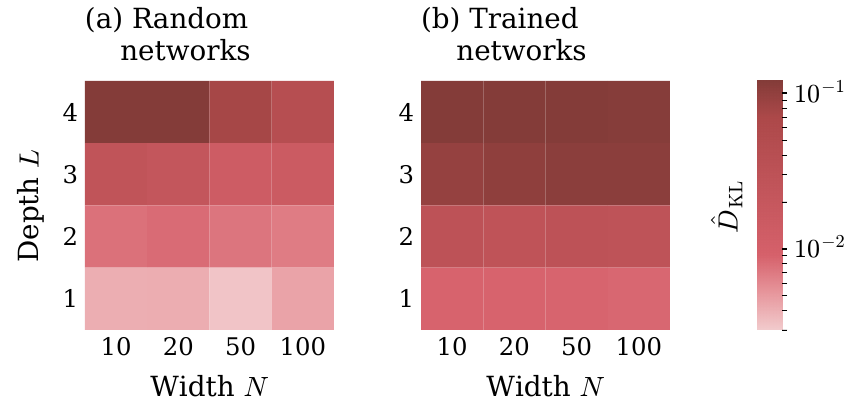}\caption{\textbf{Deviation between theoretical and empirical output distribution
for (a) random and (b) trained networks}, measured across $10^{2}$
different network realizations using the normalized Kullback-Leibler
divergence $\hat{D}_{\mathrm{KL}}(p_{\text{emp.}}\Vert p_{\text{theo.}})$.
On average, the trained networks achieve performance values of $P=96.52\%\pm0.15\%$
compared to $P_{\text{opt}}=97.5\%$.\label{fig:xor_kl_div_quad}
Networks were trained to perform the XOR task described in \prettyref{subsec:XOR-problem-setup}.
Other parameters: $\ensuremath{\phi(z)=z+\alpha\,z^{2}}$.}
\end{figure}

\section{Data sample generation for MNIST based on Gaussian approximation
of input distribution\label{app:mnist_sample_generation}}

\setcounter{equation}{0}\renewcommand\theequation{E\arabic{equation}}In
\prettyref{subsec:mnist_correlations} of the main text, we discuss
training networks to solve MNIST using $R_{\text{emp,\,MSE}}$ (Eq.~\eqref{eq:loss_data})
with data samples drawn from the Gaussian approximation of the input
distribution. For this Gaussian approximation, means $\hat{\mu}_{x}^{t}$
and covariances $\hat{\Sigma}_{x}^{t}$ for each class $t$ are estimated
empirically from the training data set where we flattened the $28\times28$
images into $784$-dimensional vectors. Due to lack of variability
in some pixel values at the image edges, the resulting covariances
$\hat{\Sigma}_{x}^{t}$ are not positive definite, but only positive
semi-definite.

To account for the zero eigenvalues of the covariance, data samples
are generated based on a principal component analysis of the covariance
matrix. For each class $t$, we decompose the covariance matrix as
\begin{equation}
\hat{\Sigma}_{x}^{t}=VDV^{\mathsf{T}}
\end{equation}
with $V=(v_{1}\vert\dots\vert v_{N_{0}})$ containing the unit-length
eigenvectors $v_{i}$ and $D=\text{diag}(\lambda_{1},\dots,\lambda_{N_{0}})$
containing the corresponding eigenvalues $\lambda_{i}$ of $\hat{\Sigma}_{x}^{t}$,
which we assume to be ordered according to their size, $\lambda_{1}\geq\dots\geq\lambda_{N_{0}}\geq0$.
We set a threshold $\vartheta_{\text{PCA}}>0$ that defines a subspace
$U$ spanned by the eigenvectors $\{v_{i}\}_{i=1,\dots,N_{\text{PCA}}}$
for which $\lambda_{i}>\vartheta_{\text{PCA}}$. Data samples $\hat{x}^{(d)}\vert_{U}$
are then generated with respect to this subspace $U$ and projected
back to the input space $\R^{N_{0}}$ according to
\begin{align}
\hat{x}^{(d)}\vert_{U} & \sim\mathcal{N}(0,\text{diag}(\lambda_{1},\dots,\lambda_{N_{\text{PCA}}})),\\
\hat{x}^{(d)} & =\hat{\mu}_{x}^{t}+V\left(\begin{array}{c}
\hat{x}^{(d)}\vert_{U}\\
0
\end{array}\right).
\end{align}
For all experiments in \prettyref{subsec:mnist_correlations} of
the main text, we choose $\vartheta_{\text{PCA}}=10^{-2}$, corresponding
to $N_{\text{PCA}}$ between $103$ and $234$ for the different classes
$t$. Since for all classes $t$ the magnitude of the largest eigenvalue
is of order 1, this choice of $\vartheta_{\text{PCA}}$ ensures including
relevant eigenvectors while excluding noise due to finite numerical
precision. Since the MNIST training data set contains $60,\!000$
samples, to allow for a fair comparison between training on Gaussian
samples and on the original images, we generated a similarly-sized
training data set of $D=60,\!000$ Gaussian samples.

\section{Problem setup for inclusion of higher-order statistics in the main
text\label{app:supplement_problem_alternating_hills}}

\setcounter{equation}{0}\renewcommand\theequation{F\arabic{equation}}The
problem studied in \prettyref{subsec:problem_alternating_hills} of
the main text is constructed as follows. We define two classes, $t=\pm1$,
each composed of two Gaussian components $+$ and $-$, with the following
means:
\begin{align}
\mu_{x}^{t=+1,\,-} & =(-0.5,0)^{\intercal}\,, & \mu_{x}^{t=-1,\,-} & =(-1.5,0)^{\intercal}\,;\\
\mu_{x}^{t=+1,\,+} & =(1.5,0)^{\intercal}\,, & \mu_{x}^{t=-1,\,+} & =(0.5,0)^{\intercal}\,.
\end{align}
 Covariances are isotropic throughout with 
\begin{equation}
\Sigma_{x}^{t,\,\pm}=0.05\,\mathbb{I}\,.
\end{equation}
 The outer components $(t=-1,\,-)$ and $(t=+1,\,+)$ are weighed
by $p_{\text{outer}}=\frac{1}{8}$, while the inner components $(t=-1,\,+)$
and $(t=+1,\,-)$ are weighed by $p_{\text{inner}}=\frac{3}{8}$,
as illustrated in \prettyref{fig:problem_alternating_hills}\textbf{a,b}
of the main text. A data sample $x^{(d)}$ is assigned a target label
$t^{(d)}\in\{\pm1\}$ based on the mixture component it is drawn from.
Distribution parameters are chosen such that the class-conditional
means and covariances of the data are identical, 
\begin{align}
\mu_{x}^{t=\pm1} & =\left(\begin{array}{c}
0\\
0
\end{array}\right)\,, & \Sigma_{x}^{t=\pm1}= & \left(\begin{array}{cc}
0.8 & 0\\
0 & 0.5
\end{array}\right)\,,
\end{align}
 while the third-order correlations differ 
\begin{equation}
G_{x,\,(i,j,k)}^{(3),\,t=\pm1}=\pm\,0.75\,\delta_{ij}\delta_{jk}\delta_{ki}\delta_{i1}\,.
\end{equation}
We use training and test data sets of size $D=10^{4}$.

\section{Depth scales of information propagation\label{app:Depth-scales}}

\setcounter{equation}{0}\renewcommand\theequation{G\arabic{equation}}We
here discuss the relation between the presented work and \citet{Poole16_3360}.
To formalize this relation, we define as $z_{\theta k}^{ld}$ the
pre-activation of neuron $k$ in layer $l$ for a given data sample
$x^{(d)}$ in a network with parameters $\theta$. \citeauthor{Poole16_3360}
study ensembles of networks across random realizations of network
parameters $\theta$. The family of distributions they study, expressed
in terms of pre-activations, is thus
\begin{align}
\tilde{p}_{\{d\}}^{l}(\{z_{k}\}) & =\langle\prod_{k,d}\delta(z_{k}^{d}-z_{k\theta}^{ld})\rangle_{\theta},\label{eq:poole}
\end{align}
which is one distribution jointly for all pre-activations $\{z_{k}^{d}\}_{d=1,\ldots,D;\,k=1,\ldots,N}$
for a given set of $P$ data samples $x^{(d)}$ and for each given
layer $l$. In the limit of wide networks, they find that $\tilde{p}$
factorizes across different neuron indices, so that $z_{i}^{ld}$
and $z_{k}^{ld^{\prime}}$ are independent for different $i\neq k$.
Further, these variables are centered Gaussian, so that a single covariance
matrix is sufficient to describe their statistics. In this limit,
it is therefore sufficient to study the joint statistics of all pairs
of networks corresponding to all pairs of inputs $d,d^{\prime}$
\begin{align}
\tilde{p}_{dd^{\prime}}^{l}(z,z^{\prime}) & =\langle\delta(z-z_{\theta}^{ld})\,\delta(z^{\prime}-z_{\theta}^{ld^{\prime}})\rangle_{\theta}.\label{eq:pairwise_gauss}
\end{align}
Correlation functions in their work a priori thus quantify fluctuations
across realizations of network parameters. Their mean-field theory
for deep feed-forward networks is identical to the classical mean-field
theory of random recurrent networks \citep{molgedey92_3717}, because
for recurrent networks with discrete-time updates the equal time statistics
is identical to the equal-layer statistics of a deep network \citep{Segadlo22_103401}.

In the presented work, instead, we study individual networks defined
by one fixed set of parameters $\theta$ across the distribution $p(x)$
of data samples $x^{(d)}$. Correlations in our work thus quantify
the variability of the network state across different data points.
Formally, the family of distributions we study is

\begin{align}
p_{\theta}^{l}(\{z_{k}\}) & =\langle\prod_{k}\delta(z_{k}-z_{\theta}^{ld})\rangle_{d},\label{eq:p_our}
\end{align}
which for each given $l$ and $\theta$ is one joint distribution
of all neurons $k$. Importantly, the distribution is across the ensemble
of data points $d$.

One formal difference is thus the expectation across $\theta$ in
\eqref{eq:poole} versus the expectation over $d$ in \eqref{eq:p_our}.
Wide networks, however, tend to be self-averaging. This means that
the ensemble across parameters $\theta$ studied by \citeauthor{Poole16_3360}
shows a concentration on a single typical behavior that one finds
in any of its (likely) individual realizations. Formally this means
that the empirical distribution of $(z_{k},z_{k}^{\prime})$ across
neurons $k$ for any random choice of parameters $\theta$ takes on
the same form as $\tilde{p}$, so that \eqref{eq:pairwise_gauss}
for $N$ large approaches the empirical average over neuron activations,
\begin{align*}
\tilde{p}_{dd^{\prime}}^{l}(z,z^{\prime}) & \stackrel{\text{self-averaging}}{\simeq}N^{-1}\sum_{k}\,\delta(z-z_{\theta k}^{ld})\,\delta(z^{\prime}-z_{\theta k}^{ld^{\prime}}),\qquad\forall\theta.
\end{align*}
A way to show this is by a saddle point approximation of the moment-generating
function after the disorder average across $\theta$ \citep[e.g,][]{Schuecker16b_arxiv,Crisanti18_062120,Helias20_970,Bordelon22_arxiv,Segadlo22_103401}.

To derive the result by \citet{Poole16_3360} or \citet{molgedey92_3717}
in our notation, we start with the expression for the pre-activations
$z_{i}^{l+1}=\sum_{k}W_{ik}^{l+1}y_{k}^{l}+b_{i}^{l+1}$ (see \eqref{eq:pre_activations}).
For Gaussian distributed $W_{ik}^{l+1}$ and $b_{i}^{l+1}$, pre-activations
for one fixed data sample $x^{(d)}$ (suppressing the superscript
$d$ for brevity in the following) become Gaussian as well, with mean
and covariance

\begin{align}
M_{z^{l+1},\,i} & :=\bigl\langle\sum_{k}W_{ik}^{l+1}y_{k}^{l}+b_{i}^{l+1}\bigr\rangle_{W,b}\nonumber \\
 & =\sum_{k}\left\langle W_{ik}^{l+1}\right\rangle _{W^{l+1}}\left\langle y_{k}^{l}\right\rangle _{W,b}+\left\langle b_{i}^{l+1}\right\rangle _{b^{l+1}}\;=\;0\,,\label{eq:mean1}\\
S_{z^{l+1},ij} & :=\bigl\langle\sum_{k,m}W_{ik}^{l+1}W_{jm}^{l+1}y_{k}^{l}y_{m}^{l}+b_{i}^{l+1}b_{j}^{l+1}\bigr\rangle_{W,b}\nonumber \\
 & =\sum_{k,m}\left\langle W_{ik}^{l+1}W_{jm}^{l+1}\right\rangle _{W^{l+1}}\left\langle y_{k}^{l}y_{m}^{l}\right\rangle _{W,b}+\left\langle b_{i}^{l+1}b_{j}^{l+1}\right\rangle _{b^{l+1}}\nonumber \\
 & =\delta_{ij}\,\Big(\sigma_{w}^{2}\,\left\langle y^{l}y^{l}\right\rangle _{W,b}+\sigma_{b}^{2}\Big)\nonumber \\
 & =:\delta_{ij}\,S_{z^{l+1}}\,,\label{eq:var1}
\end{align}
with 
\begin{equation}
S_{z^{l+1}}=\sigma_{w}^{2}\,\langle\phi(z^{l})\phi(z^{l})\rangle_{z^{l+1}\sim\N(0,S_{z^{l}})}+\sigma_{b}^{2}\,,\label{eq:single_system_var}
\end{equation}
 and where we used the mapping by the activation function \eqref{eq:mu_y}.
To determine the statistics $\left\langle y_{k}^{l}\right\rangle _{W,b}$
and $\left\langle y_{k}^{l}y_{k}^{l}\right\rangle _{W,b}$, we simultaneously
performed an average over weights and biases in layers $l^{\prime}\leq l$.
These statistics are identical across neurons, so we write $\left\langle y_{k}^{l}\right\rangle _{W,b}=\left\langle y^{l}\right\rangle _{W,b}$
and $\left\langle y_{k}^{l}y_{k}^{l}\right\rangle _{W,b}=\left\langle y^{l}y^{l}\right\rangle _{W,b}$.
\prettyref{eq:mean1} and \prettyref{eq:var1} show that correlations
among different neurons vanish on average across networks.

The covariance between pre-activations of a pair of networks for two
different inputs $x^{(d)}$ and $x^{(d^{\prime})}$ analogously becomes
\begin{equation}
\begin{aligned} & S_{z^{l+1,d}z^{l+1,d^{\prime}}}\\
 & =\sigma_{w}^{2}\,\langle\phi(z^{l,d})\phi(z^{l,d^{\prime}})\rangle_{(z^{l,d},z^{l,d^{\prime}})\sim\N\big(0,\{S_{z^{l,d}z^{l,d^{\prime}}}\}\big)}+\sigma_{b}^{2}\,,
\end{aligned}
\label{eq:mapping_cov_across}
\end{equation}
where $\N\big(0,\{S_{z^{l,d}z^{l,d^{\prime}}}\}\big)$ is meant as
the Gaussian distribution for the pair $(z^{l,d},z^{l,d^{\prime}})$
with covariance matrix $\left(\begin{array}{cc}
S_{z^{l,d}} & S_{z^{l,d}z^{l,d^{\prime}}}\\
S_{z^{l,d^{\prime}}z^{l,d}} & S_{z^{l,d^{\prime}}}
\end{array}\right)$.

We may make a connection to our results by considering a pair of inputs
$x^{(d)}$ and $x^{(d^{\prime})}$. These inputs are presented to
the network as $y^{0d}$ and $y^{0d^{\prime}}$. The theory by \citeauthor{Poole16_3360}
yields a measure of the overlap $O_{dd^{\prime}}^{l}$ of network
states after $l$ layers as the solution of the joint iterative equations
derived above. In the limit of wide networks, this overlap becomes
self-averaging, so it concentrates around its mean value across $y$,
\begin{align}
O_{dd^{\prime}}^{l} & :=N^{-1}\sum_{k}y_{k}^{ld}y_{k}^{ld^{\prime}}\label{eq:overlap}\\
 & \simeq\langle y^{ld}y^{ld^{\prime}}\rangle_{W,b}\nonumber \\
 & =\langle\phi(z^{l,d})\phi(z^{l,d^{\prime}})\rangle_{(z^{l,d},z^{l,d^{\prime}})\sim\N\big(0,\{S_{z^{l,d}z^{l,d^{\prime}}}\}\big)}\nonumber \\
 & =\sigma_{w}^{-2}\,\big(S_{z^{l+1,d}z^{l+1,d^{\prime}}}-\sigma_{b}^{2}\big)\,.\nonumber 
\end{align}
To show the simplest possible link between the depth scales studied
in \citeauthor{Poole16_3360}, we consider the case of a deep untrained
network. The statistics of $z^{l}$ and $y^{l}$ decay to a fixed
point, so that we can consider the autostatistics to become constant
for a large enough $l$

\begin{align}
S_{z^{l,d}} & =A_{0}\,,\quad\forall d,\label{eq:homogeneity-1}\\
S_{z^{l,d}z^{l,d^{\prime}}} & =C_{0},\quad\forall d\neq d^{\prime},\nonumber 
\end{align}
where $A_{0}$ is the stationary solution of \eqref{eq:single_system_var}
\begin{align}
A_{0} & =\sigma_{w}^{2}\,\langle\phi(z)\phi(z)\rangle_{z\sim\N(0,A_{0})}+\sigma_{b}^{2}\,\label{eq:stationarity-1}
\end{align}
and $C_{0}$ the stationary solution of \eqref{eq:mapping_cov_across}$$C_{0}=\sigma_{w}^{2}\,\langle\phi(z_{1})\phi(z_{2})\rangle_{(z_{1},z_{2})\sim\N\left(0,\left[\begin{smallmatrix}A_{0} & C_{0}\\C_{0} & A_{0}\end{smallmatrix}\right]\right)}+\sigma_{b}^{2}\,.$$Now
consider pairs of inputs $(y^{0d},y^{0d^{\prime}})$ for which the
statistics of pre-activations differ only little from this fixed-point
statistics: assume that any data point has variance $S_{z^{l}}=A_{0}$
and for any pair $(d,d^{\prime})$ of data points we may express the
covariance of pre-activations in the first layer as
\begin{align}
S_{z^{1,d}z^{1,d^{\prime}}} & =C_{0}+\delta C_{dd^{\prime}}^{1},\label{eq:homogeneous_var}
\end{align}
where $\delta C_{dd^{\prime}}^{1}\ll C_{0}$. Based on these assumptions,
we now compute decay constants with $l$.

Linearizing the iteration \eqref{eq:mapping_cov_across}, one obtains
for the propagation of $\delta C_{dd^{\prime}}^{l}$ 
\begin{align*}
\delta C_{dd^{\prime}}^{l+1} & =\sigma_{w}^{2}\,\langle\phi^{\prime}(z^{l,d})\phi^{\prime}(z^{l,d^{\prime}})\rangle\,\delta C_{dd^{\prime}}^{l}+\order\Big[\big(\delta C_{dd^{\prime}}^{l}\big)^{2}\Big]\,.
\end{align*}
Here, we made use of Price's theorem \citep[Appendix A]{Price58_69,PapoulisProb4th,Schuecker16b_arxiv}
$\partial\langle\phi(z)\phi(z^{\prime})\rangle/\partial\Sigma_{zz^{\prime}}=\langle\phi^{\prime}(z)\phi^{\prime}(z^{\prime})\rangle$
where $\phi^{\prime}=d\phi/dz$ and $\Sigma_{zz^{\prime}}$ is the
covariance of $z$ and $z^{\prime}$. For stationary statistics across
layers \eqref{eq:stationarity-1} and under the homogeneity assumption
across data samples \eqref{eq:homogeneity-1}, one thus has \begin{align*}
\langle\phi^{\prime}(z^{d})\phi^{\prime}(z^{d^{\prime}})\rangle
  &= \langle\phi^{\prime}(z_{1})\phi^{\prime}(z_{2})\rangle
      _{\!\!(z_{1},z_{2})\sim\N\left(0,\left[\begin{smallmatrix}A_{0} & C_{0}\\C_{0} & A_{0}\end{smallmatrix}\right]\right)}\;\;\forall d,d^{\prime} \\
  &=:\langle\phi^{\prime}\phi^{\prime}\rangle.
\end{align*}One then obtains an exponential evolution with layer index 
\begin{align}
\delta C_{dd^{\prime}}^{l+1} & =\big(\sigma_{w}^{2}\,\langle\phi^{\prime}\phi^{\prime}\rangle\big)^{l}\,\delta C_{dd^{\prime}}^{1}\label{eq:variation_propagation}\\
 & =e^{-\frac{l}{\xi}}\,\delta C_{dd^{\prime}}^{1},\nonumber 
\end{align}
with a depth scale
\begin{align}
\xi & =-1/\ln\big[\sigma_{w}^{2}\,\langle\phi^{\prime}\phi^{\prime}\rangle\big].\label{eq:depth_scale}
\end{align}
So this equation gives rise to the depth scales studied in \citeauthor{Poole16_3360}
for network ensembles. This scale $\xi^{-1}$ corresponds to the Lyapunov
exponent computed in \citeauthor{molgedey92_3717}. In particular,
at the transition to chaos, namely at the point in parameter space
$(\sigma_{w},\sigma_{b})$ for which $\sigma_{w}^{2}\,\langle\phi^{\prime}\phi^{\prime}\rangle=1$,
the depth scale diverges. The overlap of activations \eqref{eq:overlap}
shows the same depth scale, because its variation is linearly related
to $\delta C_{dd^{\prime}}^{l}$ as $\delta C_{dd^{\prime}}^{l}=\sigma_{w}^{2}\,\delta O_{dd^{\prime}}^{l-1}$,
so
\begin{align}
\delta O_{dd^{\prime}}^{l} & =\big(\sigma_{w}^{2}\,\langle\phi^{\prime}\phi^{\prime}\rangle\big)^{l}\,\delta O_{dd^{\prime}}^{0}.\label{eq:deltaO}
\end{align}
Both the covariance of pre-activations ($S_{z^{1,d}z^{1,d^{\prime}}}$)
and the overlaps of activations ($O_{dd^{\prime}}^{l}=\delta O_{dd^{\prime}}^{l}+O_{0}$)
therefore decay to fixed points -- respectively $C_{0}$ and $O_{0}$
-- that are related by $O_{0}=\sigma_{w}^{-2}(C_{0}-\sigma_{b}^{2})$.

We can relate these results to our work on single networks by re-expressing
the overlap $O_{dd'}^{l}$ in terms of the probability distribution
across different data samples. From \eqref{eq:overlap} it follows
that
\begin{align}
\mathrlap{\frac{1}{D\,(D-1)}\sum_{(d\neq d^{\prime})=1}^{D}O_{dd^{\prime}}^{l}}\qquad\label{eq:overlaps_averaged}\\
 & \simeq N^{-1}\,\sum_{k=1}^{N}\,\left[\!\frac{1}{D}\sum_{d=1}^{D}y_{k}^{ld}\!\right]\,\left[\!\frac{1}{D}\sum_{d^{\prime}=1}^{D}y_{k}^{ld^{\prime}}\!\right]+\order(D^{-1})\\
 & \!\!\!\!\stackrel{D\gg1}{\simeq}N^{-1}\,\sum_{k=1}^{N}\,\langle y_{k}^{ld}\rangle_{d}^{2}\\
 & \simeq N^{-1}\,\sum_{k=1}^{N}\big(\mu_{y^{l},\{k\}}\big)^{2}\,.\label{eq:O_our_theory}
\end{align}
Here $\mu_{y^{l},\{k\}}$ denotes the mean post-activation of neuron
$k$ in layer $l$, taken over the ensemble of all data points $d$.
This is obtained by iterating Eqs. \eqref{eq:stat_pre_act} and \eqref{eq:stat_post_act}.

\begin{figure}
\centering{}\includegraphics{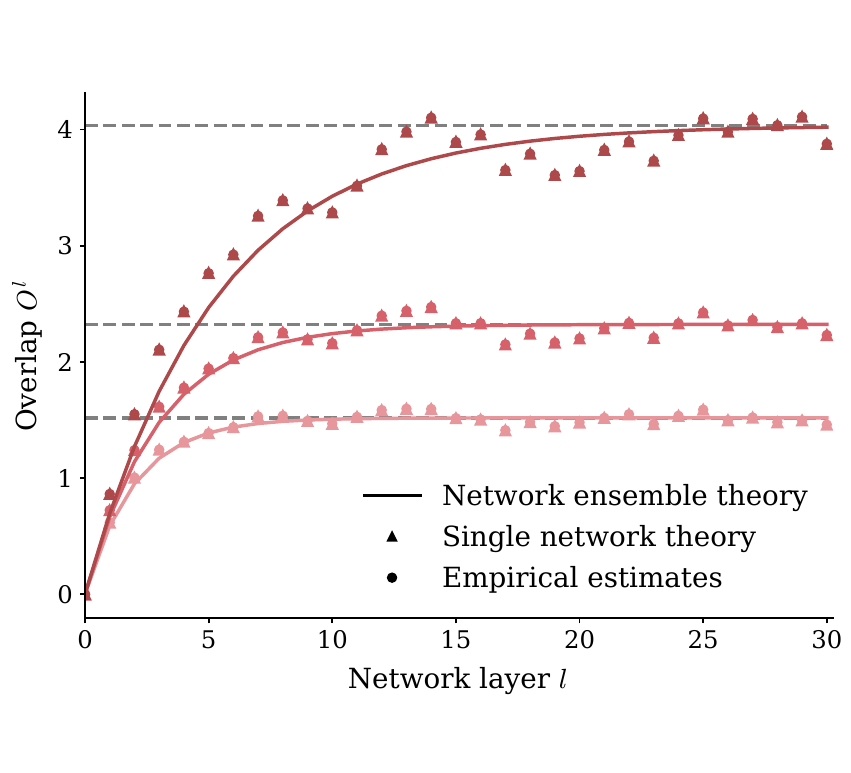}\caption{\textbf{Depth scale of cumulant propagation in a randomly initialized
network.} Evolution of overlaps $O^{l}$ in \eqref{eq:overlap} and
\eqref{eq:O_our_theory} as a function of the layer $l$. Solid curves
show the predicted decay as $e^{-l/\xi}$, with length scale $\xi$
given by \eqref{eq:depth_scale} from \citeauthor{Poole16_3360} for
network ensembles. Dashed lines indicate the fixed points $O_{0}$
to which the overlaps converge. Empirical estimates of the overlaps
$O_{dd^{\prime}}^{l}$ in \eqref{eq:overlaps_averaged} are shown
as dots. Values of the overlaps based on the cumulant propagation
for single networks derived in \eqref{eq:O_our_theory} are shown
as triangles. Empirical estimates and the values based on data statistics
match closely so that the symbols overlap. To isolate the depth scale
of the overlaps, input data $\left\{ x^{d}\right\} _{d}$ are drawn
from a Gaussian $\mathcal{N}(0,A_{0})$, with $A_{0}$ given by \eqref{eq:stationarity-1}.
Other parameters: $\sigma_{w}=\sigma_{b}\in[0.76,0.81,0.85]$ (from
light to dark colors), $\alpha=0.1$. \label{fig:Depth-scale}}
\end{figure}

We show in \prettyref{fig:Depth-scale} that predictions of \eqref{eq:O_our_theory}
are indeed consistent with the depth scale obtained from \citeauthor{Poole16_3360}'s
theory \eqref{eq:depth_scale}. Moreover, since we derived our theory
for single networks, \eqref{eq:O_our_theory} also captures variability
due to particular network realizations. Interestingly, while for
network ensembles the depth scale $\xi$ describes the evolution of
the second moments, expression \eqref{eq:O_our_theory} shows that
for single networks $\xi$ describes the evolution of the squared
means across data samples.

\end{appendices}

\bibliographystyle{apsrev4-2_prx}
\bibliography{brain,add_to_brain}

\begin{thebibliography}{59}%
\makeatletter
\providecommand \@ifxundefined [1]{%
 \@ifx{#1\undefined}
}%
\providecommand \@ifnum [1]{%
 \ifnum #1\expandafter \@firstoftwo
 \else \expandafter \@secondoftwo
 \fi
}%
\providecommand \@ifx [1]{%
 \ifx #1\expandafter \@firstoftwo
 \else \expandafter \@secondoftwo
 \fi
}%
\providecommand \natexlab [1]{#1}%
\providecommand \enquote  [1]{``#1''}%
\providecommand \bibnamefont  [1]{#1}%
\providecommand \bibfnamefont [1]{#1}%
\providecommand \citenamefont [1]{#1}%
\providecommand \href@noop [0]{\@secondoftwo}%
\providecommand \href [0]{\begingroup \@sanitize@url \@href}%
\providecommand \@href[1]{\@@startlink{#1}\@@href}%
\providecommand \@@href[1]{\endgroup#1\@@endlink}%
\providecommand \@sanitize@url [0]{\catcode `\\12\catcode `\$12\catcode
  `\&12\catcode `\#12\catcode `\^12\catcode `\_12\catcode `\%12\relax}%
\providecommand \@@startlink[1]{}%
\providecommand \@@endlink[0]{}%
\providecommand \url  [0]{\begingroup\@sanitize@url \@url }%
\providecommand \@url [1]{\endgroup\@href {#1}{\urlprefix }}%
\providecommand \urlprefix  [0]{URL }%
\providecommand \Eprint [0]{\href }%
\providecommand \doibase [0]{https://doi.org/}%
\providecommand \selectlanguage [0]{\@gobble}%
\providecommand \bibinfo  [0]{\@secondoftwo}%
\providecommand \bibfield  [0]{\@secondoftwo}%
\providecommand \translation [1]{[#1]}%
\providecommand \BibitemOpen [0]{}%
\providecommand \bibitemStop [0]{}%
\providecommand \bibitemNoStop [0]{.\EOS\space}%
\providecommand \EOS [0]{\spacefactor3000\relax}%
\providecommand \BibitemShut  [1]{\csname bibitem#1\endcsname}%
\let\auto@bib@innerbib\@empty
\bibitem [{\citenamefont {Krizhevsky}\ \emph {et~al.}(2012)\citenamefont
  {Krizhevsky}, \citenamefont {Sutskever},\ and\ \citenamefont
  {Hinton}}]{Krizhevsky12_1097}%
  \BibitemOpen
  \bibfield  {author} {\bibinfo {author} {\bibfnamefont {A.}~\bibnamefont
  {Krizhevsky}}, \bibinfo {author} {\bibfnamefont {I.}~\bibnamefont
  {Sutskever}},\ and\ \bibinfo {author} {\bibfnamefont {G.~E.}\ \bibnamefont
  {Hinton}},\ }\bibfield  {title} {\bibinfo {title} {Imagenet classification
  with deep convolutional neural networks},\ }in\ \href
  {https://proceedings.neurips.cc/paper/2012/file/c399862d3b9d6b76c8436e924a68c45b-Paper.pdf}
  {\emph {\bibinfo {booktitle} {Adv. Neural Inf. Process. Syst.}}},\
  Vol.~\bibinfo {volume} {25},\ \bibinfo {editor} {edited by\ \bibinfo {editor}
  {\bibfnamefont {F.}~\bibnamefont {Pereira}}, \bibinfo {editor} {\bibfnamefont
  {C.~J.~C.}\ \bibnamefont {Burges}}, \bibinfo {editor} {\bibfnamefont
  {L.}~\bibnamefont {Bottou}},\ and\ \bibinfo {editor} {\bibfnamefont {K.~Q.}\
  \bibnamefont {Weinberger}}}\ (\bibinfo  {publisher} {Curran Associates,
  Inc.},\ \bibinfo {year} {2012})\ pp.\ \bibinfo {pages}
  {1097--1105}\BibitemShut {NoStop}%
\bibitem [{\citenamefont {Silver}\ \emph {et~al.}(2016)\citenamefont {Silver},
  \citenamefont {Huang}, \citenamefont {Maddison}, \citenamefont {Guez},
  \citenamefont {Sifre}, \citenamefont {Van Den~Driessche}, \citenamefont
  {Schrittwieser}, \citenamefont {Antonoglou}, \citenamefont {Panneershelvam},
  \citenamefont {Lanctot} \emph {et~al.}}]{Silver16_484}%
  \BibitemOpen
  \bibfield  {author} {\bibinfo {author} {\bibfnamefont {D.}~\bibnamefont
  {Silver}}, \bibinfo {author} {\bibfnamefont {A.}~\bibnamefont {Huang}},
  \bibinfo {author} {\bibfnamefont {C.~J.}\ \bibnamefont {Maddison}}, \bibinfo
  {author} {\bibfnamefont {A.}~\bibnamefont {Guez}}, \bibinfo {author}
  {\bibfnamefont {L.}~\bibnamefont {Sifre}}, \bibinfo {author} {\bibfnamefont
  {G.}~\bibnamefont {Van Den~Driessche}}, \bibinfo {author} {\bibfnamefont
  {J.}~\bibnamefont {Schrittwieser}}, \bibinfo {author} {\bibfnamefont
  {I.}~\bibnamefont {Antonoglou}}, \bibinfo {author} {\bibfnamefont
  {V.}~\bibnamefont {Panneershelvam}}, \bibinfo {author} {\bibfnamefont
  {M.}~\bibnamefont {Lanctot}}, \emph {et~al.},\ }\bibfield  {title} {\bibinfo
  {title} {Mastering the game of go with deep neural networks and tree
  search},\ }\href {https://doi.org/https://doi.org/10.1038/nature16961}
  {\bibfield  {journal} {\bibinfo  {journal} {Nature}\ }\textbf {\bibinfo
  {volume} {529}},\ \bibinfo {pages} {484} (\bibinfo {year}
  {2016})}\BibitemShut {NoStop}%
\bibitem [{\citenamefont {Bishop}(2006)}]{Bishop06}%
  \BibitemOpen
  \bibfield  {author} {\bibinfo {author} {\bibfnamefont {C.~M.}\ \bibnamefont
  {Bishop}},\ }\href@noop {} {\emph {\bibinfo {title} {Pattern Recognition and
  Machine Learning}}}\ (\bibinfo  {publisher} {Springer-Verlag New York,
  Inc.},\ \bibinfo {address} {Secaucus, NJ, USA},\ \bibinfo {year}
  {2006})\BibitemShut {NoStop}%
\bibitem [{\citenamefont {Bahri}\ \emph {et~al.}(2020)\citenamefont {Bahri},
  \citenamefont {Kadmon}, \citenamefont {Pennington}, \citenamefont
  {Schoenholz}, \citenamefont {Sohl-Dickstein},\ and\ \citenamefont
  {Ganguli}}]{Bahri20_501}%
  \BibitemOpen
  \bibfield  {author} {\bibinfo {author} {\bibfnamefont {Y.}~\bibnamefont
  {Bahri}}, \bibinfo {author} {\bibfnamefont {J.}~\bibnamefont {Kadmon}},
  \bibinfo {author} {\bibfnamefont {J.}~\bibnamefont {Pennington}}, \bibinfo
  {author} {\bibfnamefont {S.~S.}\ \bibnamefont {Schoenholz}}, \bibinfo
  {author} {\bibfnamefont {J.}~\bibnamefont {Sohl-Dickstein}},\ and\ \bibinfo
  {author} {\bibfnamefont {S.}~\bibnamefont {Ganguli}},\ }\bibfield  {title}
  {\bibinfo {title} {Statistical mechanics of deep learning},\ }\href
  {https://doi.org/10.1146/annurev-conmatphys-031119-050745} {\bibfield
  {journal} {\bibinfo  {journal} {Annu. Rev. Condens. Matter Phys.}\ }\textbf
  {\bibinfo {volume} {11}},\ \bibinfo {pages} {501} (\bibinfo {year}
  {2020})}\BibitemShut {NoStop}%
\bibitem [{\citenamefont {Lin}\ \emph {et~al.}(2017)\citenamefont {Lin},
  \citenamefont {Tegmark},\ and\ \citenamefont {Rolnick}}]{Lin17_1223}%
  \BibitemOpen
  \bibfield  {author} {\bibinfo {author} {\bibfnamefont {H.~W.}\ \bibnamefont
  {Lin}}, \bibinfo {author} {\bibfnamefont {M.}~\bibnamefont {Tegmark}},\ and\
  \bibinfo {author} {\bibfnamefont {D.}~\bibnamefont {Rolnick}},\ }\bibfield
  {title} {\bibinfo {title} {Why does deep and cheap learning work so well?},\
  }\href {https://doi.org/10.1007/s10955-017-1836-5} {\bibfield  {journal}
  {\bibinfo  {journal} {J. Stat. Phys.}\ }\textbf {\bibinfo {volume} {168}},\
  \bibinfo {pages} {1223} (\bibinfo {year} {2017})}\BibitemShut {NoStop}%
\bibitem [{\citenamefont {Shwartz-Ziv}\ and\ \citenamefont
  {Tishby}(2017)}]{Shwartz-Ziv17_arxiv}%
  \BibitemOpen
  \bibfield  {author} {\bibinfo {author} {\bibfnamefont {R.}~\bibnamefont
  {Shwartz-Ziv}}\ and\ \bibinfo {author} {\bibfnamefont {N.}~\bibnamefont
  {Tishby}},\ }\bibfield  {title} {\bibinfo {title} {Opening the black box of
  deep neural networks via information},\ }\href
  {http://arxiv.org/abs/1703.00810} {\bibfield  {journal} {\bibinfo  {journal}
  {ArXiv}\ } (\bibinfo {year} {2017})},\ \Eprint
  {https://arxiv.org/abs/1703.00810} {1703.00810} \BibitemShut {NoStop}%
\bibitem [{\citenamefont {Jacot}\ \emph {et~al.}(2018)\citenamefont {Jacot},
  \citenamefont {Gabriel},\ and\ \citenamefont {Hongler}}]{Jacot18_8580}%
  \BibitemOpen
  \bibfield  {author} {\bibinfo {author} {\bibfnamefont {A.}~\bibnamefont
  {Jacot}}, \bibinfo {author} {\bibfnamefont {F.}~\bibnamefont {Gabriel}},\
  and\ \bibinfo {author} {\bibfnamefont {C.}~\bibnamefont {Hongler}},\
  }\bibfield  {title} {\bibinfo {title} {Neural tangent kernel: Convergence and
  generalization in neural networks},\ }in\ \href
  {https://proceedings.neurips.cc/paper/2018/file/5a4be1fa34e62bb8a6ec6b91d2462f5a-Paper.pdf}
  {\emph {\bibinfo {booktitle} {Advances in Neural Information Processing
  Systems 31}}}\ (\bibinfo {year} {2018})\ pp.\ \bibinfo {pages}
  {8580--8589}\BibitemShut {NoStop}%
\bibitem [{\citenamefont {Saxe}\ \emph {et~al.}(2019)\citenamefont {Saxe},
  \citenamefont {McClelland},\ and\ \citenamefont {Ganguli}}]{Saxe19_11537}%
  \BibitemOpen
  \bibfield  {author} {\bibinfo {author} {\bibfnamefont {A.~M.}\ \bibnamefont
  {Saxe}}, \bibinfo {author} {\bibfnamefont {J.~L.}\ \bibnamefont
  {McClelland}},\ and\ \bibinfo {author} {\bibfnamefont {S.}~\bibnamefont
  {Ganguli}},\ }\bibfield  {title} {\bibinfo {title} {A mathematical theory of
  semantic development in deep neural networks},\ }\href
  {https://doi.org/10.1073/pnas.1820226116} {\bibfield  {journal} {\bibinfo
  {journal} {Proc. Natl. Acad. Sci. USA}\ }\textbf {\bibinfo {volume} {116}},\
  \bibinfo {pages} {11537} (\bibinfo {year} {2019})}\BibitemShut {NoStop}%
\bibitem [{\citenamefont {Cohen}\ \emph {et~al.}(2021)\citenamefont {Cohen},
  \citenamefont {Malka},\ and\ \citenamefont {Ringel}}]{Cohen21_023034}%
  \BibitemOpen
  \bibfield  {author} {\bibinfo {author} {\bibfnamefont {O.}~\bibnamefont
  {Cohen}}, \bibinfo {author} {\bibfnamefont {O.}~\bibnamefont {Malka}},\ and\
  \bibinfo {author} {\bibfnamefont {Z.}~\bibnamefont {Ringel}},\ }\bibfield
  {title} {\bibinfo {title} {Learning curves for overparametrized deep neural
  networks: A field theory perspective},\ }\href
  {https://doi.org/10.1103/PhysRevResearch.3.023034} {\bibfield  {journal}
  {\bibinfo  {journal} {Phys. Rev. Res.}\ }\textbf {\bibinfo {volume} {3}},\
  \bibinfo {pages} {023034} (\bibinfo {year} {2021})}\BibitemShut {NoStop}%
\bibitem [{\citenamefont {Neal}(1996)}]{Neal96}%
  \BibitemOpen
  \bibfield  {author} {\bibinfo {author} {\bibfnamefont {R.~M.}\ \bibnamefont
  {Neal}},\ }\href {https://doi.org/10.1007/978-1-4612-0745-0} {\emph {\bibinfo
  {title} {Bayesian Learning for Neural Networks}}}\ (\bibinfo  {publisher}
  {Springer New York},\ \bibinfo {year} {1996})\BibitemShut {NoStop}%
\bibitem [{\citenamefont {Williams}(1998)}]{Williams98_1203}%
  \BibitemOpen
  \bibfield  {author} {\bibinfo {author} {\bibfnamefont {C.~K.}\ \bibnamefont
  {Williams}},\ }\bibfield  {title} {\bibinfo {title} {Computation with
  infinite neural networks},\ }\href
  {https://doi.org/10.1162/089976698300017412} {\bibfield  {journal} {\bibinfo
  {journal} {Neural Comput.}\ }\textbf {\bibinfo {volume} {10}},\ \bibinfo
  {pages} {1203} (\bibinfo {year} {1998})}\BibitemShut {NoStop}%
\bibitem [{\citenamefont {Lee}\ \emph {et~al.}(2018)\citenamefont {Lee},
  \citenamefont {Sohl-Dickstein}, \citenamefont {Pennington}, \citenamefont
  {Novak}, \citenamefont {Schoenholz},\ and\ \citenamefont {Bahri}}]{Lee18}%
  \BibitemOpen
  \bibfield  {author} {\bibinfo {author} {\bibfnamefont {J.}~\bibnamefont
  {Lee}}, \bibinfo {author} {\bibfnamefont {J.}~\bibnamefont {Sohl-Dickstein}},
  \bibinfo {author} {\bibfnamefont {J.}~\bibnamefont {Pennington}}, \bibinfo
  {author} {\bibfnamefont {R.}~\bibnamefont {Novak}}, \bibinfo {author}
  {\bibfnamefont {S.}~\bibnamefont {Schoenholz}},\ and\ \bibinfo {author}
  {\bibfnamefont {Y.}~\bibnamefont {Bahri}},\ }\bibfield  {title} {\bibinfo
  {title} {Deep neural networks as gaussian processes},\ }in\ \href
  {https://openreview.net/forum?id=B1EA-M-0Z} {\emph {\bibinfo {booktitle}
  {International Conference on Learning Representations}}}\ (\bibinfo {year}
  {2018})\BibitemShut {NoStop}%
\bibitem [{\citenamefont {Garriga-Alonso}\ \emph {et~al.}(2019)\citenamefont
  {Garriga-Alonso}, \citenamefont {Rasmussen},\ and\ \citenamefont
  {Aitchison}}]{Garriga19}%
  \BibitemOpen
  \bibfield  {author} {\bibinfo {author} {\bibfnamefont {A.}~\bibnamefont
  {Garriga-Alonso}}, \bibinfo {author} {\bibfnamefont {C.~E.}\ \bibnamefont
  {Rasmussen}},\ and\ \bibinfo {author} {\bibfnamefont {L.}~\bibnamefont
  {Aitchison}},\ }\bibfield  {title} {\bibinfo {title} {Deep convolutional
  networks as shallow gaussian processes},\ }in\ \href
  {https://openreview.net/forum?id=Bklfsi0cKm} {\emph {\bibinfo {booktitle}
  {International Conference on Learning Representations}}}\ (\bibinfo {year}
  {2019})\BibitemShut {NoStop}%
\bibitem [{\citenamefont {Rasmussen}\ and\ \citenamefont
  {Williams}(2006)}]{WilliamsRasmussen06}%
  \BibitemOpen
  \bibfield  {author} {\bibinfo {author} {\bibfnamefont {C.}~\bibnamefont
  {Rasmussen}}\ and\ \bibinfo {author} {\bibfnamefont {C.}~\bibnamefont
  {Williams}},\ }\href@noop {} {\emph {\bibinfo {title} {Gaussian Processes for
  Machine Learning}}},\ Adaptive Computation and Machine Learning\ (\bibinfo
  {publisher} {MIT Press},\ \bibinfo {address} {Cambridge, MA, USA},\ \bibinfo
  {year} {2006})\ p.\ \bibinfo {pages} {248}\BibitemShut {NoStop}%
\bibitem [{\citenamefont {Poole}\ \emph {et~al.}(2016)\citenamefont {Poole},
  \citenamefont {Lahiri}, \citenamefont {Raghu}, \citenamefont
  {Sohl-Dickstein},\ and\ \citenamefont {Ganguli}}]{Poole16_3360}%
  \BibitemOpen
  \bibfield  {author} {\bibinfo {author} {\bibfnamefont {B.}~\bibnamefont
  {Poole}}, \bibinfo {author} {\bibfnamefont {S.}~\bibnamefont {Lahiri}},
  \bibinfo {author} {\bibfnamefont {M.}~\bibnamefont {Raghu}}, \bibinfo
  {author} {\bibfnamefont {J.}~\bibnamefont {Sohl-Dickstein}},\ and\ \bibinfo
  {author} {\bibfnamefont {S.}~\bibnamefont {Ganguli}},\ }\bibfield  {title}
  {\bibinfo {title} {Exponential expressivity in deep neural networks through
  transient chaos},\ }in\ \href
  {https://proceedings.neurips.cc/paper/2016/file/148510031349642de5ca0c544f31b2ef-Paper.pdf}
  {\emph {\bibinfo {booktitle} {Advances in Neural Information Processing
  Systems 29}}}\ (\bibinfo {year} {2016})\BibitemShut {NoStop}%
\bibitem [{\citenamefont {Raghu}\ \emph {et~al.}(2017)\citenamefont {Raghu},
  \citenamefont {Poole}, \citenamefont {Kleinberg}, \citenamefont {Ganguli},\
  and\ \citenamefont {Sohl-Dickstein}}]{Raghu17_2847}%
  \BibitemOpen
  \bibfield  {author} {\bibinfo {author} {\bibfnamefont {M.}~\bibnamefont
  {Raghu}}, \bibinfo {author} {\bibfnamefont {B.}~\bibnamefont {Poole}},
  \bibinfo {author} {\bibfnamefont {J.}~\bibnamefont {Kleinberg}}, \bibinfo
  {author} {\bibfnamefont {S.}~\bibnamefont {Ganguli}},\ and\ \bibinfo {author}
  {\bibfnamefont {J.}~\bibnamefont {Sohl-Dickstein}},\ }\bibfield  {title}
  {\bibinfo {title} {On the expressive power of deep neural networks},\ }in\
  \href {https://proceedings.mlr.press/v70/raghu17a.html} {\emph {\bibinfo
  {booktitle} {Proceedings of the 34th International Conference on Machine
  Learning}}},\ \bibinfo {series} {Proceedings of Machine Learning Research},
  Vol.~\bibinfo {volume} {70},\ \bibinfo {editor} {edited by\ \bibinfo {editor}
  {\bibfnamefont {D.}~\bibnamefont {Precup}}\ and\ \bibinfo {editor}
  {\bibfnamefont {Y.~W.}\ \bibnamefont {Teh}}}\ (\bibinfo  {publisher} {PMLR},\
  \bibinfo {year} {2017})\ pp.\ \bibinfo {pages} {2847--2854}\BibitemShut
  {NoStop}%
\bibitem [{\citenamefont {Schoenholz}\ \emph {et~al.}(2017)\citenamefont
  {Schoenholz}, \citenamefont {Gilmer}, \citenamefont {Ganguli},\ and\
  \citenamefont {Sohl-Dickstein}}]{Schoenholz17}%
  \BibitemOpen
  \bibfield  {author} {\bibinfo {author} {\bibfnamefont {S.~S.}\ \bibnamefont
  {Schoenholz}}, \bibinfo {author} {\bibfnamefont {J.}~\bibnamefont {Gilmer}},
  \bibinfo {author} {\bibfnamefont {S.}~\bibnamefont {Ganguli}},\ and\ \bibinfo
  {author} {\bibfnamefont {J.}~\bibnamefont {Sohl-Dickstein}},\ }\bibfield
  {title} {\bibinfo {title} {Deep information propagation},\ }in\ \href
  {https://openreview.net/forum?id=H1W1UN9gg} {\emph {\bibinfo {booktitle}
  {International Conference on Learning Representations}}}\ (\bibinfo {year}
  {2017})\BibitemShut {NoStop}%
\bibitem [{\citenamefont {Kleinert}(1989)}]{Kleinert89}%
  \BibitemOpen
  \bibfield  {author} {\bibinfo {author} {\bibfnamefont {H.}~\bibnamefont
  {Kleinert}},\ }\href@noop {} {\emph {\bibinfo {title} {Gauge fields in
  condensed matter, Vol. I , SUPERFLOW AND VORTEX LINES Disorder Fields, Phase
  Transitions}}}\ (\bibinfo  {publisher} {World Scientific},\ \bibinfo {year}
  {1989})\BibitemShut {NoStop}%
\bibitem [{\citenamefont {Zinn-Justin}(1996)}]{ZinnJustin96}%
  \BibitemOpen
  \bibfield  {author} {\bibinfo {author} {\bibfnamefont {J.}~\bibnamefont
  {Zinn-Justin}},\ }\href@noop {} {\emph {\bibinfo {title} {Quantum field
  theory and critical phenomena}}}\ (\bibinfo  {publisher} {Clarendon Press,
  Oxford},\ \bibinfo {year} {1996})\BibitemShut {NoStop}%
\bibitem [{\citenamefont {Hertz}\ \emph {et~al.}(2017)\citenamefont {Hertz},
  \citenamefont {Roudi},\ and\ \citenamefont {Sollich}}]{Hertz16_033001}%
  \BibitemOpen
  \bibfield  {author} {\bibinfo {author} {\bibfnamefont {J.~A.}\ \bibnamefont
  {Hertz}}, \bibinfo {author} {\bibfnamefont {Y.}~\bibnamefont {Roudi}},\ and\
  \bibinfo {author} {\bibfnamefont {P.}~\bibnamefont {Sollich}},\ }\bibfield
  {title} {\bibinfo {title} {Path integral methods for the dynamics of
  stochastic and disordered systems},\ }\href
  {http://stacks.iop.org/1751-8121/50/i=3/a=033001} {\bibfield  {journal}
  {\bibinfo  {journal} {J. Phys. A}\ }\textbf {\bibinfo {volume} {50}},\
  \bibinfo {pages} {033001} (\bibinfo {year} {2017})}\BibitemShut {NoStop}%
\bibitem [{\citenamefont {Helias}\ and\ \citenamefont
  {Dahmen}(2020)}]{Helias20_970}%
  \BibitemOpen
  \bibfield  {author} {\bibinfo {author} {\bibfnamefont {M.}~\bibnamefont
  {Helias}}\ and\ \bibinfo {author} {\bibfnamefont {D.}~\bibnamefont
  {Dahmen}},\ }\href {https://doi.org/10.1007/978-3-030-46444-8} {\emph
  {\bibinfo {title} {Statistical Field Theory for Neural Networks}}}\ (\bibinfo
   {publisher} {Springer International Publishing},\ \bibinfo {year} {2020})\
  p.\ \bibinfo {pages} {203}\BibitemShut {NoStop}%
\bibitem [{\citenamefont {LeCun}\ \emph {et~al.}(2010)\citenamefont {LeCun},
  \citenamefont {Cortes},\ and\ \citenamefont {Burges}}]{lecun2010mnist}%
  \BibitemOpen
  \bibfield  {author} {\bibinfo {author} {\bibfnamefont {Y.}~\bibnamefont
  {LeCun}}, \bibinfo {author} {\bibfnamefont {C.}~\bibnamefont {Cortes}},\ and\
  \bibinfo {author} {\bibfnamefont {C.~J.~C.}\ \bibnamefont {Burges}},\
  }\bibfield  {title} {\bibinfo {title} {{MNIST} handwritten digit database},\
  }\href {http://yann.lecun.com/exdb/mnist/} {\bibfield  {journal} {\bibinfo
  {journal} {ATT Labs}\ } (\bibinfo {year} {2010})}\BibitemShut {NoStop}%
\bibitem [{\citenamefont {Vapnik}(1992)}]{Vapnik92_831}%
  \BibitemOpen
  \bibfield  {author} {\bibinfo {author} {\bibfnamefont {V.}~\bibnamefont
  {Vapnik}},\ }\bibfield  {title} {\bibinfo {title} {Principles of risk
  minimization for learning theory},\ }in\ \href
  {https://proceedings.neurips.cc/paper/1991/file/ff4d5fbbafdf976cfdc032e3bde78de5-Paper.pdf}
  {\emph {\bibinfo {booktitle} {Adv. Neural Inf. Process. Syst.}}},\
  Vol.~\bibinfo {volume} {4},\ \bibinfo {editor} {edited by\ \bibinfo {editor}
  {\bibfnamefont {J.}~\bibnamefont {Moody}}, \bibinfo {editor} {\bibfnamefont
  {S.}~\bibnamefont {Hanson}},\ and\ \bibinfo {editor} {\bibfnamefont {R.~P.}\
  \bibnamefont {Lippmann}}}\ (\bibinfo  {publisher} {Morgan-Kaufmann},\
  \bibinfo {year} {1992})\ pp.\ \bibinfo {pages} {831--838}\BibitemShut
  {NoStop}%
\bibitem [{\citenamefont {Vapnik}(1998)}]{Vapnik98_learningtheory}%
  \BibitemOpen
  \bibfield  {author} {\bibinfo {author} {\bibfnamefont {V.~N.}\ \bibnamefont
  {Vapnik}},\ }\href@noop {} {\emph {\bibinfo {title} {{Statistical Learning
  Theory}}}}\ (\bibinfo  {publisher} {Wiley},\ \bibinfo {address} {Hoboken, NJ,
  USA},\ \bibinfo {year} {1998})\BibitemShut {NoStop}%
\bibitem [{\citenamefont {Kohavi}\ and\ \citenamefont
  {Wolpert}(1996)}]{Kohavi96_275}%
  \BibitemOpen
  \bibfield  {author} {\bibinfo {author} {\bibfnamefont {R.}~\bibnamefont
  {Kohavi}}\ and\ \bibinfo {author} {\bibfnamefont {D.~H.}\ \bibnamefont
  {Wolpert}},\ }\bibfield  {title} {\bibinfo {title} {Bias plus variance
  decomposition for zero-one loss functions},\ }in\ \href@noop {} {\emph
  {\bibinfo {booktitle} {Proceedings of the Thirteenth International Conference
  on Machine Learning}}},\ Vol.~\bibinfo {volume} {96}\ (\bibinfo {year}
  {1996})\ pp.\ \bibinfo {pages} {275--283}\BibitemShut {NoStop}%
\bibitem [{\citenamefont {Kingma}\ and\ \citenamefont
  {Ba}(2015)}]{Kingma15_iclr}%
  \BibitemOpen
  \bibfield  {author} {\bibinfo {author} {\bibfnamefont {D.~P.}\ \bibnamefont
  {Kingma}}\ and\ \bibinfo {author} {\bibfnamefont {J.~L.}\ \bibnamefont
  {Ba}},\ }\bibfield  {title} {\bibinfo {title} {Adam: A method for stochastic
  gradient descent},\ }in\ \href {http://arxiv.org/abs/1412.6980} {\emph
  {\bibinfo {booktitle} {International Conference on Learning
  Representations}}}\ (\bibinfo {year} {2015})\BibitemShut {NoStop}%
\bibitem [{\citenamefont {Loshchilov}\ and\ \citenamefont
  {Hutter}(2019)}]{Loshchilov19}%
  \BibitemOpen
  \bibfield  {author} {\bibinfo {author} {\bibfnamefont {I.}~\bibnamefont
  {Loshchilov}}\ and\ \bibinfo {author} {\bibfnamefont {F.}~\bibnamefont
  {Hutter}},\ }\bibfield  {title} {\bibinfo {title} {Decoupled weight decay
  regularization},\ }in\ \href {https://openreview.net/forum?id=Bkg6RiCqY7}
  {\emph {\bibinfo {booktitle} {International Conference on Learning
  Representations}}}\ (\bibinfo {year} {2019})\BibitemShut {NoStop}%
\bibitem [{\citenamefont {Paszke}\ \emph {et~al.}(2019)\citenamefont {Paszke},
  \citenamefont {Gross}, \citenamefont {Massa}, \citenamefont {Lerer},
  \citenamefont {Bradbury}, \citenamefont {Chanan}, \citenamefont {Killeen},
  \citenamefont {Lin}, \citenamefont {Gimelshein}, \citenamefont {Antiga},
  \citenamefont {Desmaison}, \citenamefont {Kopf}, \citenamefont {Yang},
  \citenamefont {DeVito}, \citenamefont {Raison}, \citenamefont {Tejani},
  \citenamefont {Chilamkurthy}, \citenamefont {Steiner}, \citenamefont {Fang},
  \citenamefont {Bai},\ and\ \citenamefont {Chintala}}]{Paszke19_8024}%
  \BibitemOpen
  \bibfield  {author} {\bibinfo {author} {\bibfnamefont {A.}~\bibnamefont
  {Paszke}}, \bibinfo {author} {\bibfnamefont {S.}~\bibnamefont {Gross}},
  \bibinfo {author} {\bibfnamefont {F.}~\bibnamefont {Massa}}, \bibinfo
  {author} {\bibfnamefont {A.}~\bibnamefont {Lerer}}, \bibinfo {author}
  {\bibfnamefont {J.}~\bibnamefont {Bradbury}}, \bibinfo {author}
  {\bibfnamefont {G.}~\bibnamefont {Chanan}}, \bibinfo {author} {\bibfnamefont
  {T.}~\bibnamefont {Killeen}}, \bibinfo {author} {\bibfnamefont
  {Z.}~\bibnamefont {Lin}}, \bibinfo {author} {\bibfnamefont {N.}~\bibnamefont
  {Gimelshein}}, \bibinfo {author} {\bibfnamefont {L.}~\bibnamefont {Antiga}},
  \bibinfo {author} {\bibfnamefont {A.}~\bibnamefont {Desmaison}}, \bibinfo
  {author} {\bibfnamefont {A.}~\bibnamefont {Kopf}}, \bibinfo {author}
  {\bibfnamefont {E.}~\bibnamefont {Yang}}, \bibinfo {author} {\bibfnamefont
  {Z.}~\bibnamefont {DeVito}}, \bibinfo {author} {\bibfnamefont
  {M.}~\bibnamefont {Raison}}, \bibinfo {author} {\bibfnamefont
  {A.}~\bibnamefont {Tejani}}, \bibinfo {author} {\bibfnamefont
  {S.}~\bibnamefont {Chilamkurthy}}, \bibinfo {author} {\bibfnamefont
  {B.}~\bibnamefont {Steiner}}, \bibinfo {author} {\bibfnamefont
  {L.}~\bibnamefont {Fang}}, \bibinfo {author} {\bibfnamefont {J.}~\bibnamefont
  {Bai}},\ and\ \bibinfo {author} {\bibfnamefont {S.}~\bibnamefont
  {Chintala}},\ }\bibfield  {title} {\bibinfo {title} {Pytorch: An imperative
  style, high-performance deep learning library},\ }in\ \href
  {https://proceedings.neurips.cc/paper/2019/file/bdbca288fee7f92f2bfa9f7012727740-Paper.pdf}
  {\emph {\bibinfo {booktitle} {Adv. Neural Inf. Process. Syst.}}},\
  Vol.~\bibinfo {volume} {32},\ \bibinfo {editor} {edited by\ \bibinfo {editor}
  {\bibfnamefont {H.}~\bibnamefont {Wallach}}, \bibinfo {editor} {\bibfnamefont
  {H.}~\bibnamefont {Larochelle}}, \bibinfo {editor} {\bibfnamefont
  {A.}~\bibnamefont {Beygelzimer}}, \bibinfo {editor} {\bibfnamefont
  {F.}~\bibnamefont {d\textquotesingle Alch\'{e}-Buc}}, \bibinfo {editor}
  {\bibfnamefont {E.}~\bibnamefont {Fox}},\ and\ \bibinfo {editor}
  {\bibfnamefont {R.}~\bibnamefont {Garnett}}}\ (\bibinfo  {publisher} {Curran
  Associates, Inc.},\ \bibinfo {year} {2019})\ pp.\ \bibinfo {pages}
  {8024--8035}\BibitemShut {NoStop}%
\bibitem [{\citenamefont {Krizhevsky}(2009)}]{Krizhevsky09_thesis}%
  \BibitemOpen
  \bibfield  {author} {\bibinfo {author} {\bibfnamefont {A.}~\bibnamefont
  {Krizhevsky}},\ }\emph {\bibinfo {title} {Learning multiple layers of
  features from tiny images}},\ \href
  {https://www.cs.toronto.edu/~kriz/learning-features-2009-TR.pdf} {Master's
  thesis},\ \bibinfo  {school} {Department of Computer Science, University of
  Toronto} (\bibinfo {year} {2009})\BibitemShut {NoStop}%
\bibitem [{\citenamefont {Zagoruyko}\ and\ \citenamefont
  {Komodakis}(2016)}]{Zagoruyko16_87}%
  \BibitemOpen
  \bibfield  {author} {\bibinfo {author} {\bibfnamefont {S.}~\bibnamefont
  {Zagoruyko}}\ and\ \bibinfo {author} {\bibfnamefont {N.}~\bibnamefont
  {Komodakis}},\ }\bibfield  {title} {\bibinfo {title} {Wide residual
  networks},\ }in\ \href {https://doi.org/10.5244/C.30.87} {\emph {\bibinfo
  {booktitle} {Proceedings of the British Machine Vision Conference (BMVC)}}},\
  \bibinfo {editor} {edited by\ \bibinfo {editor} {\bibfnamefont {E.~R.~H.}\
  \bibnamefont {Richard C.~Wilson}}\ and\ \bibinfo {editor} {\bibfnamefont
  {W.~A.~P.}\ \bibnamefont {Smith}}}\ (\bibinfo  {publisher} {BMVA Press},\
  \bibinfo {year} {2016})\ pp.\ \bibinfo {pages} {87.1--87.12}\BibitemShut
  {NoStop}%
\bibitem [{\citenamefont {Cybenko}(1989)}]{Cybenko1989}%
  \BibitemOpen
  \bibfield  {author} {\bibinfo {author} {\bibfnamefont {G.}~\bibnamefont
  {Cybenko}},\ }\bibfield  {title} {\bibinfo {title} {Approximation by
  superpositions of a sigmoidal function},\ }\href
  {https://doi.org/10.1007/bf02551274} {\bibfield  {journal} {\bibinfo
  {journal} {Math. Control Signals. Syst.}\ }\textbf {\bibinfo {volume} {2}},\
  \bibinfo {pages} {303} (\bibinfo {year} {1989})}\BibitemShut {NoStop}%
\bibitem [{\citenamefont {Leshno}\ \emph {et~al.}(1993)\citenamefont {Leshno},
  \citenamefont {Lin}, \citenamefont {Pinkus},\ and\ \citenamefont
  {Schocken}}]{Leshno93_861}%
  \BibitemOpen
  \bibfield  {author} {\bibinfo {author} {\bibfnamefont {M.}~\bibnamefont
  {Leshno}}, \bibinfo {author} {\bibfnamefont {V.~Y.}\ \bibnamefont {Lin}},
  \bibinfo {author} {\bibfnamefont {A.}~\bibnamefont {Pinkus}},\ and\ \bibinfo
  {author} {\bibfnamefont {S.}~\bibnamefont {Schocken}},\ }\bibfield  {title}
  {\bibinfo {title} {Multilayer feedforward networks with a nonpolynomial
  activation function can approximate any function},\ }\href
  {https://doi.org/10.1016/S0893-6080(05)80131-5} {\bibfield  {journal}
  {\bibinfo  {journal} {Neural Netw.}\ }\textbf {\bibinfo {volume} {6}},\
  \bibinfo {pages} {861} (\bibinfo {year} {1993})}\BibitemShut {NoStop}%
\bibitem [{\citenamefont {Pinkus}(1999)}]{Pinkus99_143}%
  \BibitemOpen
  \bibfield  {author} {\bibinfo {author} {\bibfnamefont {A.}~\bibnamefont
  {Pinkus}},\ }\bibfield  {title} {\bibinfo {title} {Approximation theory of
  the mlp model in neural networks},\ }\href
  {https://doi.org/10.1017/S0962492900002919} {\bibfield  {journal} {\bibinfo
  {journal} {Acta Numer.}\ }\textbf {\bibinfo {volume} {8}},\ \bibinfo {pages}
  {143} (\bibinfo {year} {1999})}\BibitemShut {NoStop}%
\bibitem [{\citenamefont {MacKay}(2003)}]{Mackay2003}%
  \BibitemOpen
  \bibfield  {author} {\bibinfo {author} {\bibfnamefont {D.~J.}\ \bibnamefont
  {MacKay}},\ }\href@noop {} {\emph {\bibinfo {title} {Information theory,
  inference and learning algorithms}}}\ (\bibinfo  {publisher} {Cambridge
  university press},\ \bibinfo {year} {2003})\BibitemShut {NoStop}%
\bibitem [{\citenamefont {Williams}\ and\ \citenamefont
  {Barber}(1998)}]{Williams98}%
  \BibitemOpen
  \bibfield  {author} {\bibinfo {author} {\bibfnamefont {C.~K.~I.}\
  \bibnamefont {Williams}}\ and\ \bibinfo {author} {\bibfnamefont
  {D.}~\bibnamefont {Barber}},\ }\bibfield  {title} {\bibinfo {title} {Bayesian
  classification with gaussian processes},\ }\href
  {https://doi.org/10.1109/34.735807} {\bibfield  {journal} {\bibinfo
  {journal} {IEEE Trans. Pattern Anal. Mach. Intel.}\ }\textbf {\bibinfo
  {volume} {20}},\ \bibinfo {pages} {1342} (\bibinfo {year}
  {1998})}\BibitemShut {NoStop}%
\bibitem [{\citenamefont {Williams}\ and\ \citenamefont
  {Rasmussen}(2006)}]{Williams06}%
  \BibitemOpen
  \bibfield  {author} {\bibinfo {author} {\bibfnamefont {C.~K.}\ \bibnamefont
  {Williams}}\ and\ \bibinfo {author} {\bibfnamefont {C.~E.}\ \bibnamefont
  {Rasmussen}},\ }\href@noop {} {\emph {\bibinfo {title} {Gaussian Processes
  for Machine Learning}}},\ \bibinfo {edition} {1st}\ ed.\ (\bibinfo
  {publisher} {MIT Press},\ \bibinfo {address} {Cambridge},\ \bibinfo {year}
  {2006})\BibitemShut {NoStop}%
\bibitem [{\citenamefont {Dyer}\ and\ \citenamefont
  {Gur-Ari}(2020)}]{Dyer20_ICLR}%
  \BibitemOpen
  \bibfield  {author} {\bibinfo {author} {\bibfnamefont {E.}~\bibnamefont
  {Dyer}}\ and\ \bibinfo {author} {\bibfnamefont {G.}~\bibnamefont {Gur-Ari}},\
  }\bibfield  {title} {\bibinfo {title} {Asymptotics of wide networks from
  feynman diagrams},\ }in\ \href {https://openreview.net/forum?id=S1gFvANKDS}
  {\emph {\bibinfo {booktitle} {International Conference on Learning
  Representations}}}\ (\bibinfo {year} {2020})\BibitemShut {NoStop}%
\bibitem [{\citenamefont {Naveh}\ \emph {et~al.}(2021)\citenamefont {Naveh},
  \citenamefont {Ben~David}, \citenamefont {Sompolinsky},\ and\ \citenamefont
  {Ringel}}]{Naveh21_064301}%
  \BibitemOpen
  \bibfield  {author} {\bibinfo {author} {\bibfnamefont {G.}~\bibnamefont
  {Naveh}}, \bibinfo {author} {\bibfnamefont {O.}~\bibnamefont {Ben~David}},
  \bibinfo {author} {\bibfnamefont {H.}~\bibnamefont {Sompolinsky}},\ and\
  \bibinfo {author} {\bibfnamefont {Z.}~\bibnamefont {Ringel}},\ }\bibfield
  {title} {\bibinfo {title} {Predicting the outputs of finite deep neural
  networks trained with noisy gradients},\ }\href
  {https://doi.org/10.1103/PhysRevE.104.064301} {\bibfield  {journal} {\bibinfo
   {journal} {Phys. Rev. E}\ }\textbf {\bibinfo {volume} {104}},\ \bibinfo
  {pages} {064301} (\bibinfo {year} {2021})}\BibitemShut {NoStop}%
\bibitem [{\citenamefont {Yaida}(2020)}]{Yaida20}%
  \BibitemOpen
  \bibfield  {author} {\bibinfo {author} {\bibfnamefont {S.}~\bibnamefont
  {Yaida}},\ }\bibfield  {title} {\bibinfo {title} {Non-{G}aussian processes
  and neural networks at finite widths},\ }in\ \href
  {http://proceedings.mlr.press/v107/yaida20a.html} {\emph {\bibinfo
  {booktitle} {Proceedings of The First Mathematical and Scientific Machine
  Learning Conference}}},\ \bibinfo {series} {Proceedings of Machine Learning
  Research}, Vol.\ \bibinfo {volume} {107},\ \bibinfo {editor} {edited by\
  \bibinfo {editor} {\bibfnamefont {J.}~\bibnamefont {Lu}}\ and\ \bibinfo
  {editor} {\bibfnamefont {R.}~\bibnamefont {Ward}}}\ (\bibinfo  {publisher}
  {PMLR},\ \bibinfo {address} {Princeton University, Princeton, NJ, USA},\
  \bibinfo {year} {2020})\ pp.\ \bibinfo {pages} {165--192}\BibitemShut
  {NoStop}%
\bibitem [{\citenamefont {Deco}\ and\ \citenamefont {Brauer}(1994)}]{Deco94}%
  \BibitemOpen
  \bibfield  {author} {\bibinfo {author} {\bibfnamefont {G.}~\bibnamefont
  {Deco}}\ and\ \bibinfo {author} {\bibfnamefont {W.}~\bibnamefont {Brauer}},\
  }\bibfield  {title} {\bibinfo {title} {Higher order statistical decorrelation
  without information loss},\ }in\ \href
  {https://proceedings.neurips.cc/paper/1994/hash/892c91e0a653ba19df81a90f89d99bcd-Abstract.html}
  {\emph {\bibinfo {booktitle} {Proceedings of the 7th {International}
  {Conference} on {Neural} {Information} {Processing} {Systems}}}},\ \bibinfo
  {series and number} {{NIPS}'94}\ (\bibinfo  {publisher} {MIT Press},\
  \bibinfo {address} {Cambridge, MA, USA},\ \bibinfo {year} {1994})\ pp.\
  \bibinfo {pages} {247--254}\BibitemShut {NoStop}%
\bibitem [{\citenamefont {Goldt}\ \emph {et~al.}(2020)\citenamefont {Goldt},
  \citenamefont {M\'{e}zard}, \citenamefont {Krzakala},\ and\ \citenamefont
  {Zdeborov\'{a}}}]{Goldt20_prx}%
  \BibitemOpen
  \bibfield  {author} {\bibinfo {author} {\bibfnamefont {S.}~\bibnamefont
  {Goldt}}, \bibinfo {author} {\bibfnamefont {M.}~\bibnamefont {M\'{e}zard}},
  \bibinfo {author} {\bibfnamefont {F.}~\bibnamefont {Krzakala}},\ and\
  \bibinfo {author} {\bibfnamefont {L.}~\bibnamefont {Zdeborov\'{a}}},\
  }\bibfield  {title} {\bibinfo {title} {{Modeling the Influence of Data
  Structure on Learning in Neural Networks: The Hidden Manifold Model}},\
  }\href {https://doi.org/10.1103/PhysRevX.10.041044} {\bibfield  {journal}
  {\bibinfo  {journal} {Phys. Rev. X}\ }\textbf {\bibinfo {volume} {10}},\
  \bibinfo {pages} {041044} (\bibinfo {year} {2020})}\BibitemShut {NoStop}%
\bibitem [{\citenamefont {Goldt}\ \emph {et~al.}(2022)\citenamefont {Goldt},
  \citenamefont {Loureiro}, \citenamefont {Reeves}, \citenamefont {Krzakala},
  \citenamefont {Mezard},\ and\ \citenamefont {Zdeborova}}]{Goldt22_426}%
  \BibitemOpen
  \bibfield  {author} {\bibinfo {author} {\bibfnamefont {S.}~\bibnamefont
  {Goldt}}, \bibinfo {author} {\bibfnamefont {B.}~\bibnamefont {Loureiro}},
  \bibinfo {author} {\bibfnamefont {G.}~\bibnamefont {Reeves}}, \bibinfo
  {author} {\bibfnamefont {F.}~\bibnamefont {Krzakala}}, \bibinfo {author}
  {\bibfnamefont {M.}~\bibnamefont {Mezard}},\ and\ \bibinfo {author}
  {\bibfnamefont {L.}~\bibnamefont {Zdeborova}},\ }\bibfield  {title} {\bibinfo
  {title} {The gaussian equivalence of generative models for learning with
  shallow neural networks},\ }in\ \href
  {https://proceedings.mlr.press/v145/goldt22a.html} {\emph {\bibinfo
  {booktitle} {Proceedings of the 2nd Mathematical and Scientific Machine
  Learning Conference}}},\ \bibinfo {series} {Proceedings of Machine Learning
  Research}, Vol.\ \bibinfo {volume} {145},\ \bibinfo {editor} {edited by\
  \bibinfo {editor} {\bibfnamefont {J.}~\bibnamefont {Bruna}}, \bibinfo
  {editor} {\bibfnamefont {J.}~\bibnamefont {Hesthaven}},\ and\ \bibinfo
  {editor} {\bibfnamefont {L.}~\bibnamefont {Zdeborova}}}\ (\bibinfo
  {publisher} {PMLR},\ \bibinfo {year} {2022})\ pp.\ \bibinfo {pages}
  {426--471}\BibitemShut {NoStop}%
\bibitem [{\citenamefont {Loureiro}\ \emph {et~al.}(2022)\citenamefont
  {Loureiro}, \citenamefont {Gerbelot}, \citenamefont {Cui}, \citenamefont
  {Goldt}, \citenamefont {Krzakala}, \citenamefont {M\`ezard},\ and\
  \citenamefont {Zdeborov\'a}}]{Loureiro22_114001}%
  \BibitemOpen
  \bibfield  {author} {\bibinfo {author} {\bibfnamefont {B.}~\bibnamefont
  {Loureiro}}, \bibinfo {author} {\bibfnamefont {C.}~\bibnamefont {Gerbelot}},
  \bibinfo {author} {\bibfnamefont {H.}~\bibnamefont {Cui}}, \bibinfo {author}
  {\bibfnamefont {S.}~\bibnamefont {Goldt}}, \bibinfo {author} {\bibfnamefont
  {F.}~\bibnamefont {Krzakala}}, \bibinfo {author} {\bibfnamefont
  {M.}~\bibnamefont {M\`ezard}},\ and\ \bibinfo {author} {\bibfnamefont
  {L.}~\bibnamefont {Zdeborov\'a}},\ }\bibfield  {title} {\bibinfo {title}
  {Learning curves of generic features maps for realistic datasets with a
  teacher-student model},\ }\href {https://doi.org/10.1088/1742-5468/ac9825}
  {\bibfield  {journal} {\bibinfo  {journal} {J. Stat. Mech. Theory Exp.}\
  }\textbf {\bibinfo {volume} {2022}},\ \bibinfo {pages} {114001} (\bibinfo
  {year} {2022})}\BibitemShut {NoStop}%
\bibitem [{\citenamefont {Yang}\ and\ \citenamefont {Hu}(2021)}]{Yang21_11727}%
  \BibitemOpen
  \bibfield  {author} {\bibinfo {author} {\bibfnamefont {G.}~\bibnamefont
  {Yang}}\ and\ \bibinfo {author} {\bibfnamefont {E.~J.}\ \bibnamefont {Hu}},\
  }\bibfield  {title} {\bibinfo {title} {Tensor programs iv: Feature learning
  in infinite-width neural networks},\ }in\ \href
  {https://proceedings.mlr.press/v139/yang21c.html} {\emph {\bibinfo
  {booktitle} {Proceedings of the 38th International Conference on Machine
  Learning}}},\ \bibinfo {series} {Proceedings of Machine Learning Research},
  Vol.\ \bibinfo {volume} {139},\ \bibinfo {editor} {edited by\ \bibinfo
  {editor} {\bibfnamefont {M.}~\bibnamefont {Meila}}\ and\ \bibinfo {editor}
  {\bibfnamefont {T.}~\bibnamefont {Zhang}}}\ (\bibinfo  {publisher} {PMLR},\
  \bibinfo {year} {2021})\ pp.\ \bibinfo {pages} {11727--11737}\BibitemShut
  {NoStop}%
\bibitem [{\citenamefont {Fang}\ \emph {et~al.}(2021)\citenamefont {Fang},
  \citenamefont {Lee}, \citenamefont {Yang},\ and\ \citenamefont
  {Zhang}}]{Fang21_1887}%
  \BibitemOpen
  \bibfield  {author} {\bibinfo {author} {\bibfnamefont {C.}~\bibnamefont
  {Fang}}, \bibinfo {author} {\bibfnamefont {J.}~\bibnamefont {Lee}}, \bibinfo
  {author} {\bibfnamefont {P.}~\bibnamefont {Yang}},\ and\ \bibinfo {author}
  {\bibfnamefont {T.}~\bibnamefont {Zhang}},\ }\bibfield  {title} {\bibinfo
  {title} {Modeling from features: a mean-field framework for
  over-parameterized deep neural networks},\ }in\ \href
  {https://proceedings.mlr.press/v134/fang21a.html} {\emph {\bibinfo
  {booktitle} {Proceedings of Thirty Fourth Conference on Learning Theory}}},\
  \bibinfo {series} {Proceedings of Machine Learning Research}, Vol.\ \bibinfo
  {volume} {134}\ (\bibinfo  {publisher} {PMLR},\ \bibinfo {year} {2021})\ pp.\
  \bibinfo {pages} {1887--1936}\BibitemShut {NoStop}%
\bibitem [{\citenamefont {Seddik}\ \emph {et~al.}(2020)\citenamefont {Seddik},
  \citenamefont {Louart}, \citenamefont {Tamaazousti},\ and\ \citenamefont
  {Couillet}}]{Seddik20_8573}%
  \BibitemOpen
  \bibfield  {author} {\bibinfo {author} {\bibfnamefont {M.~E.~A.}\
  \bibnamefont {Seddik}}, \bibinfo {author} {\bibfnamefont {C.}~\bibnamefont
  {Louart}}, \bibinfo {author} {\bibfnamefont {M.}~\bibnamefont
  {Tamaazousti}},\ and\ \bibinfo {author} {\bibfnamefont {R.}~\bibnamefont
  {Couillet}},\ }\bibfield  {title} {\bibinfo {title} {{Random Matrix Theory
  Proves that Deep Learning Representations of GAN-data Behave as Gaussian
  Mixtures}},\ }in\ \href {http://proceedings.mlr.press/v119/seddik20a.html}
  {\emph {\bibinfo {booktitle} {{International Conference on Machine
  Learning}}}}\ (\bibinfo  {publisher} {PMLR},\ \bibinfo {year} {2020})\ pp.\
  \bibinfo {pages} {8573--8582}\BibitemShut {NoStop}%
\bibitem [{\citenamefont {Huang}(2018)}]{Huang18_062313}%
  \BibitemOpen
  \bibfield  {author} {\bibinfo {author} {\bibfnamefont {H.}~\bibnamefont
  {Huang}},\ }\bibfield  {title} {\bibinfo {title} {Mechanisms of
  dimensionality reduction and decorrelation in deep neural networks},\ }\href
  {https://doi.org/10.1103/PhysRevE.98.062313} {\bibfield  {journal} {\bibinfo
  {journal} {Phys. Rev. E}\ }\textbf {\bibinfo {volume} {98}},\ \bibinfo
  {pages} {062313} (\bibinfo {year} {2018})}\BibitemShut {NoStop}%
\bibitem [{\citenamefont {Zhou}\ and\ \citenamefont
  {Huang}(2021)}]{Zhou21_012315}%
  \BibitemOpen
  \bibfield  {author} {\bibinfo {author} {\bibfnamefont {J.}~\bibnamefont
  {Zhou}}\ and\ \bibinfo {author} {\bibfnamefont {H.}~\bibnamefont {Huang}},\
  }\bibfield  {title} {\bibinfo {title} {Weakly correlated synapses promote
  dimension reduction in deep neural networks},\ }\href
  {https://doi.org/10.1103/PhysRevE.103.012315} {\bibfield  {journal} {\bibinfo
   {journal} {Phys. Rev. E}\ }\textbf {\bibinfo {volume} {103}},\ \bibinfo
  {pages} {012315} (\bibinfo {year} {2021})}\BibitemShut {NoStop}%
\bibitem [{\citenamefont {Roberts}\ \emph {et~al.}(2022)\citenamefont
  {Roberts}, \citenamefont {Yaida},\ and\ \citenamefont {Hanin}}]{Roberts22}%
  \BibitemOpen
  \bibfield  {author} {\bibinfo {author} {\bibfnamefont {D.~A.}\ \bibnamefont
  {Roberts}}, \bibinfo {author} {\bibfnamefont {S.}~\bibnamefont {Yaida}},\
  and\ \bibinfo {author} {\bibfnamefont {B.}~\bibnamefont {Hanin}},\ }\href
  {https://doi.org/10.1017/9781009023405} {\emph {\bibinfo {title} {The
  Principles of Deep Learning Theory}}}\ (\bibinfo  {publisher} {Cambridge
  University Press},\ \bibinfo {year} {2022})\BibitemShut {NoStop}%
\bibitem [{\citenamefont {He}\ \emph {et~al.}(2016)\citenamefont {He},
  \citenamefont {Zhang}, \citenamefont {Ren},\ and\ \citenamefont
  {Sun}}]{He16_CVPR}%
  \BibitemOpen
  \bibfield  {author} {\bibinfo {author} {\bibfnamefont {K.}~\bibnamefont
  {He}}, \bibinfo {author} {\bibfnamefont {X.}~\bibnamefont {Zhang}}, \bibinfo
  {author} {\bibfnamefont {S.}~\bibnamefont {Ren}},\ and\ \bibinfo {author}
  {\bibfnamefont {J.}~\bibnamefont {Sun}},\ }\bibfield  {title} {\bibinfo
  {title} {Deep residual learning for image recognition},\ }in\ \href
  {https://ieeexplore.ieee.org/document/7780459} {\emph {\bibinfo {booktitle}
  {Proceedings of the IEEE Conference on Computer Vision and Pattern
  Recognition (CVPR)}}}\ (\bibinfo {year} {2016})\BibitemShut {NoStop}%
\bibitem [{\citenamefont {Price}(1958)}]{Price58_69}%
  \BibitemOpen
  \bibfield  {author} {\bibinfo {author} {\bibfnamefont {R.}~\bibnamefont
  {Price}},\ }\bibfield  {title} {\bibinfo {title} {A useful theorem for
  nonlinear devices having gaussian inputs},\ }\href@noop {} {\bibfield
  {journal} {\bibinfo  {journal} {IRE Trans. Inf. Theory}\ }\textbf {\bibinfo
  {volume} {4}},\ \bibinfo {pages} {69} (\bibinfo {year} {1958})}\BibitemShut
  {NoStop}%
\bibitem [{\citenamefont {Papoulis}\ and\ \citenamefont
  {Pillai}(2002)}]{PapoulisProb4th}%
  \BibitemOpen
  \bibfield  {author} {\bibinfo {author} {\bibfnamefont {A.}~\bibnamefont
  {Papoulis}}\ and\ \bibinfo {author} {\bibfnamefont {S.~U.}\ \bibnamefont
  {Pillai}},\ }\href@noop {} {\emph {\bibinfo {title} {Probability, Random
  Variables, and Stochastic Processes}}},\ \bibinfo {edition} {4th}\ ed.\
  (\bibinfo  {publisher} {McGraw-Hill},\ \bibinfo {address} {Boston},\ \bibinfo
  {year} {2002})\BibitemShut {NoStop}%
\bibitem [{\citenamefont {Schuecker}\ \emph {et~al.}(2016)\citenamefont
  {Schuecker}, \citenamefont {Goedeke}, \citenamefont {Dahmen},\ and\
  \citenamefont {Helias}}]{Schuecker16b_arxiv}%
  \BibitemOpen
  \bibfield  {author} {\bibinfo {author} {\bibfnamefont {J.}~\bibnamefont
  {Schuecker}}, \bibinfo {author} {\bibfnamefont {S.}~\bibnamefont {Goedeke}},
  \bibinfo {author} {\bibfnamefont {D.}~\bibnamefont {Dahmen}},\ and\ \bibinfo
  {author} {\bibfnamefont {M.}~\bibnamefont {Helias}},\ }\bibfield  {title}
  {\bibinfo {title} {Functional methods for disordered neural networks},\
  }\bibfield  {journal} {\bibinfo  {journal} {ArXiv}\ }\href
  {https://doi.org/10.48550/arXiv.1605.06758} {10.48550/arXiv.1605.06758}
  (\bibinfo {year} {2016}),\ \bibinfo {note} {1605.06758
  [cond-mat.dis-nn]}\BibitemShut {NoStop}%
\bibitem [{\citenamefont {Blinnikov}\ and\ \citenamefont
  {Moessner}(1998)}]{Blinnikov98_193}%
  \BibitemOpen
  \bibfield  {author} {\bibinfo {author} {\bibfnamefont {S.}~\bibnamefont
  {Blinnikov}}\ and\ \bibinfo {author} {\bibfnamefont {R.}~\bibnamefont
  {Moessner}},\ }\bibfield  {title} {\bibinfo {title} {Expansions for nearly
  gaussian distributions},\ }\href {https://doi.org/10.1051/aas:1998221}
  {\bibfield  {journal} {\bibinfo  {journal} {Astron. Astrophys. Suppl. Ser.}\
  }\textbf {\bibinfo {volume} {130}},\ \bibinfo {pages} {193} (\bibinfo {year}
  {1998})}\BibitemShut {NoStop}%
\bibitem [{\citenamefont {Gardiner}(1985)}]{Gardiner85}%
  \BibitemOpen
  \bibfield  {author} {\bibinfo {author} {\bibfnamefont {C.~W.}\ \bibnamefont
  {Gardiner}},\ }\href@noop {} {\emph {\bibinfo {title} {Handbook of Stochastic
  Methods for Physics, Chemistry and the Natural Sciences}}},\ \bibinfo
  {edition} {2nd}\ ed.,\ \bibinfo {series} {Springer Series in Synergetics}\
  No.~\bibinfo {number} {13}\ (\bibinfo  {publisher} {Springer-Verlag},\
  \bibinfo {address} {Berlin},\ \bibinfo {year} {1985})\BibitemShut {NoStop}%
\bibitem [{\citenamefont {Molgedey}\ \emph {et~al.}(1992)\citenamefont
  {Molgedey}, \citenamefont {Schuchhardt},\ and\ \citenamefont
  {Schuster}}]{molgedey92_3717}%
  \BibitemOpen
  \bibfield  {author} {\bibinfo {author} {\bibfnamefont {L.}~\bibnamefont
  {Molgedey}}, \bibinfo {author} {\bibfnamefont {J.}~\bibnamefont
  {Schuchhardt}},\ and\ \bibinfo {author} {\bibfnamefont {H.}~\bibnamefont
  {Schuster}},\ }\bibfield  {title} {\bibinfo {title} {Suppressing chaos in
  neural networks by noise},\ }\href
  {https://doi.org/10.1103/PhysRevLett.69.3717} {\bibfield  {journal} {\bibinfo
   {journal} {Phys. Rev. Lett.}\ }\textbf {\bibinfo {volume} {69}},\ \bibinfo
  {pages} {3717} (\bibinfo {year} {1992})}\BibitemShut {NoStop}%
\bibitem [{\citenamefont {Segadlo}\ \emph {et~al.}(2022)\citenamefont
  {Segadlo}, \citenamefont {Epping}, \citenamefont {van Meegen}, \citenamefont
  {Dahmen}, \citenamefont {Kr{\"a}mer},\ and\ \citenamefont
  {Helias}}]{Segadlo22_103401}%
  \BibitemOpen
  \bibfield  {author} {\bibinfo {author} {\bibfnamefont {K.}~\bibnamefont
  {Segadlo}}, \bibinfo {author} {\bibfnamefont {B.}~\bibnamefont {Epping}},
  \bibinfo {author} {\bibfnamefont {A.}~\bibnamefont {van Meegen}}, \bibinfo
  {author} {\bibfnamefont {D.}~\bibnamefont {Dahmen}}, \bibinfo {author}
  {\bibfnamefont {M.}~\bibnamefont {Kr{\"a}mer}},\ and\ \bibinfo {author}
  {\bibfnamefont {M.}~\bibnamefont {Helias}},\ }\bibfield  {title} {\bibinfo
  {title} {Unified field theoretical approach to deep and recurrent neuronal
  networks},\ }\href@noop {} {\bibfield  {journal} {\bibinfo  {journal} {J.
  Stat. Mech. Theory Exp.}\ }\textbf {\bibinfo {volume} {2022}},\ \bibinfo
  {pages} {103401} (\bibinfo {year} {2022})}\BibitemShut {NoStop}%
\bibitem [{\citenamefont {Crisanti}\ and\ \citenamefont
  {Sompolinsky}(2018)}]{Crisanti18_062120}%
  \BibitemOpen
  \bibfield  {author} {\bibinfo {author} {\bibfnamefont {A.}~\bibnamefont
  {Crisanti}}\ and\ \bibinfo {author} {\bibfnamefont {H.}~\bibnamefont
  {Sompolinsky}},\ }\bibfield  {title} {\bibinfo {title} {Path integral
  approach to random neural networks},\ }\href
  {https://doi.org/10.1103/PhysRevE.98.062120} {\bibfield  {journal} {\bibinfo
  {journal} {Phys. Rev. E}\ }\textbf {\bibinfo {volume} {98}},\ \bibinfo
  {pages} {062120} (\bibinfo {year} {2018})}\BibitemShut {NoStop}%
\bibitem [{\citenamefont {Bordelon}\ and\ \citenamefont
  {Pehlevan}(2022)}]{Bordelon22_arxiv}%
  \BibitemOpen
  \bibfield  {author} {\bibinfo {author} {\bibfnamefont {B.}~\bibnamefont
  {Bordelon}}\ and\ \bibinfo {author} {\bibfnamefont {C.}~\bibnamefont
  {Pehlevan}},\ }\bibfield  {title} {\bibinfo {title} {Self-consistent
  dynamical field theory of kernel evolution in wide neural networks},\ }\href
  {http://arxiv.org/abs/arXiv:2205.09653} {\bibfield  {journal} {\bibinfo
  {journal} {arXiv preprint arXiv:2205.09653}\ } (\bibinfo {year}
  {2022})}\BibitemShut {NoStop}%
\end{thebibliography}%

\end{document}